\documentclass[a4paper,11pt]{article}
\pdfoutput=1 

\usepackage{jheppub} 

\usepackage[T1]{fontenc} 
\usepackage{subfig}
\usepackage{multirow}
\usepackage{amsfonts}
\usepackage{float,subfloat}
\usepackage{array}
\usepackage{mathrsfs}
\usepackage{slashed}
\usepackage{tabularx}

\usepackage{tikz-feynman}
\usepackage{tikz}
\usepackage{tkz-euclide}
\usetikzlibrary{shapes.misc}
\usetikzlibrary{decorations.pathmorphing} 
\tikzset{
	v/.style={decorate, decoration={snake, segment length=3mm, amplitude=0.75mm}, draw=ColorT},
	f/.style={draw=ColorT, postaction={decorate},
		decoration={markings,mark=at position .6 with {\arrow[very thick]{latex}}}},
	fb/.style={draw=ColorT, postaction={decorate},
		decoration={markings,mark=at position .4 with {\arrowreversed[very thick]{latex}}}},
	fnar/.style={draw=black, postaction={decorate},
		decoration={markings,mark=at position .6 with {\arrow[very thick]{latex}}}},
	g/.style={decorate, draw=ColorT,
		decoration={coil,amplitude=3pt, segment length=3.5pt}},
	s/.style={dashed,draw=blue, postaction={decorate},
		decoration={markings,mark=at position .55 with {\arrow[very thick]{latex}}}},
	sb/.style={dashed,draw=blue, postaction={decorate},
		decoration={markings,mark=at position .45 with {\arrowreversed[draw=blue,very thick]{latex}}}},
	a/.style={double distance=2pt,dashed,draw=red, postaction={decorate},
		decoration={markings,mark=at position .55 with {\arrow[very thick]{latex}}}},
	ab/.style={double distance=2pt,dashed,draw=red, postaction={decorate},
		decoration={markings,mark=at position .5 with {\arrowreversed[draw=red,very thick]{latex}}}},
	snar/.style={draw=black,line width =1.2pt},
	gen/.style={draw=ColorT,line width =0.8pt},
	cross/.style={cross out, draw=black, minimum size=2*(#1-\pgflinewidth), inner sep=0pt, outer sep=0pt},
	cross/.default={3pt},
	none/.style={draw=white}}
	
\newcommand \bfp {{\bf{p}}}

\newcommand{\be}{\begin{equation}}
\newcommand{\ee}{\end{equation}}
\newcommand{\bes}{\begin{equation*}}
\newcommand{\ees}{\end{equation*}}
\newcommand{\Eq}[1]{Eq.~\eqref{#1}}
\newcommand{\Eqs}[2]{Eqs.~\eqref{#1} and \eqref{#2}}
\newcommand{\Tab}[1]{Tab.~\ref{#1}}
\newcommand{\Sec}[1]{Sec.~\ref{#1}}
\newcommand{\App}[1]{App.~\ref{#1}}
\newcommand{\Fig}[1]{Fig.~\ref{#1}}
\newcommand{\Figs}[2]{Figs.~\ref{#1} and \ref{#2}}

\newcommand{\MeV}{\text{MeV}}
\newcommand{\GeV}{\text{GeV}}

\newcommand \p {{\prime}}

\newcommand{\lag}{\mathscr{L}}

\newcommand{\wc}{\mathcal{C}}
\newcommand{\M}{\mathcal{M}}
\newcommand{\avgM}{\overline{|\mathcal{M}|^2}}

\newcommand{\ZT}{$\mathbb{Z}_3$ }
\newcommand{\SMG}{$SU(3)_c\otimes SU(2)_L\otimes U(1)_Y$}
\newcommand{\dd}{\overleftrightarrow{\partial}}
\newcommand{\dD}{\overleftrightarrow{D}}
\newcommand{\Mpl}{M_{\rm Pl}}

\newcommand{\ttiny}[1]{\textup{\tiny{#1}}}

\newcommand{\vmol}{v_\textup{\tiny{Møl}}}
\newcommand{\algn}[1]{\begin{aligned} #1\end{aligned}}
\newcommand{\obar}[1]{\mkern 1.5mu\overline{\mkern-1.5mu#1\mkern-1.5mu}\mkern 1.5mu}
\definecolor{ColorF}{RGB}{136, 8, 8}
\definecolor{ColorT}{RGB}{20,140,10}
\definecolor{Purple}{rgb}{0.5, 0.2, 0.5}

\title{\boldmath Axion Portal to Scalar Dark Matter: \\ Unveiling Stabilizing Symmetry Footprints}

\author[a,b]{Francesco D'Eramo}
\author[a,b]{, Tommaso Sassi}
\affiliation[a]{Dipartimento di Fisica e Astronomia, Universit\`a degli Studi di Padova, \\ Via Marzolo 8, 35131 Padova, Italy}
\affiliation[b]{Istituto Nazionale di Fisica Nucleare (INFN), Sezione di Padova, \\ Via Marzolo 8, 35131 Padova, Italy}
\emailAdd{francesco.deramo@pd.infn.it}
\emailAdd{tommaso.sassi@phd.unipd.it}

\abstract{We investigate the role of an axion-like particle (ALP) as a portal between the dark and visible sectors. Unlike conventional studies, which typically assume fermionic dark matter (DM), we explore the phenomenological implications of scalar DM within this ALP portal framework. A key challenge arises from the fact that the interaction between the ALP spacetime derivative and the spin-one current of the scalar DM can be a redundant operator, which may be removed via a field redefinition. However, this interaction reveals a profound connection to the underlying global symmetry that stabilizes the DM particle. We choose a non-Abelian discrete symmetry, ensuring the persistence of the DM-ALP interactions, and in doing so, unveil a rich phenomenology. Working within a general effective field theory approach, we identify the following hallmark features of our scenario: (i) a relic density determined by semi-annihilations, with an abundance independent of ALP couplings to the visible sector; (ii) direct detection rates naturally suppressed; (iii) indirect detection spectra enriched relative to pure annihilation scenarios, with rates also independent of ALP couplings to visible particles. Lastly, we discuss potential microscopic origins for this framework and highlight the broader implications of our results.}

\begin{document} 
\maketitle
\flushbottom

\section{Introduction}

The robust observational evidence for dark matter (DM) at vastly different length scales is one of the strongest motivations supporting the need for physics beyond the Standard Model (SM)~\cite{Bertone:2004pz,Cirelli:2024ssz,Bozorgnia:2024pwk}. Understanding its composition and properties remains one of the most pressing open questions in fundamental physics.

Axion-like particles (ALPs) naturally emerge in theoretical frameworks addressing various shortcomings of the SM, including the particle physics origin of DM. They typically arise as the low-energy remnants of global symmetries that are spontaneously broken at energy scales beyond current experimental reach. A well-known example is the Peccei-Quinn symmetry~\cite{Peccei:1977np,Peccei:1977ur}, which gives rise to the QCD axion as its low-energy residual~\cite{Wilczek:1977pj,Weinberg:1977ma}. While originally introduced to solve the strong CP problem~\cite{Peccei:2006as}, this framework has the added advantage of providing a cold DM candidate without requiring additional ingredients~\cite{Preskill:1982cy,Abbott:1982af,Dine:1982ah}. Other scenarios featuring ALPs in the low-energy spectrum include models addressing the gauge hierarchy problem, such as supersymmetric theories~\cite{Nelson:1993nf,Bagger:1994hh}, as well as string theory constructions, where light ALPs are generically expected~\cite{Svrcek:2006yi,Arvanitaki:2009fg}.

In this work, we investigate the role of ALPs in DM phenomenology. Following Occam’s razor, we assume that a single new particle is sufficient to account for the observed DM abundance. This new particle is taken to be a singlet under the SM gauge group, requiring additional beyond-the-SM degrees of freedom to mediate its interactions with the visible sector. We assume that this mediating role, or \textit{portal}, is played by the ALP field itself. Examples of axion portal scenarios studied in the literature include cases where the DM particle is a Dirac or Majorana fermion~\cite{Nomura:2008ru,Gola:2021abm,Bharucha:2022aa,Ghosh:2023wi,Fitzpatrick:2023xks,Dror:2023fyd,Armando:2023zwz,Allen:2024ndv} or a spin-one vector boson~\cite{Kaneta:2016wvf,Kaneta:2017wfh}.

The main purpose of this paper is to investigate the distinct features of \textit{scalar DM} via the ALP portal. We aim to keep the discussion as general as possible and do not commit to any explicit microscopic descriptions valid at arbitrarily high energies, but rather adopt a bottom-up approach. Indeed, the relevant phenomenology does not necessitate knowledge of the high-energy dynamics that generate the ALP mass and interactions. For this reason, we focus on a low-energy effective field theory (EFT) describing ALP interactions with DM and SM fields. A comprehensive overview of the EFT for ALP interactions with SM fields can be found in Refs.~\cite{Bauer:2017ris,Bauer:2020jbp,Bauer:2021mvw,Arias-Aragon:2022aa}. Our underlying assumption is that the UV complete theory features a spontaneous breaking of a global symmetry. Consequently, the low-energy spectrum includes Nambu-Goldstone bosons, with their count matching the total number of broken generators. For simplicity, we consider only one Nambu-Goldstone $\varphi$ endowed with the shift symmetry $\varphi \rightarrow \varphi\,+\,{\rm const}$. This symmetry is assumed to be softly broken by unspecified high-energy dynamics, giving a nonzero mass $m_\varphi$. To introduce a DM candidate within our framework, we also include a massive complex scalar field $S$ in the EFT spectrum. Both $\varphi$ and $S$ are taken to be singlets under the SM gauge group \SMG.

Adding a scalar DM candidate features some subtleties and requires careful formulation. The dominant couplings between DM and the ALP are dimension 5 contact interactions. Consistent with the shift symmetry, the ALP can enter only through its spacetime derivatives, and the leading Lagrangian for the two new bosonic degrees of freedom reads
\be
\mathcal{L}_{S \varphi} =  \frac 12 (\partial_\mu\varphi)^2 + |\partial_\mu S|^2  - \frac 12 m_\varphi^2\varphi^2 + \wc_S \frac{\partial_\mu \varphi}{2 f_\varphi} S^\dagger i\dd^\mu S + \mathcal{O}\left( \frac{1}{f_\varphi^2} \right)  \ .
\label{eq:Lintro}
\ee 
The double derivative reads $S^\dagger \dd^\mu S=S^\dagger (\partial^\mu S) - (\partial^\mu S^\dagger ) S$. The dimension 5 operator is suppressed by the ALP decay constant $f_\varphi$ and is proportional to a model-dependent dimensionless Wilson coefficient $\wc_S$. This interaction has a similar structure to the well-known dimension 5 operator connecting the ALP field with the SM Higgs doublet $H$, with the only difference being the ordinary spacetime derivative acting on $S$ and $S^\dag$ replaced by a gauge covariant derivative. Famously, this operator is redundant and can be eliminated via suitable field redefinitions~\cite{Georgi:1986df}. One may wonder why this is not the same for the case in \Eq{eq:Lintro}, or in other words, one may question whether scalar DM candidates can have physical dimension 5 interactions with the ALP. This is what we need to clarify next.

A good starting point is to review how to remove the ALP coupling to the SM Higgs doublet. First, we rotate the field $H$ via an ALP-dependent phase and at the same time redefine SM fermion fields to leave the Yukawa operators invariant. Once we transform the Higgs kinetic term, we produce additional dimension 5 interactions that are equal and opposite to the ALP couplings, which are therefore removed. It is crucial to appreciate that the Higgs field enters the scalar potential only via the gauge invariant combination $H^\dag H$; the field redefinitions have no effect on potential terms. If the DM field enters the scalar potential only through the combination $S^\dag S$, the conclusion would be the same, and the dimension 5 interaction in \Eq{eq:Lintro} would be redundant. However, we are not forced into this case since gauge invariance plays no role here, with the DM field being a singlet. The discrete symmetry stabilizing the DM is what matters here, and if we manage to find an option that allows DM potential terms beyond the sole combination $S^\dag S$, then the ALP-DM interaction in \Eq{eq:Lintro} is physical.

The primary goal of this paper is to identify a class of models---captured by the same low-energy EFT---in which the dimension-5 interaction between the ALP and the DM candidate in \Eq{eq:Lintro} represents a physical effect. In other words, we seek scalar potentials $V_S(S)$ for the DM field that are not invariant under the field rephasing which would otherwise eliminate the derivative interaction. This requirement is deeply connected to the nature of the symmetry that stabilizes the DM candidate. We find that non-Abelian discrete symmetries are particularly well-suited for this purpose. At the same time, our choice must be compatible with phenomenological constraints, as both DM direct detection experiments and ALP searches place stringent bounds on the strength of ALP interactions. In this work, we focus on a model with a $\mathbb{Z}_3$ stabilizing symmetry under which $S$ is the only charged field, allowing for scalar potential terms proportional to $S^3$. While the ALP derivative interaction can still be removed through a field redefinition, this procedure induces new ALP-DM interactions in the scalar potential. As we will explore in detail throughout the paper, this setup offers several advantages: DM direct detection rates are naturally suppressed, and the DM relic abundance becomes independent of the ALP-SM couplings, provided these are large enough to maintain thermal equilibrium in the early universe.

The first consequence is that the DM relic density cannot be set by annihilations to SM final states since the matrix element for these processes is vanishing. Amplitudes for processes where the dimension 5 interaction in \Eq{eq:Lintro} contributes with both DM fields on external legs are always vanishing. We need at least one of them to be off-shell. A viable option would be to use this interaction twice to induce the DM annihilation $S S^\star \rightarrow \varphi \varphi$.\footnote{We denote the DM particle and antiparticle with $S$ and $S^\star$, respectively.} However, the amplitude for this process would be suppressed by two powers of the ALP decay constant and of the same order as contributions from dimension 6 operators. We need to go beyond the interactions in \Eq{eq:Lintro}. At the end of this paper, we will present a UV complete theory where the two contributions are equal and opposite, and therefore the amplitude for DM annihilation to ALP pairs is vanishing. All of these considerations are much simpler in the non-derivative basis, where the ALP-DM interactions are in the scalar potential. Throughout this work, we construct the EFT both in the derivative and non-derivative bases and perform all calculations for physical observable quantities in both bases. We check that the final results do not depend on the basis, as it should be.

We need to establish what sets the DM relic density. The $\mathbb{Z}_3$ symmetry allows for another reaction with amplitude suppressed by only one power of $f_\varphi$: \textit{semi-annihilation}. In general, semi-annihilations take the form $\psi_i \psi_j \rightarrow \psi_k \chi$~\cite{DEramo:2010keq}. Here, $\psi_i$ are DM particles and $\chi$ is either a SM degree of freedom or a new state that decays to SM particles. DM stability requires $m_k < m_i + m_j$ (up to $m_\chi$ corrections), and this does not pose any threat in our single-component DM scenario, where the semi-annihilation takes the form $S S \rightarrow S^\star \varphi$. The phenomenology of semi-annihilating DM has been investigated in other contexts~\cite{DEramo:2010keq,Batell:2010bp,Adulpravitchai:2011ei,Belanger:2012vp,DEramo:2012fou,Belanger:2012zr,Belanger:2014bga,Cai:2015zza,Arcadi:2017vis,Balkin:2018tma,Ghosh:2020lma,Miyagi:2022gvy,Dominguez:2024gxh}.

\begin{table}
    \centering
    \renewcommand{\arraystretch}{1.4} 
    \begin{tabular}{|c|c|c|c|c|}
        \hline
         & \textbf{Freeze-Out} & \textbf{Indirect} & \textbf{Direct} & \textbf{Collider} \\
        \hline
        \textbf{Fermion ($\mathbb{Z}_2$)}
        & \multicolumn{2}{c|}{\begin{tikzpicture}[baseline=(current bounding box.center), scale=0.7]
                \draw[fnar] (-0.75,1) node[above ] {$\chi$} -- (0,0);
                \draw[fnar] (0,0) -- (-0.75,-1) node[below ] {$\bar\chi$};
                \draw[a] (0,0) -- (1.25,0) node[midway,below] {$\color{red}{\varphi}$};
                \draw[gen] (1.25,0) -- (2,-1) node[below ] {\color{ColorT}{SM}};
                \draw[gen] (1.25,0) -- (2,1) node[above ] {\color{ColorT}{SM}};
            \end{tikzpicture}}
            &\begin{tikzpicture}[baseline=(current bounding box.center), scale=0.7]
         \draw[fnar](-1.,0.75) node[above]{$\chi$}--(0,0)[baseline];
         \draw[fnar] (0,0) -- (1,0.75)node[above]{$\bar\chi$};
         \draw[a] (0,0) -- (0,-1)node[midway,left]{$\color{red}{\varphi}$};
         \draw[gen] (0,-1) -- (1,-1.75)node[below]{\color{ColorT}{SM}};
         \draw[gen] (-1,-1.75)node[below]{\color{ColorT}{SM}} -- (0,-1);
        \end{tikzpicture} & 
        \begin{tikzpicture}[baseline=(current bounding box.center), scale=0.7]
            \draw[gen] (-0.75,1) node[above ] {\color{ColorT}{SM}} -- (0,0);
            \draw[gen] (0,0) -- (-0.75,-1) node[below ] {\color{ColorT}{SM}};
            \draw[a] (0,0) -- (1.25,0) node[midway,below] {$\color{red}{\varphi}$};
            \draw[fnar] (1.25,0) -- (2,-1) node[below] {$\chi$};
            \draw[fnar]  (2,1)node[above ] {$\bar\chi$}--(1.25,0)  ;
        \end{tikzpicture} \\
        \hline
        \textbf{Scalar ($\mathbb{Z}_3$)} & \begin{tikzpicture}[baseline=(current bounding box.center), scale=0.7]
         \draw[s](-1.,1.25) node[above]{$\color{blue}{S}$}--(0,0);
         \draw[a] (0,0) -- (1,-1.25)node[below ]{$\color{red}{\varphi}$};
         \draw[sb] (0,0) -- (1,1.25)node[above]{$\color{blue}{S^\star}$};
         \draw[s] (-1,-1.25)node[below]{$\color{blue}{S}$} -- (0,0);
          \end{tikzpicture} 
          &\begin{tikzpicture}[baseline=(current bounding box.center), scale=0.6]
         \draw[s](-1,0.75) node[above ]{$\color{blue}{S}$}--(0,0);
        \draw[a] (0,0) -- (1,-0.75)node[below ]{$\color{red}{\varphi}$};
         \draw[sb] (0,0) -- (1,0.75)node[above ]{$\color{blue}{S^\star}$};
         \draw[s] (-1,-0.75)node[below ]{$\color{blue}{S}$} -- (0,0);
	 \draw[a] (-1.25,-4.25)node[below]{$\color{red}{\varphi}$}--(0,-4.25);
         \draw[gen] (0,-4.25) -- (1,-3.5)node[above]{\color{ColorT}{SM}};
         \draw[gen] (0,-4.25)-- (1,-5)node[below ]{\color{ColorT}{SM}} ;
         \path (0,-2.15) node[] {\color{red}{\footnotesize followed by}}; 
          \end{tikzpicture} 
          &\begin{tikzpicture}[baseline=(current bounding box.center), scale=0.7]
         \draw[s](-1.,1) node[above]{$\color{blue}{\footnotesize \text{boosted} \, S}$}--(0,0)[baseline];
         \draw[sb] (0,0) -- (1,1)node[above right]{$\color{blue}{S^\star}$};
         \draw[sb] (0,0) -- (1,0.5)node[right]{$\color{blue}{S^\star}$};
         \draw[a] (0,0) -- (0,-1)node[midway,left]{$\color{red}{\varphi}$};
         \draw[gen] (0,-1) -- (1,-1.75)node[below]{\color{ColorT}{SM}};
         \draw[gen] (-1,-1.75)node[below]{\color{ColorT}{SM}} -- (0,-1);
        \end{tikzpicture}  &
         \begin{tikzpicture}[baseline=(current bounding box.center), scale=0.7]
         \draw[a] (0,0) -- (1.25,0) node[midway,below] {$\color{red}{\varphi}$};
         \draw[gen](-0.75,1) node[above ] {\color{ColorT}{SM}} -- (0,0);
         \draw[gen](0,0) -- (-0.75,-1) node[below ] {\color{ColorT}{SM}};
         \draw[sb]  (2.25,0)node[right]{$\color{blue}{S}$}-- (1.25,0) ;
         \draw[sb]  (2.25,-1)node[below]{$\color{blue}{S}$}-- (1.25,0) ;
         \draw[sb]  (2.25,1)node[above]{$\color{blue}{S}$}-- (1.25,0) ;
         \end{tikzpicture}\\
        \hline
    \end{tabular}
    \caption{Phenomenology for ALP portal scenarios where the DM particle is a fermion or a scalar. The former has been studied in the literature, the latter is the focus of this work.}
    \label{tab:pheno}
\end{table}

We summarize the key features of our framework in \Tab{tab:pheno}. To facilitate comparison with the existing literature, the first row of the table outlines the well-established case of fermion DM stabilized by a $\mathbb{Z}_2$ symmetry. This scenario exhibits all the hallmark characteristics of WIMP-like DM. Thermal production via freeze-out and indirect detection (ID) signals are governed by DM annihilations into SM particles, with rates proportional to the product of ALP couplings to DM and SM fields. Direct detection (DD), though potentially suppressed by the Lorentz structure of the interaction, can be understood as a ninety-degree rotation of the annihilation diagram, leading to elastic scattering. Similarly, DM production at colliders corresponds to the inverse process of DM annihilation.

The second row of \Tab{tab:pheno} clearly highlights several key differences in our scenario. If the ALP is lighter than the DM particle and sufficiently coupled to the primordial bath to remain in thermal equilibrium, DM freeze-out is determined by the semi-annihilation process $S S \rightarrow S^\star \varphi$. Notably, the resulting DM relic density is independent of ALP couplings to the visible sector.\footnote{DM freeze-out through semi-annihilations can occur even if the dark sector is secluded from the visible sector, as long as it remains in thermal equilibrium with itself. The temperature of the dark sector is an additional independent parameter that must be specified. While we do not explore this option in this work, it would necessitate the assumption that inflationary reheating also populates the dark sector.} The same holds for the overall indirect detection rate: the number of DM semi-annihilation events per unit time today remains unaffected by ALP interactions with SM fields. However, the produced ALPs eventually decay, and their couplings to SM particles dictate the branching ratios, ultimately shaping the spectral features of indirect detection signals. Semi-annihilating DM naturally evades direct detection constraints. This is most transparently understood in the non-derivative basis for ALP interactions, where it becomes evident that $\varphi$-mediated interactions always involve three DM fields. Consequently, the conventional DM-nucleus scattering is kinematically forbidden for non-relativistic DM. However, such interactions could still be relevant if a subdominant DM population is sufficiently boosted~\cite{Berger:2014sqa,Berger:2019ttc,Toma:2021vlw,Aoki:2023tlb,Kamenetskaia:2025aa} to enable a $2 \to 3$ scattering event. Finally, DM production at colliders necessarily involves three DM particles in the final state, and specialized kinematic variables have been proposed to distinguish this scenario~\cite{Agashe:2010tu}.

The remaining part of this article is structured as follows. After developing the EFT in Sec.~\ref{sec:EFT}, we study the phenomenology. First, in Sec.~\ref{sec:relic}, we identify the couplings and masses that allow for the correct relic density via semi-annihilations. This analysis assumes that the ALP field $\varphi$ remains in equilibrium during freeze-out. In Sec.~\ref{sec:ALP}, we examine the validity of this assumption by identifying the required coupling strengths. Additionally, we ensure that $\varphi$ decays before the universe is one second old, preventing any disruption to standard cosmological history. Detecting the remnants of DM semi-annihilations is the most promising observational signal. Unlike conventional DM annihilation scenarios, the indirect detection spectra from the semi-annihilation process are expected to exhibit richer spectral features. In Sec.~\ref{sec:gammaray}, we focus on gamma-ray signals and present the expected spectra for different ALP scenarios. In Sec.~\ref{sec:UV}, we provide an explicit UV completion that gives rise to the broad framework explored in this work. Finally, we summarize our conclusions and discuss potential future directions in Sec.~\ref{sec:final}. A detailed explanation of our notation and conventions, along with all computational details, is provided in the appendices.

\section{ALP effective interactions}
\label{sec:EFT}
With the assumptions stated in the introduction, we pursue a bottom-up strategy and construct the EFT describing the interactions between the two new scalars and the SM. The low-energy Lagrangian can be written in the schematic form
\be
\lag_{\rm EFT} = \lag_{\rm SM} + \frac 12 (\partial_\mu\varphi)^2 + |\partial_\mu S|^2  - \frac 12 m_\varphi^2\varphi^2  + V_S(S) + V_{\rm mix}(H,S) + \mathcal{L}_{\rm INT} \ .
\label{eq:EFT}
\ee
The first term on the right-hand side corresponds to the SM Lagrangian. We include canonically normalized kinetic terms for the two new scalars. The mass term for the portal field $\varphi$ is highlighted to reflect its origin. Renormalizable potential terms beyond the SM include $V_S(S)$, which involves only the scalar $S$ and depends on the symmetry stabilizing the DM field, and $V_{\rm mix}(H,S)$, which contains the mixing terms with the Higgs doublet $H$. The latter are tightly constrained by DM searches~\cite{Arcadi:2017kky,Arcadi:2024ukq} and are assumed to be negligible in this work. Finally, higher-dimensional operators suppressed by powers of the ALP decay constant $f_\varphi$ are included in $\mathcal{L}_{\rm INT}$. The ALP decay constant is related to the EFT cutoff scale via $\Lambda_{\rm UV} \simeq 4 \pi f_\varphi$. At energies higher than this scale, the EFT in Eq.~\eqref{eq:EFT} is no longer valid to describe the underlying physics. We retain only dimension 5 operators and therefore the only interactions compatible with the imposed constraints are the ones connecting $\varphi$ separately to DM or SM fields.\footnote{The lowest dimensional contact interactions between DM and SM (other than the Higgs portal, which is neglected here) arise at dimension 6, and are therefore suppressed and beyond the scope of this work.}

The next two subsections present the explicit expressions for the dimension 5 operators. We assume the EFT cutoff is much larger than the weak scale, and provide the interactions in the electroweak unbroken phase. In \Sec{sec:EFTder}, we adopt a field basis where ALP interactions are invariant under the shift symmetry $\varphi \rightarrow \varphi  +  {\rm const}$. In \Sec{sec:EFTNder}, we show how to eliminate the derivative couplings through field redefinitions. These redefinitions also help identify the number of independent physical couplings, as discussed in \Sec{sec:EFTphys}. In \Sec{sec:DMstability}, we address the issue of the DM-stabilizing symmetry and provide the explicit expression for $V_S(S)$ respecting a $\mathbb{Z}_3$ symmetry. We also explain how the field redefinitions affect the DM potential terms in $V_S(S)$. We conclude in \Sec{sec:potential} with a list of constraints that need to be satisfied by the couplings entering the full scalar potential of the theory.

\subsection{Derivative basis}
\label{sec:EFTder}

The ALP interactions consistent with the shift symmetry take the schematic form
\be
\mathcal{L}^{(\partial)}_{\rm INT} =  \frac{\varphi}{8 \pi f_\varphi} \sum_V \wc_V \alpha_V  V_{\mu\nu} \widetilde V^{\mu\nu} + \frac{\partial_\mu \varphi}{2 f_\varphi} \left( \mathcal{S}_\varphi^\mu + \mathcal{F}_\varphi^\mu \right) + \mathcal{O}(1/f_\varphi^2) \ .
\label{eq:Lag}
\ee
Here, the superscript $(\partial)$ specifies that we provide the interactions in the derivative basis. The sum runs over the SM gauge bosons $V = (G, W, B)$; $V_{\mu\nu}$ is the corresponding field strength, $\widetilde V^{\mu\nu} \equiv \epsilon^{\mu\nu\rho\sigma} V_{\rho\sigma} / 2$ is its dual, and $\alpha_V = g_V^2 / (4 \pi)$ is the fine structure constant of the gauge group. Adjoint indices (for non-Abelian groups) are implicitly summed over. The spin-one current coupled to the ALP spacetime derivative gets contributions from scalar and fermion fields. The former contains $S$ and the Higgs doublet
\be
\mathcal{S}_\varphi^\mu = \wc_S S^\dagger i\dd^\mu S + \wc_H H^\dagger i\dD^\mu H \ .
\label{eq:JmuS}
\ee
We focus on scenarios with flavor conserving fermion currents in the field basis where the SM Yukawa matrices are diagonal. The spin-one fermion current $\mathcal{F}_\varphi^\mu$ explicitly reads
\be
\mathcal{F}_\varphi^\mu = \sum_{j=1}^3 \left[ \wc^j_{Q} \, \bar Q_L^j \gamma^\mu Q_L^j +
\wc^j_{u} \, \bar u_R^j \gamma^\mu u_R^j + \wc^j_{d} \, \bar d_R^j \gamma^\mu d_R^j+
\wc^j_{E} \, \bar E_L^j \gamma^\mu E_L^j + \wc^j_{e} \,  \bar e_R^j \gamma^\mu e_R^j
 \right] \ .
\label{eq:JmuF}
\ee
Here, the fermion fields have well-defined electroweak quantum numbers, and the index $j$ runs over the three generations. 

The dimension 5 operators in Eq.~\eqref{eq:Lag} involve 20 unknown Wilson coefficients: $3$ for the gauge bosons ($\wc_V$), $2$ for the scalars ($\wc_S$ and $\wc_H$), and $5 \times 3 = 15$ for the fermions ($\wc_\psi$). However,  not all of these coefficients are physical, as it is manifest in a different field basis.

\subsection{Non-derivative basis}
\label{sec:EFTNder}

It is possible to remove all the ALP derivative interactions via appropriate field redefinitions. The ones with scalars can be eliminated via
\be
S \rightarrow \exp\left[ i \, \wc_S \frac{\varphi}{2 f_\varphi} \right] S \ , \qquad \qquad 
H \rightarrow \exp\left[ i \, \wc_H \frac{\varphi}{2 f_\varphi} \right] H \ .
\label{eq:scalarrot}
\ee
Likewise, derivative couplings with fermions can be eliminated via chiral rotations. The specific transformations for the fermion fields are given by
\be
\begin{split}
Q_L^j \rightarrow & \, \exp\left[ i \, \wc_Q^j \frac{\varphi}{2 f_\varphi} \right] Q_L^j \ , \quad 
u_R^j \rightarrow \exp\left[ i \, \wc_u^j \frac{\varphi}{2 f_\varphi} \right] u_R^j \ , \quad
d_R^j \rightarrow \exp\left[ i \, \wc_d^j \frac{\varphi}{2 f_\varphi} \right] d_R^j \ , \\ 
E_L^j \rightarrow & \, \exp\left[ i \, \wc_E^j \frac{\varphi}{2 f_\varphi} \right] E_L^j \ , \quad 
e_R^j \rightarrow \exp\left[ i \, \wc_e^j \frac{\varphi}{2 f_\varphi} \right] e_R^j \ .
\end{split}
\label{eq:chiralrot}
\ee
Eliminating the derivative couplings is not the sole consequence of these field redefinitions. First, they modify the entire Lagrangian in Eq.~\eqref{eq:EFT} at the classical level, affecting not just the new fields but also the SM sector. In particular, the Yukawa operators are not invariant under these transformations, resulting in a modified set of dimension 5 ALP operators. Moreover, fermion chiral rotations also induce changes in the ALP interactions with gauge bosons since they are anomalous. The resulting Lagrangian after these field redefinitions take the schematic form 
\be
\mathcal{L}^{({\rm N}\partial)}_{\rm INT} = \frac{\varphi}{2 f_\varphi} \mathcal{P}_\varphi + \frac{\varphi}{8 \pi f_\varphi} \sum_V \wc^\prime_V \alpha_V  V_{\mu\nu}\tilde V^{\mu\nu} +  \mathcal{O}(1/f_\varphi^2) \ .
\label{eq:Lag2}
\ee
Here, the superscript $({\rm N}\partial)$ indicates that the interactions are expressed in the non-derivative basis. Derivative interactions with SM fermions are replaced by pseudo-Yukawa operators. The fermion current in Eq.~\eqref{eq:JmuF} is expressed in the field basis where the SM Yukawa operators are diagonal in flavor space, and therefore the ALP interactions with fermions in Eq.~\eqref{eq:Lag2} remain flavor-conserving and are proportional to the spin-zero current term
\be
\begin{split}
\mathcal{P}_\varphi = & \, - i \sum_{j = 1}^3 \left[ (\wc_u^j - \wc_Q^j - \wc_H )y_u^j \, \bar Q^j_L \tilde{H} u^j_R + 
(\wc_d^j - \wc_Q^j + \wc_H ) y_d^j \,\bar Q_L^j H d_R^j + \right. \\ & \left.  \qquad\qquad
(\wc_e^j - \wc_E^j + \wc_H ) y_e^j \,\bar E_L^j H e_R^j  \right] + {\rm h.c.} \ .
\end{split}
\label{eq:pY}
\ee
Here, $y_\psi^j$ is the diagonal element of the Yukawa matrix for the $j$-th generation of the fermion $\psi$. Gauge anomalies modify the ALP couplings to gauge bosons as follows\footnote{For a chiral rotation $\chi_L \rightarrow \exp[i \, \omega] \chi_L$ of a left-handed Weyl field $\chi_L$ charged under the representation $\mathcal{R}$ of the gauge group $\mathcal{G}$, the anomalous effect is to induce the operator $- \omega \alpha_\mathcal{G} T_\mathcal{R}/(4 \pi) \; \mathcal{G}_{\mu\nu} \tilde{\mathcal{G}^{\mu\nu}} $ with $T_\mathcal{R}$ the Dynkin index. The result is the same for a right-handed Weyl fermion up to an overall minus sign.} 
\begin{subequations}
\begin{align}
\wc^\prime_G = & \, \wc_G - \sum_{j=1}^3 \left(\wc_Q^j - \frac{1}{2} \wc_u^j - \frac{1}{2} \wc_d^j  \right) \ , \\
\wc^\prime_W = & \, \wc_W - \sum_{j=1}^3 \left(\frac{3}{2} \, \wc_Q^j + \frac{1}{2} \wc_E^j \right)    \ , \\
\wc^\prime_B = & \, \wc_B - \sum_{j=1}^3 \left( \frac{1}{6} \wc_Q^j + \frac{1}{2} \wc_E^j  - \frac{4}{3} \wc_u^j - \frac{1}{3} \wc_d^j - \wc_e^j \right)   \ .
\end{align}
\label{eq:wCVshift}
\end{subequations}
Finally, and crucially for the subsequent analysis, the DM scalar potential takes the form 
\be
V_S^{({\rm N}\partial)}(S) = V_S^{(\partial)}\left(e^{i \, \wc_S \frac{\varphi}{2 f_\varphi}} S\right) \ .
\label{eq:VStr}
\ee

\subsection{How many ALP interactions are physical?}
\label{sec:EFTphys}

The field redefinitions given in \Eqs{eq:scalarrot}{eq:chiralrot} are useful to identify the number of independent ALP couplings. Starting from the derivative basis, where we identified $20$ Wilson coefficients, we will proceed to eliminate as many as possible to isolate a set of independent operators.

The first coupling we get rid of is $\wc_H$ via the redefinition in Eq.~\eqref{eq:scalarrot}. This operation requires special attention because it also induces pseudo-Yukawa interactions for the ALP proportional to $\wc_H$ (see Eq.~\eqref{eq:pY}). However, this can be avoided at the price of rotating also the fermion fields to compensate this effect. The anomalous fermion rotations would then shift the couplings to gauge bosons similarly to Eq.~\eqref{eq:wCVshift} with the appropriate coefficients. Thus $\wc_H$ can be effectively removed while remaining in the derivative basis, with this operation simply redefining the coefficients $\wc_V$. 

The same argument could be applied to the Wilson coefficient $\wc_S$. Things look even simpler here because the DM field is not directly coupled to fermions and there is no need to redefine the coefficients $\wc_V$. The phenomenological consequences of removing $\wc_S$ could be significant as it would remove all DM dimension 5 interactions with the portal field. That said, just as the rotation of the Higgs field affects the SM Yukawa interactions (and no other terms, since it appears everywhere else as the gauge-invariant combination $|H|^2$), the rotation of $S$ can modify the scalar potential $V_S(S)$ as described by Eq.~\eqref{eq:VStr}. We defer a more detailed discussion to the next subsection and keep $\wc_S$ as a physical parameter. 

The physical interactions of SM fermions are most easily identified in the non-derivative basis. We begin with $15$ Wilson coefficients in the derivative basis. However, we notice how only $3 \times 3 = 9$ linear combinations of them appear in Eq.~\eqref{eq:pY}. This can also be derived by rewriting the spin-one current in Eq.~\eqref{eq:JmuF} as a combination of vector and axial-vector currents for Dirac fields. Still working under the assumption of flavor-diagonal couplings in the mass eigenbasis, we neglect conserved vector currents (up to anomaly corrections that are not phenomenologically relevant to our analysis) and we have  
\be
\mathcal{L}_{\rm INT} \supset \frac{\partial_\mu \varphi}{2 f_\varphi}  \left[ \sum_q \wc_q \, \bar q \gamma^\mu \gamma^5 q + \sum_\ell \wc_\ell \,  \bar \ell \gamma^\mu \gamma^5 \ell \right] \ .
\label{eq:JmuDirac}
\ee
The Wilson coefficients for the quark fields $q = \{u, d, c, s, t, b\}$ explicitly read
\begin{subequations}
\begin{align}
\{\wc_u, \wc_c, \wc_t \} = & \, \left\{ \frac{- \wc^1_{Q} + \wc^1_{u}}{2}, \frac{- \wc^2_{Q} + \wc^2_{u}}{2}, \frac{- \wc^3_{Q} + \wc^3_{u}}{2} \right\} \ , \\
\{\wc_d, \wc_s, \wc_b \} = & \, \left\{ \frac{- \wc^1_{Q} + \wc^1_{d}}{2}, \frac{- \wc^2_{Q} + \wc^2_{d}}{2}, \frac{- \wc^3_{Q} + \wc^3_{d}}{2} \right\} \ .
\end{align}
\end{subequations}
Likewise, the sum over charged leptons run over $\ell = \{e, \mu, \tau\}$ with couplings
\be
\{\wc_e, \wc_\mu, \wc_\tau \} = \left\{ \frac{- \wc^1_{E} + \wc^1_{e}}{2}, \frac{- \wc^2_{E} + \wc^2_{e}}{2}, \frac{- \wc^3_{E} + \wc^3_{e}}{2} \right\} \ .
\ee

In summary, we have identified $13$ independent dimensionless Wilson coefficients that describe dimension $5$ contact interactions of the ALP portal field. These include: one coupling with DM ($\wc_S$), $6$ with quarks ($\wc_q$), $3$ with leptons ($\wc_\ell$), and $3$ with gauge bosons ($\wc_V$). The remaining low-energy couplings are contained inside the scalar potential $V_S(S)$, whose explicit form depends on the symmetry stabilizing the DM field—a topic that we will address in the next subsection.

\subsection{On the dark matter stabilizing symmetry}
\label{sec:DMstability}

A crucial requirement for our framework is that the DM field must remain stable on cosmological scales. This requires the presence of a stabilizing symmetry in the theory that must be respected both by the scalar potential $V_S(S)$ and the DM interactions with the ALP contained in $\mathcal{L}_{\rm INT}$. In addition to ensuring stability, it is important to make sure that the couplings between $S$ and $\varphi$ are physical and cannot be removed through suitable field redefinitions. We have just seen how the interactions between spin-one currents of scalar fields and the ALP spacetime derivative can be redefined away. The only possibility for a surviving interaction between $S$ and $\varphi$ is hidden in \Eq{eq:VStr}. Specifically, we do not want $V_S(S)$ to be invariant under \Eq{eq:scalarrot}. This is a non-trivial condition, and a simple counterexample arises when the DM field enters the scalar potential solely through the combination $S^\dag S$. If this is the case, we observe that the interactions between DM and ALP are not physical, making this scenario not phenomenologically viable unless dimension 6 operators are included in the analysis.

In this study, we focus on a phenomenology governed by dimension $5$ operators, and we aim to identify symmetries that allow for such interactions. If the scalar potential contains operators that scale as $S^n$ (with $n$ a positive integer) the field redefinition in \Eq{eq:scalarrot} does not leave the potential invariant. We choose for this study the case $n =3$. In particular, we take the DM field charged under a \ZT symmetry, transforming as $S\to e^{2i\pi/3}S$. SM fields and the ALP are assumed $\mathbb{Z}_3$-neutral. Being the only $\mathbb{Z}_3$-charged field, $S$ is stable. The most general renormalizable scalar potential satisfying these requirements reads
\be
V_S^{(\partial)}(S) = m_S^2 S^\dag S + \frac{1}{3!}(A \, S^3 + A^* S^{\dagger\,3}) +  \frac{\lambda_S}{4} (S^\dag S)^2  \ .
\label{eq:potentialZ3}
\ee
This potential includes a mass term and a quartic coupling, both dependent on the combination $S^\dag S$, as well as a cubic coupling proportional to a new complex and dimensionful parameter $A$. This form of the potential is intended to be in the derivative basis, as emphasized by our notation.

It is instructive to determine the scalar potential in the non-derivative basis via \Eq{eq:scalarrot}. The mass and quartic terms are left invariant by this operation, whereas the cubic coupling is affected and acquires an explicit dependence on the ALP. This operation generates DM interactions with an arbitrary number of $\varphi$ fields with the cost of producing each of them suppressed by an inverse ALP decay constant in the transition amplitude. Here, we keep only operators with a single ALP field and this induces the quartic interaction
\be
V_S^{(N \partial)}(S)  \supset \frac{1}{3!} \left(\lambda_{S\varphi} S^3 + \lambda_{S\varphi}^* S^{\dagger\,3} \right) \varphi \ , \qquad \qquad 
\lambda_{S\varphi} \equiv i \frac{3}{2} \frac{\wc_S A}{f_\varphi} \ .
\label{eq:VND}
\ee
This dimensionless coupling $\lambda_{S\varphi}$ arises from the ratio between two dimensionful parameters: the $\mathbb{Z}_3$-preserving potential coupling $A$ and the ALP decay constant $f_\varphi$.

\subsection{Constraints on the scalar potential}
\label{sec:potential}

The quartic coupling $\lambda_{S\varphi}$ plays a crucial role in the phenomenology of the model, as it governs the transition amplitude for DM semi-annihilations. Consequently, it is important to assess how large this coupling can be without leading to any significant issues. The strength of $\lambda_{S\varphi}$ is proportional to the dimensionful parameter $A$, which is subject to specific constraints, as we will discuss in this subsection. 

We work in the derivative basis where the ALP does not appear in the scalar potential. As a result, the total scalar potential is the sum of three distinct terms
\be
V(H,S) = V_{\rm SM}(H) + V_S(S) + V_{\rm mix}(H,S)  \ .
\ee
We have already assumed that all mixing terms are negligible, and thus we ignore $V_{\rm mix}(H,S)$ in this discussion. The two other contributions, $V_{\rm SM}(H)$ and $V_S(S)$, evolve independently. The former depends solely on the Higgs doublet field and remains unchanged compared to the pure SM case, while the latter is the focus of our analysis. We require the condition that $V_S(S)$ is bounded from below, and this is true only if $\lambda_S > 0$. Moreover, it is always possible to redefine $S$ via a global rotation that leaves all the terms unaffected and makes $A$ real and positive. From now on, we assume $A^* = A > 0$ without losing any generality. 

A key requirement is that the vacuum state does not spontaneously break the \ZT symmetry, as this would destabilize the DM particle. We require the global minimum to occur at the field configuration where the vacuum expectation value (vev) of the complex scalar field vanishes, i.e., $\langle S \rangle = 0$. To find the minima of $V_S(S)$, we rewrite the field in polar coordinates, $S = s e^{i \theta}$, and derive the conditions for the stationary points
\begin{subequations}
\begin{align}
\label{eq:min1} s \left(\lambda_S s^2 + A \, s \cos(3 \theta) + 2 m_S^2  \right) = & \, 0 \ , \\
\label{eq:min2} A \, s^3 \sin(3 \theta) = & \, 0 \ .
\end{align}
\end{subequations}
One straightforward solution corresponds to the $\mathbb{Z}_3$-preserving field configuration, $\langle s \rangle = 0$. If this is not the case (i.e., $\langle s \rangle \neq 0$), we must satisfy the condition $\sin(3 \langle \theta \rangle) = 0$. The combination of Eq.~\eqref{eq:min1} with the assumption $A > 0$ implies $\cos(3 \langle \theta \rangle) = -1$. Consequently, we obtain the solutions
\begin{subequations}
\begin{align}
\label{eq:min1sol} \langle s \rangle_{\pm} = & \, \frac{A \pm \sqrt{A^2 - 8 \lambda_S m_S^2}}{2 \lambda_S}  \ , \\
\label{eq:min2sol} \langle \theta \rangle = & \, \left\{ \frac{\pi}{3}, \pi, \frac{5\pi}{3} \right\}  \ .
\end{align}
\end{subequations}
The solutions for the vevs of $s$ are real and positive as long as $A^2 \geq 8 \lambda_S m_S^2$. Upon examining the second derivatives of the potential, we find that $\langle s \rangle_{+}$ corresponds to a local minimum, while $\langle s \rangle_{-}$ corresponds to a local maximum (both threefold degenerate). The local minimum, if allowed, is particularly concerning because we do not want it to be also the global minimum. Thus we impose that the potential energy in the $\mathbb{Z}_3$-breaking vacuum states is larger than the one of the $\mathbb{Z}_3$-preserving vacuum, and we find the condition
\be
A \left( A + \sqrt{A^2 - 8 \lambda_S m_S^2 } \right) < 12 \lambda_S m_S^2 \ .
\ee
This inequality implies 
\be
A < A_{\rm max} = 3 \sqrt{\lambda_S} \, m_S\,\,.
\label{eq:Amax}
\ee

To summarize, small values of $A$ are harmless because the $\mathbb{Z}_3$-breaking vacuum cannot exist if $A \leq 2 \sqrt{2} \sqrt{\lambda_S} \, m_S$. As $A$ increases, we remain safe as long as $A \leq 3 \sqrt{\lambda_S} \, m_S$, since the $\mathbb{Z}_3$-breaking vacuum is a local, not global, minimum. Notably, the two inequalities are almost identical. In other words, once $A$ is large enough to permit a $\mathbb{Z}_3$-breaking minimum, this minimum becomes the global one for the theory. This inequality can be relaxed as long as we ensure that the $\mathbb{Z}_3$-preserving vacuum is a local metastable minimum with a lifetime longer than the age of the universe~\cite{Belanger:2012zr,Belanger:2012vp}.

The upper bound we found for $A$ allows us to constrain the quartic coupling $\lambda_{S\varphi}$
\be
\left|\lambda_{S\varphi}\right| < \lambda_{S\varphi}^{\rm max} = \frac{3}{2} \frac{\wc_S A_{\rm max}}{f_\varphi} = 
\frac{9}{2}\,\frac{\wc_S}{f_\varphi}  \, \sqrt{\lambda_S} \, m_S \ .
\label{eq:lambda_max}
\ee
The EFT description breaks down when the propagating degrees of freedom become too heavy and approach the cutoff scale $\Lambda_{\rm UV} \simeq 4 \pi f_\varphi$. We thus observe that, in the non-derivative basis, the DM interaction strength is constrained by the hierarchy between the ALP decay constant and the DM mass. 

\section{Dark matter production via semi-annihilations}
\label{sec:relic}

We examine DM production in the early universe, focusing on the parameter space where the relic density is determined by the thermal freeze-out mechanism. For this to occur, portal interactions must be strong enough to ensure thermal equilibrium between the primordial bath and $S$. While thermalization of the ALP is not strictly required for the freeze-out process—since DM can annihilate into visible particles via $\varphi$ virtual exchanges—the ALP remains the primary communication channel. Here, we assume that $\varphi$ is in equilibrium at the time of freeze-out and check the validity of this assumption in the next section when discussing specific scenarios for ALP couplings.

Departure from equilibrium occurs when the interaction rate falls below the Hubble expansion rate, causing densities to dilute and the universe to cool. Afterward, DM particles just free-stream. To investigate the decoupling epoch, we use the Boltzmann equation for the DM number density $n_S$, which takes the general form
\be
\frac{d n_S}{dt} + 3 H n_S = \sum_\alpha \mathcal{C}_\alpha \ .
\label{eq:genBE}
\ee
The evolution of the number density is driven by two factors. First, the expansion of the universe causes a dilution of particles in a fixed comoving volume. This effect is proportional to the Hubble parameter $H$, which scales with the bath temperature $T$ as follows
\be
H(T) = \frac{\pi \sqrt{g_*(T)}}{3 \sqrt{10}} \frac{T^2}{ \Mpl} \ . 
\ee 
We adopt for the effective number of degrees of freedom $g_*(T)$ contributing to the bath energy density the results of Ref.~\cite{Laine:2015kra}. Second, the number of DM particles evolves also because there are number changing processes. The sum on the right-hand side of Eq.~\eqref{eq:genBE} runs over all processes $\alpha$ (e.g., decays and scatterings) that affect the number of $S$ particles. It is important to note that Eq.~\eqref{eq:genBE} tracks the number density of DM \textit{particles} $S$. An analogous equation tracks the number density of DM \textit{antiparticles} $S^\star$. The collision terms $\mathcal{C}_\alpha$ account for the variation of the number of DM particles only, as discussed in detail in App.~\ref{app:coll}. We assume there is no significant matter/antimatter asymmetry in the dark sector (i.e., $n_S = n_{S^\star}$), so it is sufficient to track $n_S$. 

The \ZT symmetry, under which DM is the only charged field, allows semi-annihilations $SS \to S^\star \varphi$. Other allowed processes have their rates suppressed by additional powers of $f_\varphi$, as shown in \App{app:DMsupp}. Consequently, semi-annihilations are the only processes that contribute significantly to the freeze-out. The corresponding collision operator is derived in App.~\ref{app:coll}, and it reads
\be
\mathcal{C}_{SS\to S^\star\varphi} =  - \langle \sigma_{SS\to  S^\star \varphi}  \vmol \rangle \left[ n_S^2  - n_S \, n_S^{\rm eq} \right] \ .
\ee
The quantity on the right-hand side is the thermal average of the semi-annihilation cross section times the Møller velocity. Its evaluation requires an integral over the semi-annihilation cross section over the center of mass energy (see Eq.~\eqref{eq:thav}). Given the non-relativistic nature of freeze-out, the leading effects are captured by the s-wave expansion
\be
\langle \sigma_{SS\to  S^\star \varphi}  \vmol \rangle \simeq \frac{\lambda_{S\varphi}^2}{128 \pi \, m_S^2} \sqrt{9 - 10 \left(\frac{m_{\varphi}}{m_S}\right)^2 + \left(\frac{m_{\varphi}}{m_S}\right)^4} + \mathcal{O}\left( \frac{T}{m_S }\right) \ .
\label{eq:sigmasemiswave}
\ee
The above equation serves to illustrate the dependence of the relic density on key quantities. In our analysis, we use the full thermal average accounting for all partial waves. 

\begin{figure}
\centering
\subfloat[\label{fig:yield}]{\includegraphics[width=0.48\textwidth]{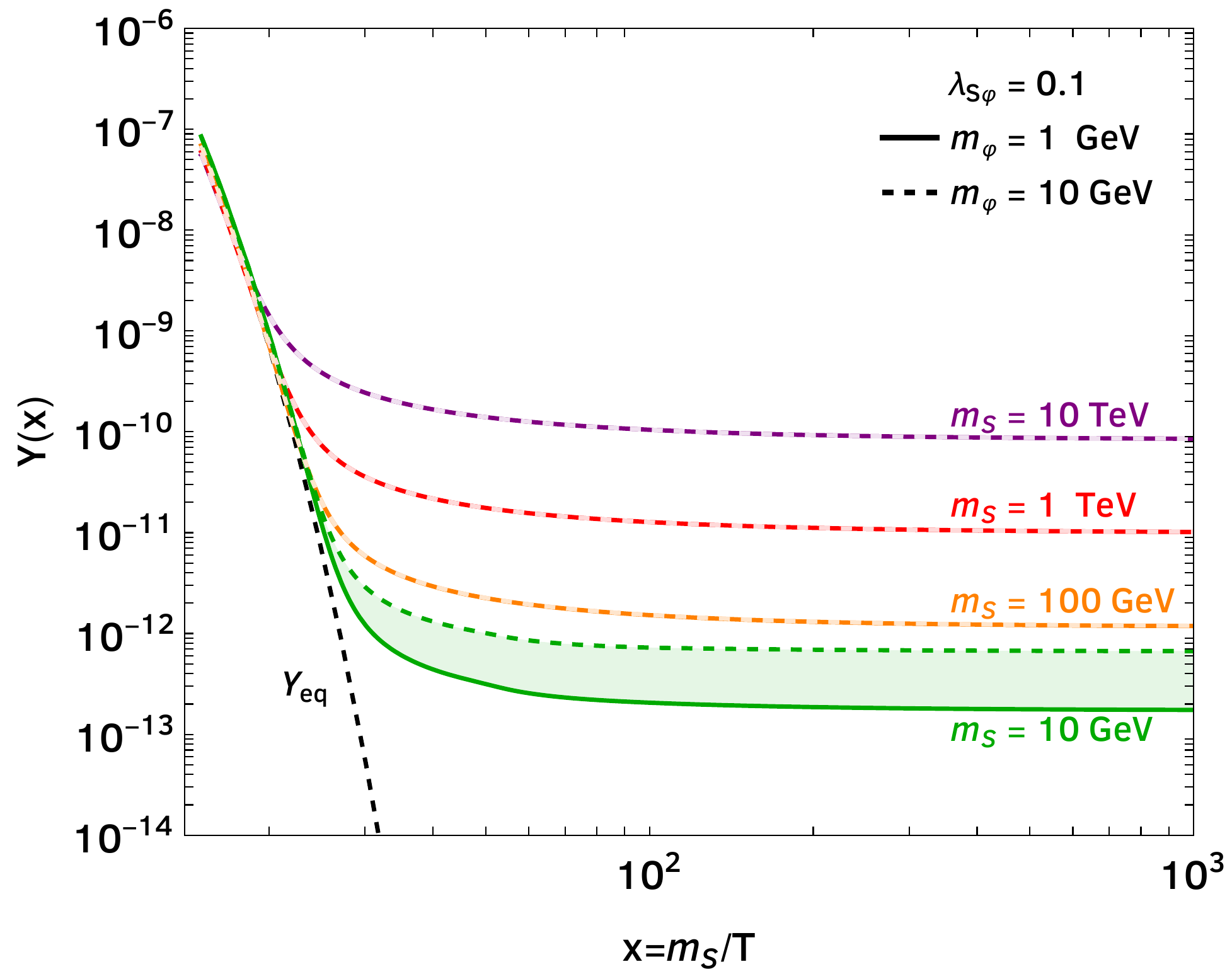}} \quad
\subfloat[\label{fig:relic_bound_lambda}]{\includegraphics[width=0.47\textwidth]{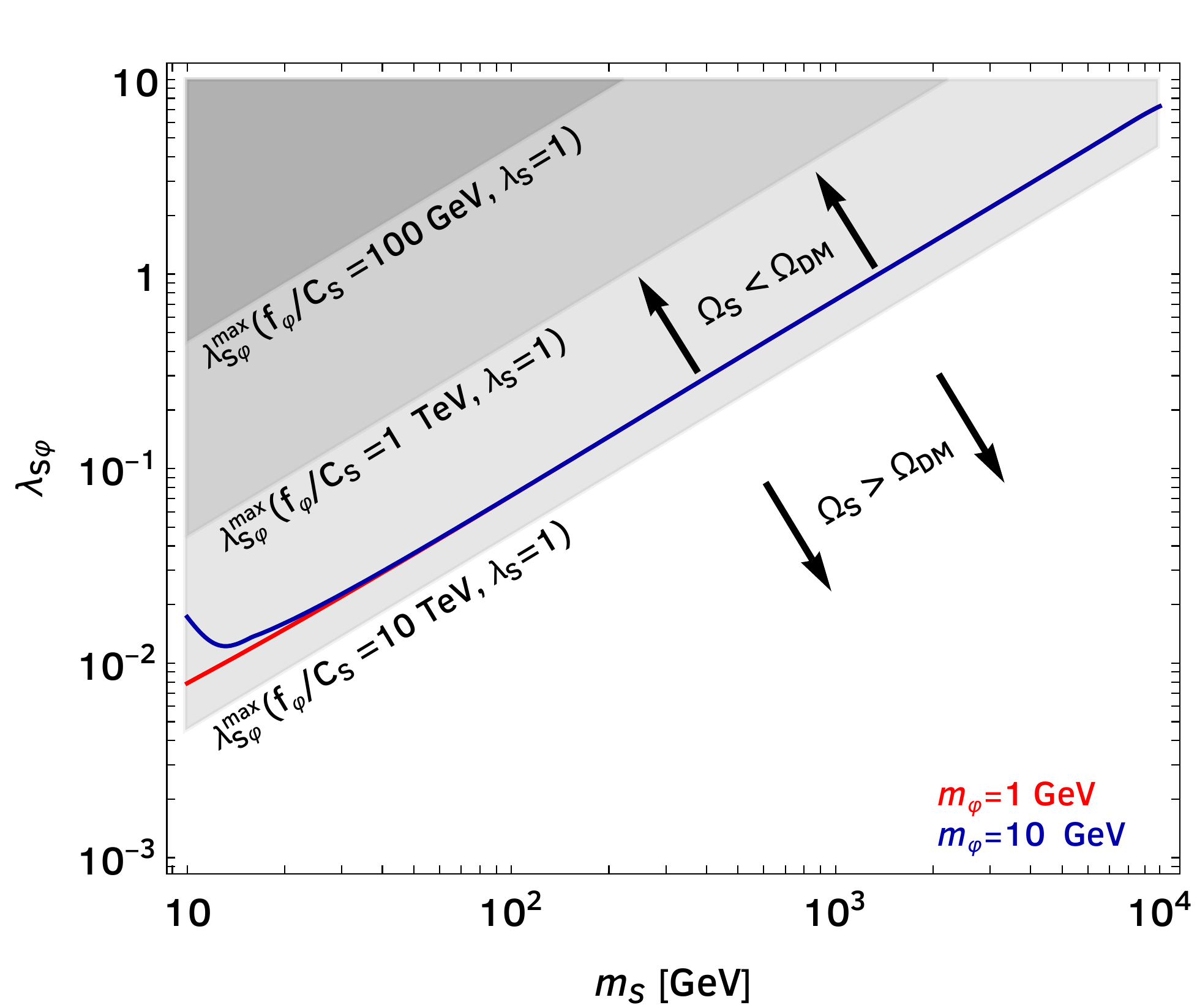}}
\caption{\textbf{Left:} Numerical solution for the DM comoving density $Y_S = n_S / s$ as a function of the inverse temperature $x = m_S / T$. The quartic coupling is fixed to $\lambda_{S\varphi} = 0.1$, the ALP mass is $m_\varphi = 1 \, {\rm GeV}$ (solid lines) and $m_\varphi = 10 \, {\rm GeV}$ (dashed lines). The DM mass takes the four possible values indicated by the different colors. The ALP mass affects the relic density only when it is nearly degenerate with the DM due to its phase space effects. \textbf{Right:} DM relic density in the $(m_S, \lambda_{S\varphi}$) plane for $m_\varphi = 1 \, {\rm GeV}$ (red line) and $m_\varphi = 10 \, {\rm GeV}$ (blue line). Solid lines identify $\Omega_S = \Omega_{\rm DM}$. We explicitly point to the regions where we achieve over and under-production. Shaded regions identify the upper bound in \Eq{eq:lambda_max} for different values of the ALP decay constant and with $\lambda_S=1$.}
\label{fig:lambdabounds}
\end{figure}

It is convenient to introduce comoving and dimensionless variables to ease the numerical analysis. We scale out the Hubble expansion using the comoving number density $Y_S \equiv n_S /s$. Here, $s = (2 \pi^2 / 45) g_{*s}(T) T^3$ is the entropy density, and $g_{*s}(T)$ is the effective number of entropic degrees of freedom taken from Ref.~\cite{Laine:2015kra}. We also trade the time variable with the inverse temperature $x \equiv m_S / T$. Using entropy conservation ($d s / dt + 3 H s = 0$), we rewrite the Boltzmann equation in terms of these dimensionless variables as 
\be
\frac{d Y_S}{d \ln x} = - \left( 1 - \frac13\frac{d \ln g_{*s}}{d \ln x} \right) \frac{s(x) \, \langle \sigma_{SS\to  S^\star \varphi}  \vmol \rangle}{H(x)} \left[Y_S^2  - Y_S \, Y_S^{\rm eq} \right] \ .
\label{eq:genBEdimless}
\ee
Numerical solutions of this equation are presented in \Fig{fig:yield}. We fix the quartic coupling controlling the semi-annihilation transition amplitude to $\lambda_{S \varphi} = 0.1$ and consider various values of the DM and ALP masses. The s-wave expansion in \Eq{eq:sigmasemiswave} provides insight into the observed behavior of these solutions. The ALP mass $m_\varphi$ appears only in the final-state phase space factor, and its impact on the relic density is significant only when $m_\varphi \simeq m_S$. In contrast, the cross section exhibits a power-law dependence on the DM mass, which explains the different behavior for the mass values considered in the plot. Specifically, a larger $m_S$ results in a smaller cross section, and this implies an earlier freeze-out. As expected, all solutions reach an asymptotic value $Y_S^\infty$ at late times. 

The current DM density expressed in terms of the $\Omega$ variables results in
\be
\Omega_S h^2 = 2 \times \frac{m_S Y_S^\infty s(t_0)}{\rho_c (t_0)/h^2} \ .
\ee
Here, the overall factor of $2$ accounts for both DM particles and antiparticles, whereas the present values of the entropy density and the critical density are~\cite{ParticleDataGroup:2024cfk} 
\begin{subequations}
\begin{align}
s(t_0) = & \, 2891.2 \, {\rm cm}^{-3} \ , \\
\rho_c(t_0) = & \, 1.05 \times 10^{-5} \, h^2 \, {\rm GeV} \, {\rm cm}^{-3} \ .
\end{align}
\end{subequations}
The relic density of $S$ and $S^\star$ matches the observed one~\cite{Planck:2018vyg} when we satisfy
\be
\Omega_S = \Omega_{\rm DM} = 0.1198 \pm 0.0012 \ .
\ee
We visualize the parameter space in \Fig{fig:relic_bound_lambda} where we consider two different values of the ALP mass and identify the corresponding relic density lines in the $(m_S, \lambda_{S\varphi})$ plane. Consistently with what we have already discussed, the detailed value of the ALP mass has an impact only in the near degenerate case and therefore in the low DM mass region. The relic density line, up to logarithmic corrections, corresponds to parameter space region with $\langle \sigma_{SS\to  S^\star \varphi}  \vmol \rangle \simeq {\rm const}$~\cite{Lee:1977ua}. The s-wave expansion in \Eq{eq:sigmasemiswave} shows how lines at constant relic density must satisfy the linear scaling $\lambda_{S \varphi} \propto m_S$. For the region above that line DM would be underproduced. If this is the case, one has to invoke non-standard cosmological histories (e.g., the relentless phase proposed in Ref.~\cite{DEramo:2017gpl} for expansions faster than usual) or simply accept that $S$ is a sub-dominant DM component. On the contrary, the parameter region below the relic density lines corresponds to DM overproduction and one has to invoke some dilution mechanism at later stages (e.g., the entropy injection discussed in Ref.~\cite{Giudice:2000ex}). The quartic coupling $\lambda_{S \varphi}$ cannot be arbitrarily large since we cannot violate the inequality given by \Eq{eq:lambda_max}. This theoretical bound is illustrated by the shaded gray region in \Fig{fig:relic_bound_lambda} that are all obtained for $\lambda_S = 1$ and for different values of $f_\varphi / \wc_S$. We notice how the correct relic density requires values of the quartic couplings compatible with this bound for $f_\varphi / \wc_S$ as large as fews TeV, and the requirement starts becoming problematic once we approach values around 10 TeV. One could relax this bound by considering a larger quartic coupling $\lambda_S$ but only up to some extent. Indeed, the maximum $\lambda_{S \varphi}$ scales as the square root of $\lambda_S$, and at some point we run into perturbative limits for the DM self-interaction. 

The semi-annihilation cross section, with its s-wave approximation given in \Eq{eq:sigmasemiswave}, depends on both the DM mass and the quartic coupling $\lambda_{S\varphi}$. The relic density analysis in this two-dimensional parameter space is shown in \Fig{fig:relic_bound_lambda}. In fact, the quartic coupling $\lambda_{S\varphi}$, as expressed in \Eq{eq:VND}, is a combination of two distinct parameters: the ALP-DM coupling $\wc_S / f_\varphi$ and the DM cubic self-interaction $A$. Therefore, the parameter space is effectively three-dimensional, as can be seen manifestly in the derivative basis. Theoretical bounds on the cubic coupling $A$ suggest that the correct relic density can be achieved as long as the ratio $f_\varphi / \wc_S$ does not exceed approximately 10 TeV. We find it valuable to further explore the relationship between relic density and the fundamental parameters of the EFT.

\begin{figure}[t]
\subfloat[\label{fig:AvsmS}]{\includegraphics[width=0.48\textwidth]{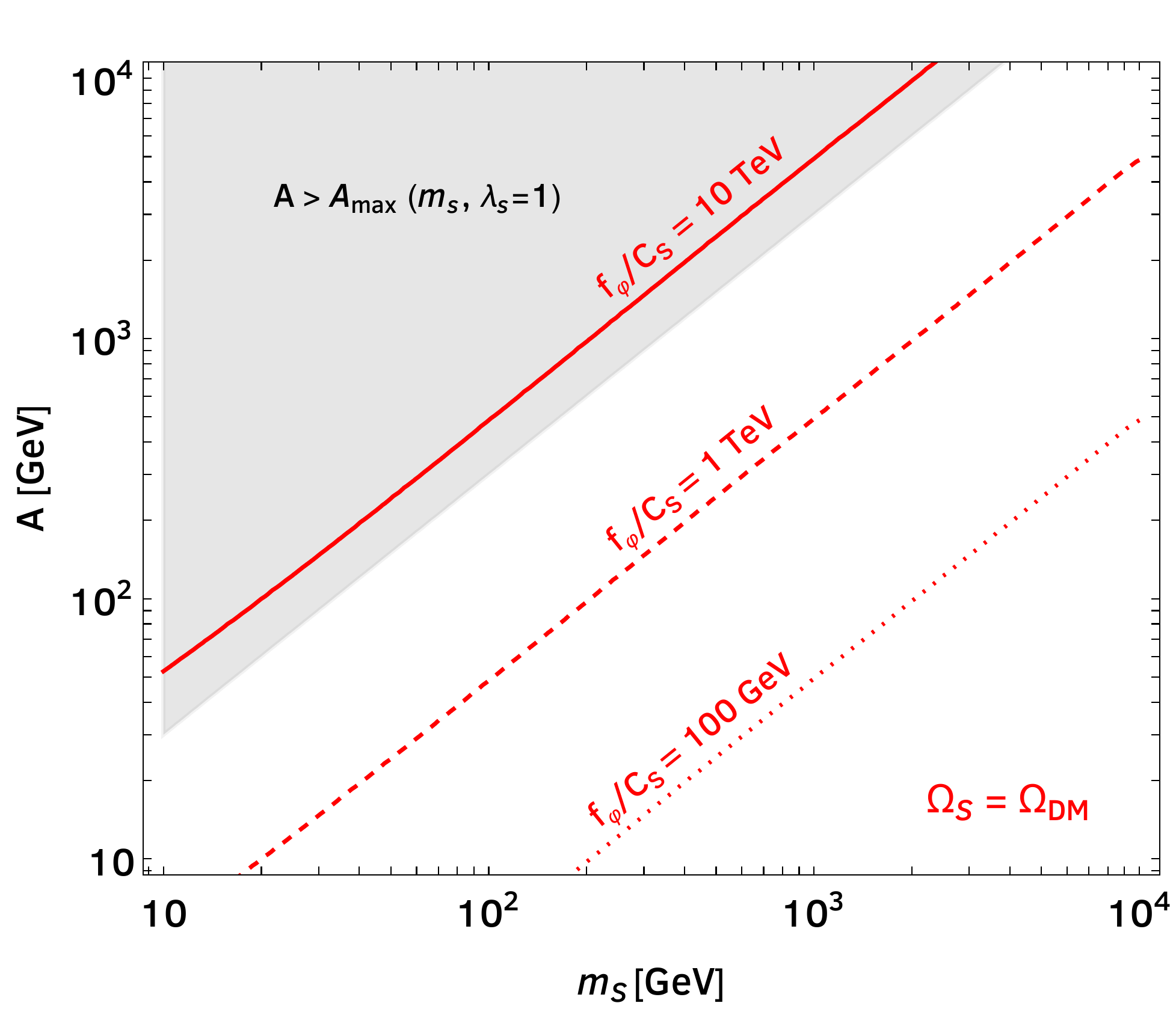}} \quad\quad
\subfloat[\label{fig:Avsf}]{\includegraphics[width=0.485\textwidth]{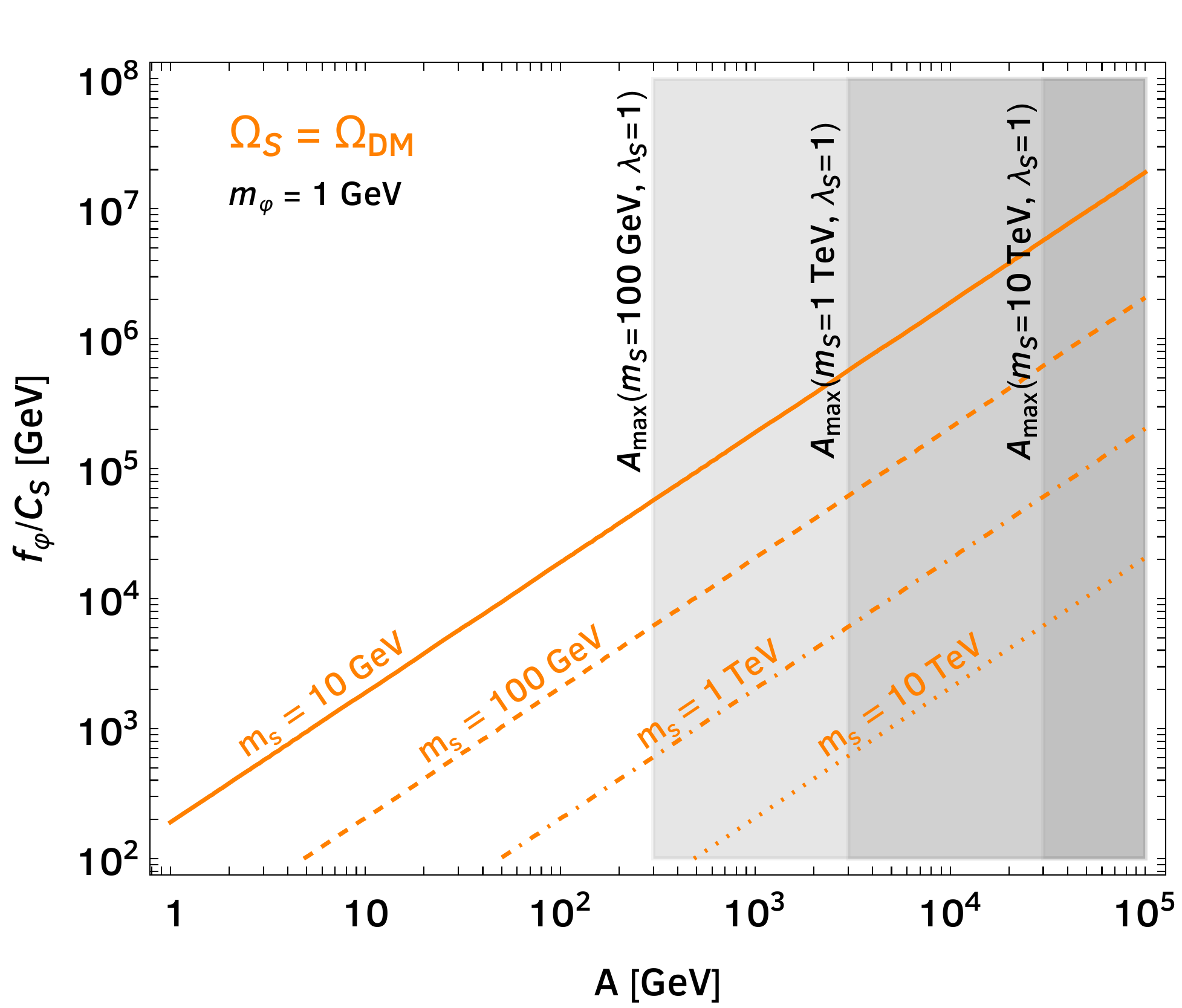}}
\caption{\textbf{Left:} Solid red lines identify points in the $(m_S,\,A)$ plane with $\Omega_S=\Omega_{DM}$ for different values of $f_\varphi/\wc_S$. \textbf{Right:} Relic density isocontours in the $(A,\,f_\varphi/\wc_S)$ plane for different DM masses. For both figures, the grey shaded area violates the bound in \Eq{eq:Amax} with $\lambda_S=1$.}
\label{fig:relic_bound_A}
\end{figure}

The plots in \Fig{fig:relic_bound_A} show two-dimensional slices of the parameter space. In \Fig{fig:AvsmS}, we explore the $(m_S, A)$ plane, where the red lines identify the regions with the correct relic density for different values of the inverse ALP-DM coupling $f_\varphi / \wc_S$. We also shade in gray the region where the coupling $A$ violates the inequality in \Eq{eq:Amax}, with $\lambda_S$ fixed at 1. The slope of the red lines follows the scaling relation $A \propto m_S$, consistent with the relic density constraint $\langle \sigma_{SS\to S^\star \varphi} \vmol \rangle \simeq {\rm const}$. The results in \Fig{fig:AvsmS} also confirm that we can push the ratio $f_\varphi / \wc_S$ only slightly below 10 TeV before being in conflict with theoretical bounds. The plot in \Fig{fig:Avsf} provides the same information, but in the $(A,\,f_\varphi/\wc_S)$ plane for different DM mass values. The slope of the relic density line again matches the expectations from the freeze-out analysis. For a fixed DM mass, the intersection between the relic density line and the boundary of the shaded gray region identifies the maximum values of $A$ and $f_\varphi / \wc_S$ allowed by the constraint in \Eq{eq:Amax}. These intersection points are located at approximately constant values of $f_\varphi / \wc_S$, around a few TeV, consistent with earlier results.

\section{ALP cosmology}
\label{sec:ALP}

In the previous section, we thoroughly analyzed how DM abundance can be achieved through the thermal freeze-out of semi-annihilations involving $S$ particles, finding that the resulting DM relic density is remarkably independent of ALP couplings to SM fields. However, this does not imply that arbitrary ALP couplings to visible matter are allowed. A key assumption in the freeze-out analysis is that the portal field $\varphi$ remains in both kinetic and chemical equilibrium with the primordial bath during the DM freeze-out epoch, which necessitates a minimal ALP-SM interaction to sustain equilibrium. In this section, we critically assess the validity of this assumption and explore ALP cosmology in a broader context. Within the framework outlined in \Sec{sec:EFT}, ALP-SM interactions include three electroweak-invariant couplings with gauge bosons and nine flavor-conserving couplings to fermions. We systematically analyze the impact of these interactions by activating a subset of ALP-SM couplings at a time and studying the resulting cosmological evolution of ALP particles.

\subsection{General analysis of ALP thermalization}

First, ALP interactions with SM fields must be sufficiently strong to maintain equilibrium between the two sectors during the freeze-out epoch. Our goal is to identify the region of parameter space where this condition is satisfied. The Boltzmann equation governing the evolution of the ALP number density $n_\varphi$ can be generally expressed as 
\be
\frac{dn_\varphi}{dt} + 3 H n_\varphi = \sum_\alpha \wc_\alpha \ .
\label{eq:ALPBE}
\ee
The collision operator on the right-hand side accounts for the sum of all number-changing processes, with its explicit form depending on the ALP-SM couplings. For the interactions considered in this work, two main classes of processes contribute to ALP production and thermalization: inverse decays and binary scatterings. As usual, there is a price to pay for each ALP on the external legs so we focus on processes with only one $\varphi$ in the final state. Using the general analysis in \App{app:coll}, the collision operators take the general form
\begin{align}
\text{Single ALP production: }  \qquad  \qquad \wc_\alpha = \gamma_\alpha \left( 1-  \frac{n_\varphi}{n_\varphi^{\rm eq} } \right) \ .
\end{align}
The quantity $\gamma_\alpha$ denotes the number of interactions per unit time and volume. Denoting a generic primordial bath particle as $\mathcal{B}_i$, the expressions for these quantites are  
\begin{subequations}
\begin{align}
\label{eq:ALPCollDecay} & \, \text{Inverse decays: }   \;\;\;
\gamma_{\mathcal{B}_1 \mathcal{B}_2 \rightarrow \varphi }  = n_\varphi^{\rm eq} \frac{K_1[m_\varphi / T]}{K_2[m_\varphi / T]}
\Gamma_{\varphi \rightarrow \mathcal{B}_1 \mathcal{B}_2}  \ , \\
\label{eq:ALPCollScat} & \,  \text{Scatterings: } \quad\quad \,
\gamma_{\mathcal{B}_1 \mathcal{B}_2 \rightarrow \mathcal{B}_3 \varphi} =  n_{\mathcal{B}_1}^{\rm eq} n_{\mathcal{B}_2}^{\rm eq} \langle \sigma_{\mathcal{B}_1 \mathcal{B}_2 \rightarrow \mathcal{B}_3 \varphi}   \vmol \rangle = n_{\mathcal{B}_3}^{\rm eq} n_{\varphi}^{\rm eq} \langle \sigma_{\mathcal{B}_3 \varphi \rightarrow \mathcal{B}_1 \mathcal{B}_2}   \vmol \rangle\ .
\end{align}
\end{subequations}
The Bessel functions in the first expression accounts for the Lorentz dilatation of the ALP lifetime. The two expressions for the scattering rate are equivalent, as one can show using the detailed balance principle, and the more convenient one depends on the mass spectrum. 

As already done for the DM freeze-out analysis, it is convenient to write the Boltzmann equation in terms of comoving and dimensionless quantities. The choice of the variable $x$ is arbitrary, and it is indeed convenient to keep $x = m_S / T$ since we are investigating ALP thermalization at the DM freeze-out epoch, $x_{\rm FO} = m_S / T_{\rm FO} \simeq 20$ (see \Fig{fig:yield}). We find 
\be
\frac{d Y_\varphi}{d \ln x} = - \left( 1 - \frac13\frac{d \ln g_{*s}}{d \ln x} \right) \frac{\sum_\alpha \gamma_\alpha(x)}{H(x) s(x)} \left( 1-  \frac{Y_\varphi}{Y_\varphi^{\rm eq} } \right)  \ .
\label{eq:BEALPdimless}
\ee
The expression above provides a convenient qualitative criterion for ALP thermalization. Specifically, thermal equilibrium is maintained as long as $\sum_\alpha \gamma_\alpha(T) \gtrsim H(T) \, s(T)$. This criterion essentially compares two time scales: the interval between two consecutive interactions and the characteristic time scale of the Hubble expansion. This approach will be adopted in this section, as our primary interest lies not in tracking the precise numerical density of the ALP, but in determining whether equilibrium is established at $T \simeq T_{\rm FO}$.

The DM mass range explored in Figs.~\ref{fig:lambdabounds} and \ref{fig:relic_bound_A} corresponds to a finite interval of freeze-out temperatures. For the largest DM mass, $m_S = 10 \, {\rm TeV}$, the freeze-out temperature is approximately $T_{\rm FO} \simeq m_S / 20 \simeq 500 \, {\rm GeV}$. For the smallest DM mass, $m_S = 10 \, {\rm GeV}$, we find $T_{\rm FO} \simeq m_S / 20 \simeq 500 \, {\rm MeV}$. Consequently, nearly the entire range of freeze-out temperatures of interest falls within the regime of the SM electroweak sector in the broken phase and with quarks still deconfined. Caution is required at the extremes of the DM mass range. For large DM masses, where freeze-out occurs during the unbroken electroweak phase, interaction rates remain calculable and are not significantly altered. However, at the lower end of the DM mass range, the freeze-out temperature approaches a regime where QCD becomes non-perturbative, preventing the primordial bath from being treated as a weakly coupled gas of quarks and gluons. Consequently, we restrict our analysis of thermalization to temperatures above the $\GeV$ scale for ALP coupled to colored SM degrees of freedom.

\subsection{Dangerous ALP decays at late times}

Ensuring that the ALP is in equilibrium at the freeze-out epoch is not sufficient to guarantee the phenomenological viability of this scenario. The interactions in \Eq{eq:Lag} induce ALP decays, necessitating a thorough examination of the cosmological history of this metastable degree of freedom. ALP particles are expected to undergo a freeze-out process similar to that of DM. While the Boltzmann equation in \Eq{eq:ALPBE} is the appropriate tool to analyze the ALP freeze-out epoch, a detailed study of this process is not required for our purposes.

If the ALP lifetime is sufficiently short—at most comparable to the age of the universe at the time of $\varphi$ freeze-out—decays of frozen-out ALPs will produce particles that rapidly thermalize with the primordial bath, ensuring a standard cosmological history. Conversely, a long-lived ALP could have significant implications for the evolution of the universe, as energy injection from its decay products could: (i) alter the primordial abundances of light nuclei produced via Big Bang Nucleosynthesis (BBN)~\cite{Kawasaki:2017bqm,Kawasaki:2020qxm}, (ii) induce spectral distortions in the black-body spectrum of the Cosmic Microwave Background (CMB)~\cite{Chluba:2013wsa,Balazs:2022tjl}, and (iii) modify the predicted CMB anisotropy spectrum~\cite{Slatyer:2016qyl,Poulin:2016anj}.

To avoid these constraints, we impose that ALP decays occur before the onset of BBN, when neutrons decouple. This corresponds to a bath temperature of approximately $\mathcal{O}(\MeV)$ and a universe age of about one second. Therefore, throughout our analysis in the following subsections, we require $\tau_\varphi \lesssim 1 \, {\rm s}$ to ensure compatibility with standard cosmology.
 
\subsection{Results I: couplings to SM gauge bosons}
\label{subsec:KSVZpheno}

The first case we study is when the ALP couples exclusively to SM gauge bosons. Such statements must be made with due caution, as the couplings depend on the field basis in which we are working. To make unambiguous statements, we define the scenario investigated here with the assumption that there exists a field basis where both derivative and non-derivative ALP couplings to SM fermions are absent. Low-energy interactions of this type naturally emerge from UV-complete constructions in which heavy fermions carrying SM gauge quantum numbers are integrated out. In particular, if $\varphi$ originates from the spontaneous breaking of an Abelian global symmetry, this symmetry must be anomalous under (at least part of) the SM gauge group. A well-known example is the Kim-Shifman-Vainshtein-Zakharov (KSVZ) framework for the QCD axion~\cite{Kim:1979if,Shifman:1979if}. In this scenario, the axion arises as the Nambu-Goldstone boson of a PQ symmetry, under which all SM fields remain neutral, while a new colored fermion, $\Psi_{\rm PQ}$, acquires mass through PQ breaking. Electroweak quantum numbers for $\Psi_{\rm PQ}$ are possible, though not mandatory. Upon integrating out $\Psi_{\rm PQ}$, the dimension-5 contact interactions with SM gauge bosons in \Eq{eq:Lag} are generated.  The coupling to gluons is essential for addressing the strong CP problem, but scenarios in which the heavy fermion only carries electroweak quantum numbers remain phenomenologically viable. Adopting a bottom-up perspective, we treat the three Wilson coefficients for gauge boson interactions in \Eq{eq:Lag} as free parameters. In the following, we focus on two examples involving ALP couplings to photons and gluons.

\paragraph{Photon coupling.} The EFT described by \Eq{eq:Lag} is valid at high energies, where the electroweak symmetry is unbroken. Consequently, the coupling to photons is a residual effect at low energies of the electroweak gauge-invariant couplings to the $SU(2)_L \times U(1)_Y$ gauge bosons. To establish a proper connection between the UV and IR regimes, one should account for the renormalization group evolution, evolving the theory down to energies on the order of the ALP mass~\cite{Bauer:2020jbp}. One-loop radiative corrections are negligible for our analysis, and we focus on the tree-level matching that generates the low-energy interaction 
\be
\mathcal{L}^{(\partial)}_{\varphi \gamma\gamma} =  \frac{\varphi}{8 \pi f_\varphi}  \wc_\gamma \alpha_{\rm em}  F_{\mu\nu} \widetilde F^{\mu\nu} \ ,
\label{eq:Lgammagamma}
\ee
where we introduce the electromagnetic fine structure constant $\alpha_{\rm em} \equiv e^2 / (4 \pi)$ and we find it convenient to define the combination of Wilson coefficients $\wc_\gamma \equiv \wc_{W}+\wc_{B}$. Once we switch on $\wc_{W}$ and $\wc_{B}$, the Lagrangian in \Eq{eq:Lag} mediates potentially ALP decays into pairs of SM electroweak gauge bosons that include also the $W$ and the $Z$. However, given the ALP mass range we focus on, the only allowed final state is the one with two photons. We also specify that the interaction is defined in the derivative basis, with the additional assumption that couplings to fermions vanish in this particular case.

At large temperatures, substantially larger than $m_\varphi$, ALPs thermalize via scatterings with the thermal bath. The calculation for the scattering rate is complicated by the long range nature of the force mediated by massless photons that leads to a problematic IR behavior. In order to carefully cure these divergences and regulate the final result, a proper and full finite $T$ thermal field theory calculations must be performed. The scattering rate at high temperatures includes contributions from all the weak bosons and the corresponding rates have been computed by Ref.~\cite{Salvio:2013iaa}. For our cases, DM freeze-out happens almost exclusively in the electroweak broken phase where the number density of $W$ and $Z$ bosons is Maxwell-Boltzmann exponentially suppressed. It is legit to keep contributions from photons only, and the leading contribution to the thermalization rate would come from Primakoff scattering~\cite{Braaten:1991dd,Bolz:2000fu,Cadamuro:2011fd} resulting in\footnote{A recent analysis that goes beyond the Hard Thermal Loop approximation can be found in Ref.~\cite{Becker:2025yvb}.} 
\be\label{eq:KSVZphoton_rate}
\gamma^{(\gamma)}_{\rm scattering} = \frac{\alpha_{\rm em}^3}{144 \pi^2} \left(\frac{\wc_{\gamma}}{f_\varphi}\right)^2 \, T^3 \,n_Q(T) \left[\log \left(\frac{T^2}{m^2_\gamma(T)}\right)+0.8194\right] \ .
\ee
The thermal bath enters via the effective  number density of charged particles $n_Q$ that interact with the ALP via Primakoff scattering. We adopt $n_Q = \sum_i Q_i^2 n_i = (\zeta(3) / \pi^2) g_Q(T) T^3$, and we employ the results for $g_Q(T)$ provided by Ref.~\cite{Caloni:2022uya}. Thermal effects generate a photon mass $m_\gamma$ that keeps into account all charged plasma particles. Its explicit expression scales as $m_\gamma \propto \sqrt{g_Q(T)}$, and we use $m^2_\gamma = g_Q(T)  \, e^2 \, T^2 / 18$. As a check, for a pure QED plasma with $g_Q = 3$ (from $e^\pm$) we recover the known result $m^2_\gamma = e^2 \, T^2 / 6$~\cite{Bellac:2011kqa}. An additional word of warning is in order here since \Eq{eq:KSVZphoton_rate} is only applicable in the massless ALP limit (i.e., $m_\varphi \ll T$). It is worth remembering the spirit of this section: we are interested in \textit{qualitatively} establishing whether the ALP reaches thermal equilibrium, and with this in mind, we can make approximations regarding the treatment of thermal effects. We keep the validity of \Eq{eq:KSVZphoton_rate} until the temperature reaches the ALP mass scale. For smaller temperatures, we manually switch off the scattering processes and retain only the inverse decay contribution. The collision operator for this latter channel is explicitly provided by \Eq{eq:ALPCollDecay} and it is proportional to the ALP decay width. The ALP lifetime in its rest frame is obtained from the decay width given \Eq{eq:gammaALPgauge} of App.~\ref{app:ALP}, and it results in
\be
\tau_{\varphi \rightarrow \gamma \gamma} = \Gamma^{-1}_{\varphi \rightarrow \gamma \gamma} \approx 0.5 \, {\rm s} \, \left(\frac{1}{\wc_\gamma}\right)^2 \left(\frac{0.01}{\alpha_{\rm em}}\right)^2  \left(\frac{1\,\GeV}{m_\varphi}\right)^3\left(\frac{f_\varphi}{10^{8} \, \GeV} \right)^2 \ .
\label{eq:alptauphoton}
\ee
We fix the running fine structure constant~\cite{L3:2000hbp} to a value that is typical for the ALP mass range under investigation. The other parameters are chosen to identify the time scale of one second. We can plug this expression into \Eq{eq:ALPCollDecay} and obtain the number of interactions per unit time and volume $\gamma_{\varphi \rightarrow \gamma \gamma}$. 

These two contributions to ALP thermalization have a drastically different temperature behavior and they are most efficient at different phases of the expansion history. To establish when a generic process $\alpha$ is efficient at thermalizing ALPs, we look at equation \Eq{eq:BEALPdimless} and consequently we try to find when the dimensionless ratio $\gamma_\alpha(T) / (H(T) s(T))$ reaches its maximum. We work always in a standard cosmological history with the primordial radiation bath dominating the energy budget, and we have $H(T) s(T) \propto T^5$. The contribution from inverse decays is obtained from \Eq{eq:ALPCollDecay} and for large temperatures ($T \gg m_\varphi$) scales as $\gamma_{\rm inv-D}(T) \propto  T^2$. On the contrary, the large temperature behavior for scattering rate in \Eq{eq:KSVZphoton_rate} reads $\gamma^{(\gamma)}_{\rm scattering}(T) \propto T^6$. Therefore, ALP thermalization from inverse decays is an IR dominated process that reaches its maximal efficiency at $T \simeq m_\varphi$, whereas the scatterings lead to a UV dominated ALP production. 

\begin{figure}[t]
\centering
\subfloat[\label{fig:boundsKSVZ_gluon}]{\includegraphics[width=0.47\textwidth]{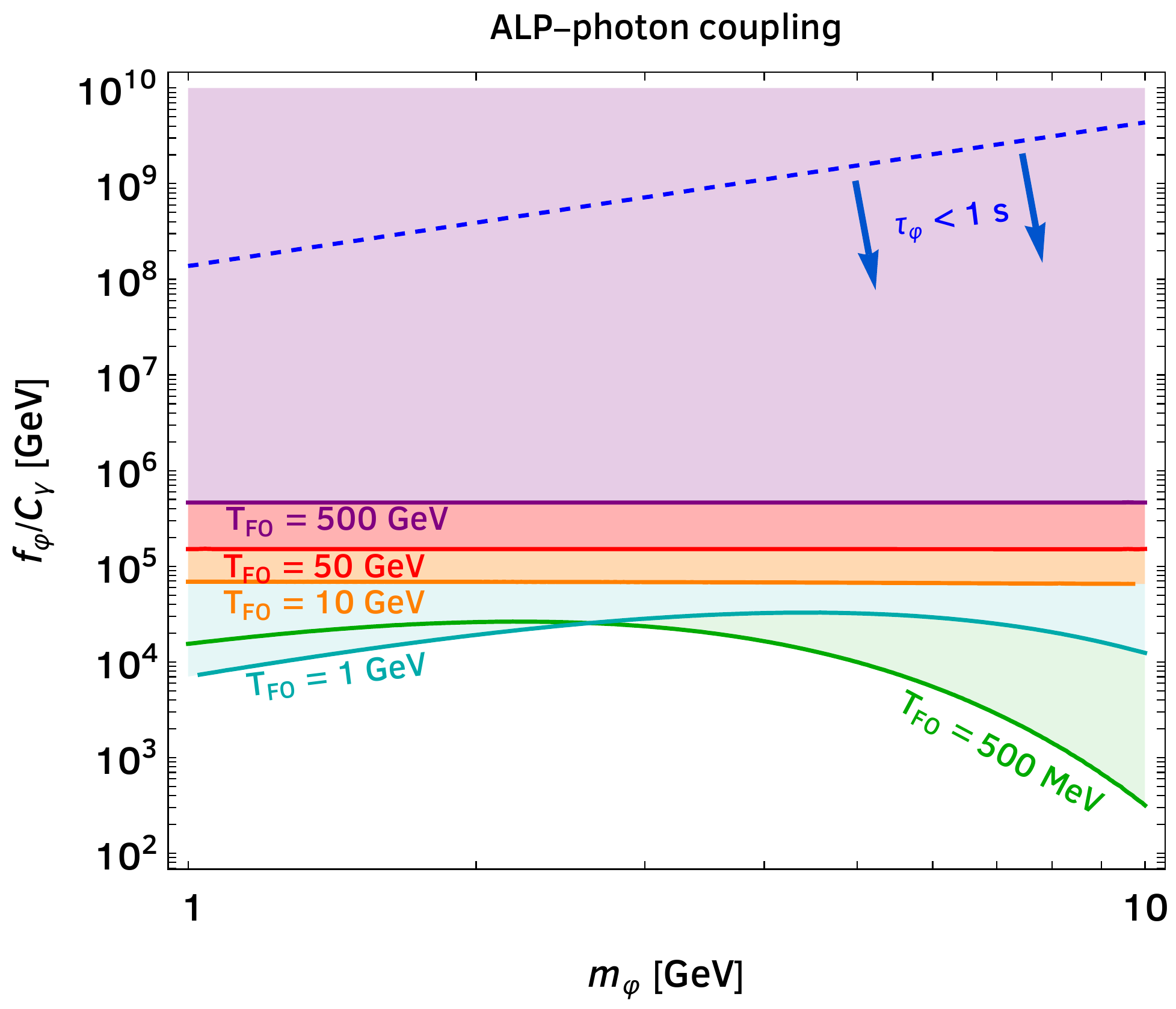}}\qquad
\subfloat[\label{fig:boundsKSVZ_gamma}]{\includegraphics[width=0.47\textwidth]{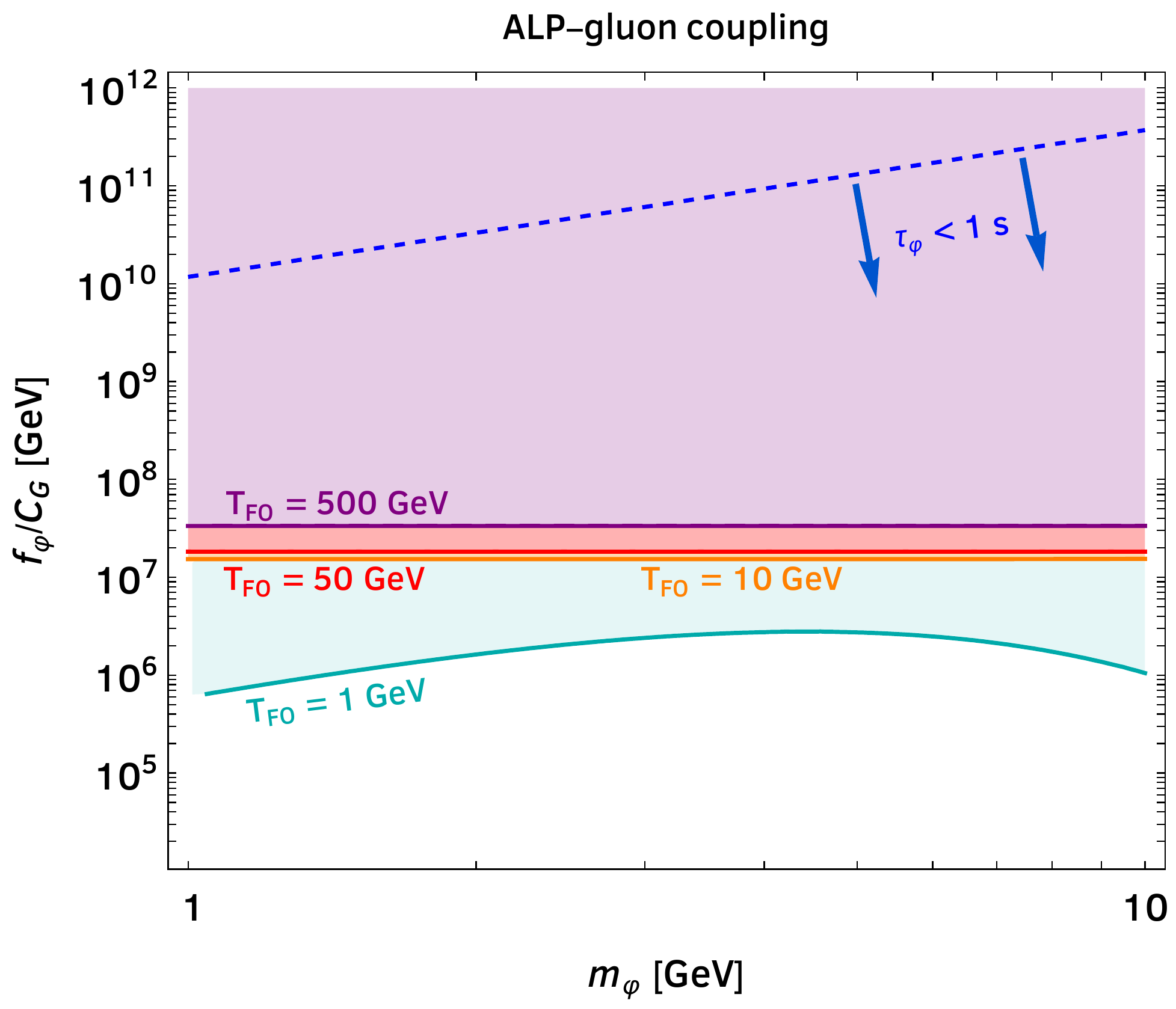}} 
\caption{\textbf{Left:} Thermalization for an ALP coupled to photons in the $(m_\varphi,\,f_\varphi / \wc_\gamma)$ plane. Different colors correspond to the various DM freeze-out temperatures specified in the figure, while the shaded regions indicate where thermalization is not achieved. The dashed blue line represents the ALP lifetime of one second, with the region below it being cosmologically safe from late-time ALP decays. \textbf{Right:} Same for an ALP coupled to gluons.}
\label{fig:KSVZbounds}
\end{figure}

The plot in \Fig{fig:boundsKSVZ_gluon} shows the viable parameter space region in the $(m_\varphi, f_\varphi / \wc_\gamma)$ plane. We select a few representative values for $T_{\rm FO}$, indicated by different colors, and shade the region where ALPs fail to achieve thermal equilibrium. This plot exhibits characteristic features that are straightforward to interpret. For $T_{\rm FO}$ values significantly larger than the ALP mass, the boundary of the thermalization region becomes independent of the ALP mass, resulting in horizontal lines. Moreover, since ALP thermalization in this case is a UV-dominated process, higher DM freeze-out temperatures allow for larger values of the ALP decay constant. The thermalization boundary deviates from horizontal when the DM freeze-out temperature approaches the ALP mass range considered. In these cases, where only inverse decays are accounted for, the boundary displays the characteristic behavior arising from the Maxwell-Boltzmann suppression of the ALP equilibrium number density. The dashed blue line identifies the parameters for which the ALP lifetime is one second. Above this line, BBN is potentially endangered. Notably, this occurs for relatively large values of the ALP decay constant, which are already excluded by the thermalization criterion.

To summarize, once a DM mass value $m_S$ is fixed, the resulting freeze-out temperature is approximately $T_{\rm FO} \simeq m_S / 20$. The unshaded region in \Fig{fig:boundsKSVZ_gluon} for that specific value of $T_{\rm FO}$ leads to a consistent DM freeze-out analysis, as the ALP is in thermal equilibrium at the freeze-out epoch. We observe that the typical values of $f_\varphi / \wc_\gamma$ that satisfy the thermalization constraints are comparable to those of $f_\varphi / \wc_S$ identified in \Sec{sec:relic}. This is noteworthy because it implies that no hierarchy is required among the dimensionless Wilson coefficients. Furthermore, these values are also compatible with bounds on the ALP-photon coupling from other searches~\cite{CMS:2018erd,ATLAS:2020hii,Belle-II:2020jti,BESIII:2022rzz,BESIII:2024hdv}. Thus, DM thermal freeze-out via semi-annihilations as described in the previous section is phenomenologically viable in the scenario where the ALP couples to photons.

\paragraph{Gluon coupling.} The other gauge boson we consider coupled to the portal field $\varphi$ is the gluon. Thermalization can be driven again by both scatterings and inverse decays, and we approximate the thermal effects as already done in the previous case. The scattering rate calculation presents the same IR peculiar behavior that was discussed for the photons. The situation is even more intricated for gluons since QCD becomes nonperturbative at low temperatures and quarks confine into hadrons. This issue has been investigated in the literature for over two decades given its robust theoretical motivation from the strong CP problem. We borrow the convention from the QCD axion literature~\cite{Salvio:2013iaa,DEramo:2021psx,DEramo:2021lgb,Bouzoud:2024bom}
\be
\gamma^{(G)}_{\rm scattering} = \frac{2 \zeta(3) \mathcal{D}_G}{\pi^3} \left( \frac{\wc_G \, \alpha_G}{8 \pi f_\varphi} \right)^2 T^6 \, F_G(T) \ .
\label{eq:KSVZgluon_rate}
\ee
The calculations for the QCD axion are only applicable in the massless ALP limit, and we will make sure to apply these results only when it is legitimate by switching if off 
as we approach temperatures around the ALP mass where IR dominated inverse decays are the dominant thermalization channel. With this warning in mind, we employ the results from Refs.~\cite{DEramo:2021psx,DEramo:2021lgb} that extended the analysis by Ref.~\cite{Salvio:2013iaa} and evaluated the function $F_G(T)$ for the temperatures that are in our range of interest. Furthermore, at the low-end of the $T_{\rm FO}$ interval of interest we are dangerously close to the QCD phase transition, and we cannot treat the quark-gluon plasma as a weakly-coupled gas of elementary particles anymore. As a consequence, we cannot push our analysis down to arbitrarily low temperatures and we will stop at $T\approx1\,\GeV$. For temperatures below the ALP mass, thermalization is controlled by ALP inverse decays with a rate that depends on the ALP lifetime
\be
\tau_{\varphi \rightarrow gg} = \Gamma^{-1}_{\varphi \rightarrow gg} \approx 0.7 \, {\rm s} \, \left(\frac{1}{\wc_G}\right)^2 \left(\frac{0.3}{\alpha_G}\right)^2  \left(\frac{1\,\GeV}{m_\varphi}\right)^3\left(\frac{f_\varphi}{10^{10} \, \GeV} \right)^2 \ .
\label{eq:alptaugluon}
\ee
We choose the representative value $\alpha_G \simeq 0.3$ for the running QCD coupling constant~\cite{Deur:2023dzc}. Furthermore, we choose the other reference values conveniently to identify the ballpark for the ALP decay constant when the lifetime is around the BBN epoch.  

The constraints on the ALP parameter space for this case are illustrated in \Fig{fig:boundsKSVZ_gamma}. We keep the same conventions as the plot for the photon case. We notice how BBN is potentially in danger for quite large values of the ALP decay constant, significantly larger than the ones identified for the DM relic density in the previous section. The remaining thing to check is whether ALP are in equilibrium at the DM freeze-out epoch. For a practical comparison, we select again a few reference values for $T_{\rm FO}$, which are identified by the different colors, and we shade the parameter space region where ALPs are not in thermal equilibrium. We conclude that DM freeze-out via semi-annihilations is phenomenologically viable for an ALP coupled to gluons. In particular, there is no need of introducing any hierarchy between dimensionless Wilson coefficients and the relevant parameter space region is not excluded by other searches for ALP coupled to gluons~\cite{Bauer:2018uxu,Bauer:2021mvw}.

\subsection{Results II: couplings to SM fermions}
\label{subsec:DFSZpheno}

The second scenario involves ALP couplings exclusively to SM fermions. Defining it requires some care, as the statement can be formulated either in the derivative or the non-derivative basis. These two approaches are not equivalent, since the basis transformation inevitably induces ALP couplings to gauge bosons, as shown in \Eq{eq:wCVshift}. We adopt the definition that the ALP couples only to SM fermions in the derivative basis. This choice is well-motivated: ALP-fermion interactions in the non-derivative basis—specifically, the pseudo-Yukawa operators proportional to the spin-0 fermion current in \Eq{eq:pY}—generate one-loop threshold corrections that induce gauge boson couplings. By working in the derivative basis, we sidestep these threshold effects~\cite{Georgi:1986df}.

Our discussion proceeds along similar lines as in \Sec{subsec:KSVZpheno}. The key input for the thermalization analysis is the collision operator appearing on the right-hand side of the Boltzmann equation in \Eq{eq:ALPBE}. Inverse decays, $\psi\bar\psi \rightarrow \varphi$, become active whenever the ALP decay is kinematically allowed. For an ALP coupled to an SM fermion $\psi$, this condition requires $m_\varphi > 2 m_\psi$, with the partial decay width given in \Eq{eq:ALPdecaypsi} of \App{app:ALP}. If the decay is kinematically forbidden, the dominant contribution to thermalization comes from scatterings. An important subtlety arises in the scattering rate calculations when ALP decays are kinematically allowed, as the intermediate fermion may be on-shell. Properly handling this contribution necessitates a careful treatment of thermal effects~\cite{Czarnecki:2011mr}. In line with the \textit{qualitative} spirit of this thermalization condition, we distinguish between the cases where the temperature is much larger or smaller than the ALP mass, retaining only the contributions we can confidently trust. Furthermore, scatterings feature notable differences compared to previous cases. Broadly speaking, there are two main classes of scattering processes, both of which rely on SM gauge interactions. The first involves fermion-antifermion annihilation in the early universe, $\psi\bar\psi \rightarrow \varphi V$, producing an ALP alongside a SM gauge boson $V$. If the ALP couples to quarks, $V$ corresponds to a gluon; if it couples to leptons, $V$ is a photon. The second type of scattering, derived via crossing symmetry, resembles Compton scattering: a SM fermion $\psi$ undergoes the interaction $\psi V \rightarrow \psi \varphi$, with an analogous contribution for antifermions. For the DM freeze-out temperatures of interest, the SM is in the electroweak broken phase with massive fermions. We cannot neglect the ALP mass $m_\varphi$ as it may be comparable to or even exceed the masses of some SM fermions involved in these scattering processes. A full calculation of the relevant cross sections is provided in \App{app:ALP}. Unlike the previous scenario, ALP thermalization through scatterings here is an IR-dominated process, much like inverse decays. A simple way to appreciate this distinction is to note that, in the non-derivative basis, ALP pseudo-Yukawa interactions with SM fermions in the broken phase correspond to renormalizable operators. To summarize, the thermalization rate can take three distinct contributions
\begin{subequations}
\begin{align}
\gamma^{(\psi)}_{\rm ID} = & \, n_\varphi^{\rm eq} \frac{K_1[m_\varphi / T]}{K_2[m_\varphi / T]} \Gamma_{\varphi \rightarrow \bar\psi \psi} \ , \\ 
\gamma^{(\psi)}_{\rm Pair} = & \, (n_{\psi}^{\rm eq})^2 \langle \sigma_{\psi \bar\psi \rightarrow V \varphi}   \vmol \rangle \ , \\
\gamma^{(\psi)}_{\rm Compton} = & \, 2 \times n_\psi^{\rm eq} n_V^{\rm eq} \langle \sigma_{\psi V \rightarrow \psi \varphi}   \vmol \rangle \ .
\end{align}
\end{subequations}
The first term accounts for inverse decays and is relevant only when they are kinematically allowed. The contribution from fermion-antifermion annihilation assumes no significant matter-antimatter asymmetry, while the factor of 2 in front of the Compton-like rate accounts for both fermions and antifermions. 

Before discussing specific cases, we would like to spend a few more words about the situation when ALP decays are kinematically forbidden. One may erroneously conclude that the ALP is stable on cosmological scales. However, radiative corrections can induce interactions with lighter SM particles and induce loop-suppressed decays. While loop corrections do induce decays to lighter SM degrees of freedom, these are important for establishing the ALP lifetime but have little impact on thermalization, as scattering processes are always active and without any loop suppression. 

For concreteness, let us consider the situation in which the portal field couples to a given SM fermion $\psi$ and the ALP is too light to decay to $\bar\psi \psi$. Two different classes of radiative corrections are in action ready to make the ALP unstable. First, triangle diagrams induce couplings to gauge bosons. We focus on the two massless gauge bosons relevant at the ALP mass scale, which are photons and gluons, and we find the induced couplings~\cite{Bauer:2017ris,Bauer:2020jbp}
\begin{subequations}
\begin{align}
\wc_\gamma = & \, 2 \, \wc_{\psi} \, N_c^{\psi} Q_\psi^2 \left( 1 - \frac{ \arcsin^2(x_\psi)}{x_\psi^2} \right) \ , \\
\wc_G = & \, \wc_{\psi} \, \left( 1 - \frac{ \arcsin^2(x_\psi)}{x_\psi^2} \right) \ .
\end{align}
\label{eq:loopINDbosons}
\end{subequations}
These one-loop generated Wilson coefficients are proportional to the coupling strength of the ALP with the original fermion $\wc_{\psi}$, its number of colors $N_c^{\psi}$, and the coupling to photons also depends on the fermion electric charge $Q_\psi$ (in units of the proton charge $e$). The function in parenthesis is evaluated for the argument $x_\psi \equiv m_\varphi / (2 m_\psi) < 1$. We notice how these couplings are phenomenologically relevant as long as the mass hierarchy is not too large. Once the fermion is too heavy, $x_\psi \ll 1$, these expressions go to zero. 

The second radiative correction we discuss is relevant when there exists a lighter fermion $\psi^\prime$ with a mass such that the decay $\varphi \rightarrow \bar\psi^\prime \psi^\prime$ is allowed. This decay will occur with a rate suppressed by a loop factor~\cite{Feng:1997tn}. In the unbroken electroweak phase of the SM, this radiative correction can be understood as a mixing between the fermion axial current and the Higgs current in \Eq{eq:JmuS}, as shown in Refs.~\cite{Crivellin:2014qxa,DEramo:2014nmf} in the context of WIMP direct detection. The resulting ALP-fermion coupling, evaluated at the rest-frame energy of the decaying ALP, is given by (see, e.g., App.~B of~\cite{DEramo:2018vss})
\be
\wc_{\psi^\prime} \simeq  - \,\wc_{\psi} \, N_c^{\psi} \frac{Y^2_\psi}{8 \pi^2} \ln\left( \frac{f_\varphi}{\Lambda_{\rm IR}} \right) \ . 
\label{eq:loopINDfermions}
\ee
We notice from the above equation that even if the interactions with the lighter fermion $\psi^\prime$ are absent at high energies, SM Yukawa interactions generate this coupling at low energies, starting from the one to the heavier fermion $\psi$. The effect is proportional to the Yukawa coupling of the fermion $\psi$. We stop the renormalization group evolution once we encounter the first relevant mass threshold at the scale $\Lambda_{\rm IR}  = {\rm max} \left\{m_\varphi, m_\psi \right\}$.

In what follows, we consider two benchmark scenarios: a leptophilic ALP coupled to electrons, and a hadrophilic $\varphi$ interacting with $b$ quarks.

\begin{figure}[t]
\centering
\subfloat[\label{fig:boundsDFSZ_electron}]{\includegraphics[width=0.48\textwidth]{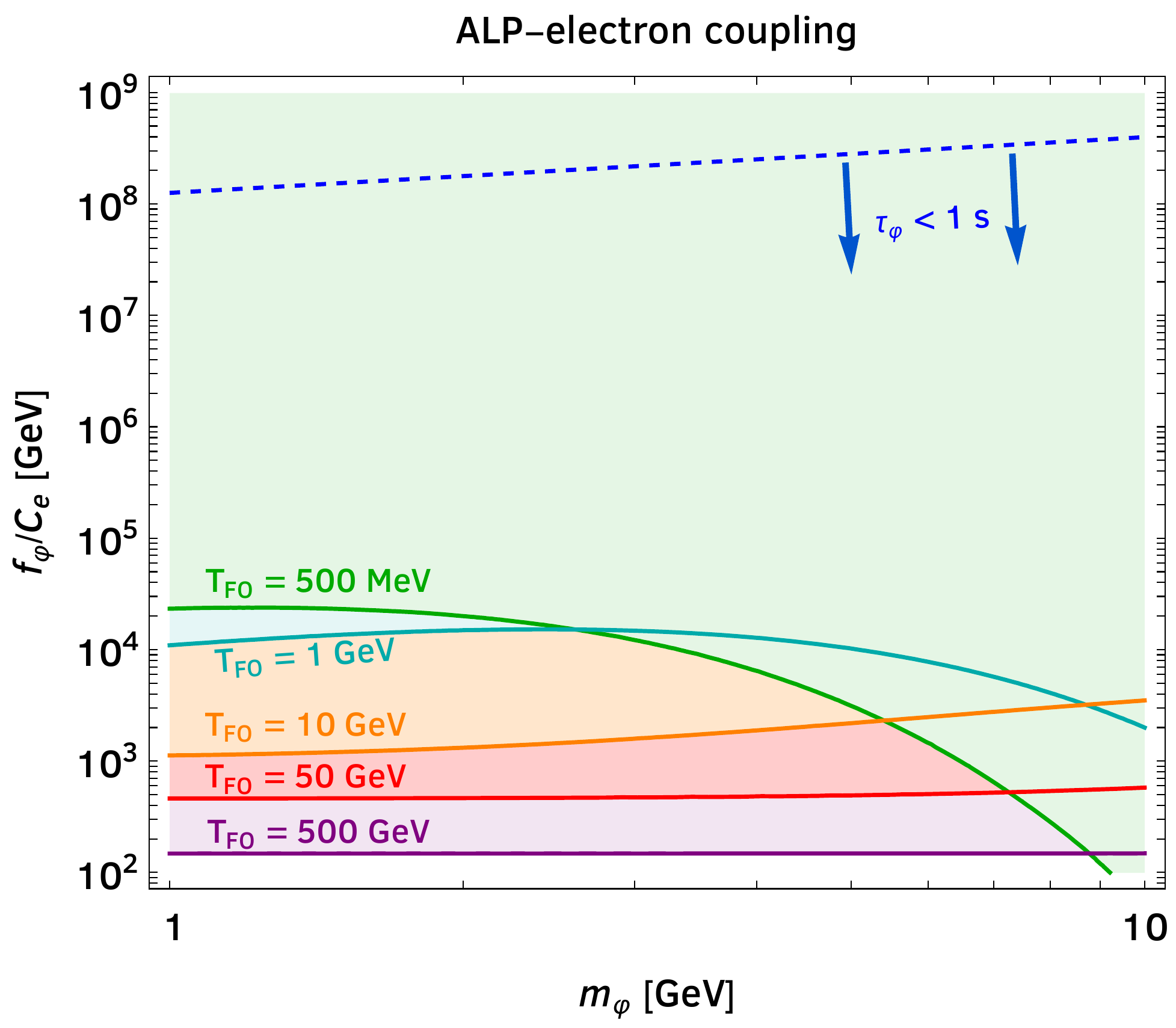}}\quad
\subfloat[\label{fig:boundsDFSZ_bottom}]{\includegraphics[width=0.49\textwidth]{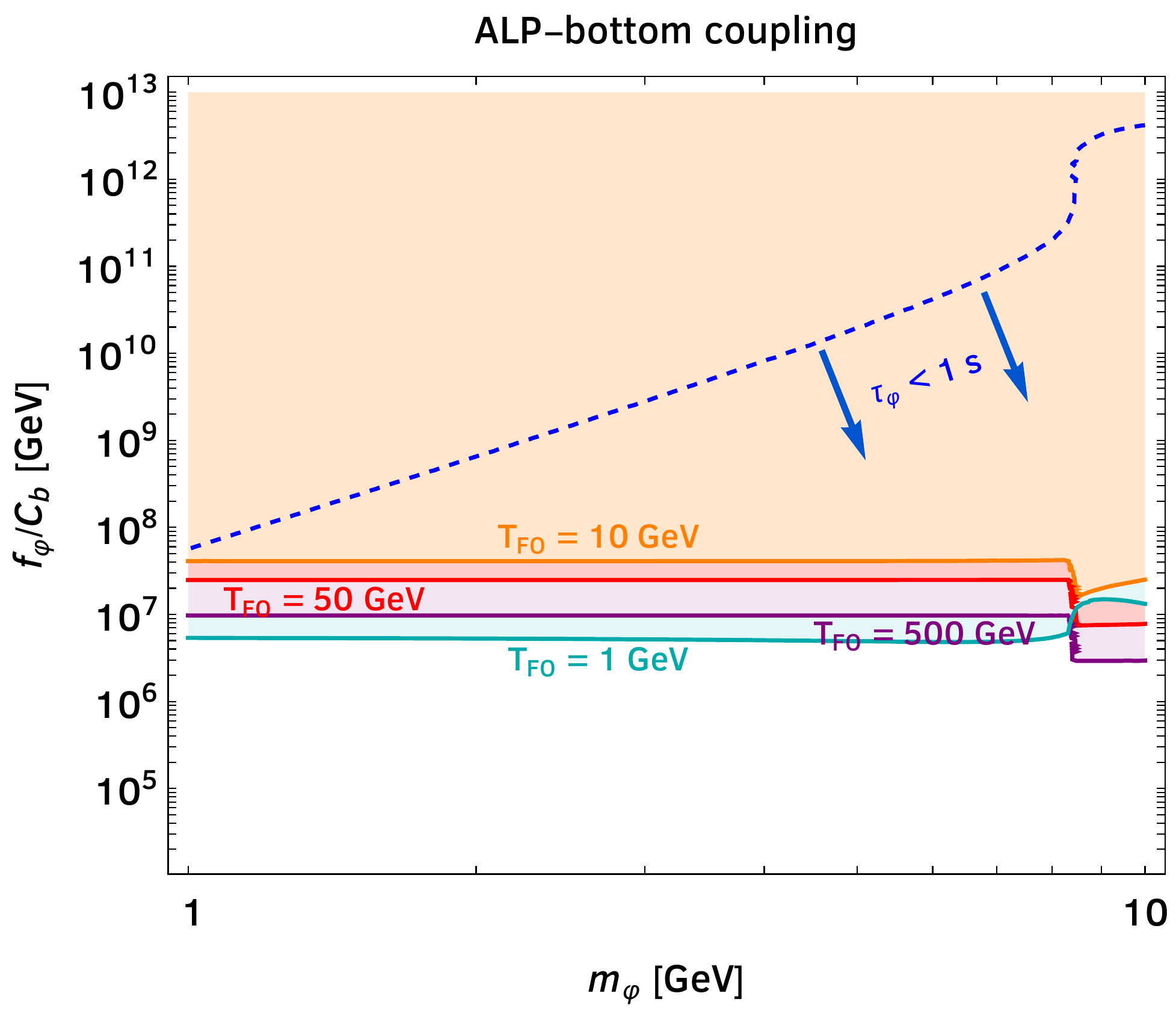}}
\caption{\textbf{Left:} ALP thermalization in the $(m_\varphi,\,f_\varphi / \wc_e)$ plane for coupling to electrons. The shaded regions identify where thermalization is not achieved at the corrisponding DM freeze-out temperatures. The dashed blue line identifies where the ALP decays after one second. \textbf{Right:} Same for an ALP coupled to bottom quarks.}
\label{fig:DFSZbounds}
\end{figure}

\paragraph{Electron coupling.} For an ALP coupled to electrons, ALP decays are always kinematically allowed in the mass range under investigation. The resulting lifetime reads
\be
\tau_{\varphi \rightarrow e^+ e^-} = \Gamma^{-1}_{\varphi \rightarrow e^+ e^-} \approx 0.6 \, {\rm s} \, \left(\frac{1}{\wc_e}\right)^2  \left(\frac{1\,\GeV}{m_\varphi}\right) \left(\frac{f_\varphi}{10^{8} \, \GeV} \right)^2 \ .
\ee
As usual, we set the parameters to identify the BBN benchmark of one second. For the reasons explained in the second paragraph of \Sec{subsec:DFSZpheno}, we keep only inverse decays for the thermalization rate when the DM freeze-out temperature approaches the ALP mass. The plot in \Fig{fig:boundsDFSZ_electron} shows the thermalization condition in the $(m_\varphi, f_\varphi / \wc_e)$ plane. As done already for the ALP coupled to gauge bosons, we fix representative values for the DM freeze-out temperature and shade away the regions where ALP interactions are too feeble to bring the field to achieve thermal equilibrium. We notice a striking difference with respect to the cases of gauge bosons. For coupling to electrons, the requirement of being in equilibrium at larger temperatures translates into the need for smaller values of the ALP decay constant (which implies larger couplings). This is exactly the opposite of what we observe in \Fig{fig:KSVZbounds}. The explanation for this different behavior is due to the aforementioned important fact that ALP scatterings mediated by couplings to SM fermions are IR dominated processes. In other words, they lose power as the temperature increases and this explains the need for stronger interaction strengths. We clearly observe this trend in \Fig{fig:boundsDFSZ_electron}, and things start changing only when the DM freeze-out temperature reaches the ALP mass with consequent effects from the Maxwell-Boltzmann suppression in their number density. 

For small DM masses, we notice how values of $f_\varphi / \wc_e$ between 1 and 10 TeV are compatible with the assumption that ALPs belong to the thermal bath at the freeze-out epoch. This easily reconciles with the relic density analysis in the previous section, where we observed that DM interactions of similar size can reproduce the observed abundance. Things start to get difficult at larger DM masses for which the freeze-out analysis is consistent only if the ratio $f_\varphi / \wc_e$ is below the TeV. This does not mean that the relic density analysis in inconsistent, we just need a small hierarchy among the Wilson coefficients, $\wc_e \gg \wc_S$. 

Finally, we notice how terrestrial bounds do not conflict with our conclusions. The BaBar collaboration has searched for a dark photon coupled to electrons~\cite{BaBar:2014zli}, and their bound can be recast for ALPs~\cite{Armando:2023zwz}. However, these constrains marginally touch the portion of parameter space we show in the figure.

\paragraph{Bottom coupling.} The last case we consider is an ALP coupled to bottom quarks. This scenario is enriched by the fact that the bottom quark mass is right in between the ALP mass range under investigation. In particular, tree-level decays to bottom/antibottom pairs are kinematically allowed only in the upper edge of the mass interval we consider. As a result, the ALP lifetime depends quite sensibly on the DM mass. For larger masses, $m_\varphi > 2 m_b$, the tree-level decays give a resulting lifetime
\be
\tau_{\varphi \rightarrow b \bar{b}} = \Gamma^{-1}_{\varphi \rightarrow b \bar{b}} \approx 1.4 \, {\rm s} \, \left(\frac{1}{\wc_b}\right)^2  \left(\frac{10 \, \GeV}{m_\varphi}\right) \left(\frac{f_\varphi}{5 \times 10^{12} \, \GeV} \right)^2 \qquad \qquad (m_\varphi > 2 m_b) \ .
\label{eq:taubottom1}
\ee
For light DM masses, $m_\varphi < 2 m_b$, this channel is kinematically forbidden and the couplings in \Eqs{eq:loopINDbosons}{eq:loopINDfermions} induced by radiative corrections set the lifetime. The dominant channel is to gluons with a resulting lifetime 
\be
\tau_{\varphi \rightarrow g g} = \Gamma^{-1}_{\varphi \rightarrow g g} \approx 0.6 \, {\rm s} \, \left(\frac{1}{\wc_b}\right)^2  \left(\frac{2 \, \GeV}{m_\varphi}\right)^7 \left(\frac{f_\varphi}{5 \times 10^{8} \, \GeV} \right)^2 \qquad \qquad (m_\varphi < 2 m_b) \ .
\label{eq:taubottom2}
\ee
We notice the prominent dependence on the ALP mass, much stronger than in \Eq{eq:taubottom1}, for a combination of two factors: the stronger dependence of the ALP width into gauge bosons as given in \Eq{eq:gammaALPgauge} and the loop function that appears in \Eq{eq:loopINDbosons}.

The plot in \Fig{fig:boundsDFSZ_bottom} illustrates this case in the $(m_\varphi, f_\varphi / \wc_b)$ plane. The dashed blue line identifying the ALP lifetime of one second changes its behavior drastically below the bottom threshold, $m_\varphi < 2 m_b$, and we observe the strong dependence on the ALP mass described by \Eq{eq:taubottom2}. The relative hierarchy of the thermalization boundaries for different freeze-out temperatures is again understood from the fact that thermalization via scatterings is an IR-dominated process. In particular, this explains why larger DM freeze-out temperatures require smaller values of $f_\varphi / \wc_b$ (i.e., stronger couplings). There is one exception for the line $T_{\rm FO} = 1 \, {\rm GeV}$, which requires the largest couplings. This seems in conflict with the IR domination. However, such a value of $T_{\rm FO}$ is smaller than both the ALP and DM masses. A thermal bath with a temperature of 1 GeV has difficulty producing ALPs with masses larger than the temperature from processes involving Maxwell-Boltzmann suppressed bottom quarks. To summarize, the ALP-bottom couplings required for a consistent relic density calculation are quite small, and this case is consistent with our analysis without any additional caveat.

\section{Signals in the gamma ray sky}
\label{sec:gammaray}

We dedicate this section to discussing a key signature of our framework. As illustrated in the second row of Tab.~\ref{tab:pheno}, the combined effect of the \ZT stabilizing symmetry and the ALP-mediated interactions makes indirect searches a promising avenue. The ALPs produced via semi-annihilations subsequently decay into SM states, with the specific coupling influencing only the branching ratio of the final states, not the overall rate. The intermediate state significantly alters the kinematics of the reaction chain, as opposed to the usual DM annihilations, leaving a distinctive imprint on the resulting spectrum of cosmic rays. We can frame our framework as a variation of the so-called \textit{cascade annihilation} scenario, where the fields of a secluded dark sector annihilate (or decay) into a number of intermediate unstable states, which eventually produce indirect signals after decaying into SM particles~\cite{Pospelov:2007mp,Fortin:2009rq,Elor:2015tva}. In this context, another example of ALP portal connections between the dark sector and the SM was investigated in Ref.~\cite{Mardon:2009rc}, with a final state multiplicity of three ALP particles in the final state. In our case, we only have one ALP particle in the final state of each semi-annihilation. The low-energy particle spectrum of our framework includes only $\varphi$ as a mediator, which can directly decay into visible sector states. We refer to this specific phenomenological chain as \emph{one-step cascade semi-annihilations}: $SS \to S^\star \varphi \to S^\star \, \text{SM}\, \text{SM}$.

The study of cascade chains is motivated by their distinctive spectral features, which enrich the range of signals typically investigated in the context of DM annihilations or  decays in astrophysical environments~\cite{Bringmann:2012ez,Slatyer:2017sev}. In particular, the presence of intermediate states naturally broadens the signal and shifts the final-state energies to lower values, generally leading to softer and weaker injection spectra. Moreover, these models have been explored as potential explanations for certain anomalies such as the gamma-ray Galactic Center excess~\cite{Elor:2015tva,Tempel:2012ey}, with an ALP portal example discussed in Ref.~\cite{Boehm:2014hva}, as well as other peculiar deviations in astrophysical observations that cannot be fully accounted for by SM physics alone~\cite{Bergstrom:2008ag,Cholis:2008wq,Bai:2012qy}.

A full computation of the final cosmic ray fluxes on Earth is beyond the scope of this paper. Nevertheless, we aim to provide the particle physics input, specifically the injection spectra defined as the differential distribution per unit energy of SM particles produced per single semi-annihilation. For illustration purposes, we will focus on photon spectra and discuss the production of $\gamma$ rays due to $\varphi$ decays into SM states, as well as the consequent photon emission in the particle shower induced by both leptonic and hadronic ALP channels.

\subsection{Kinematics of one-step cascade semi-annihilations}

We want to study the kinematics of a single semi-annihilation process $SS\to S^\star\varphi$ followed by the ALP decay into two SM states. We work in the galactic frame (GF) that roughly coincides with the DM rest frame. Indeed, DM is cold at present times and semi-annihilations occur in the non-relativistic regime with a cross section well approximated by the s-wave expansion in \Eq{eq:sigmasemiswave}. The functional form of the cross section is not important for our purposes since our goal is to determine the injection spectra per single semi-annihilation. 

For DM initial state particles at rest in the GF, the final state ALP energy and spatial momentum are given by the following expressions
\be
E^\textup{(GF)}_\varphi = \frac{3}{4} m_S \left( 1 + \frac{\epsilon_\varphi^2}{3} \right) \ , 
\qquad \qquad p^\textup{(GF)}_\varphi = m_S \sqrt{\lambda\left(1, \frac{1}{2}, \frac{\epsilon_\varphi}{2} \right)} \ ,
\ee
where we introduce the dimensionless mass ratio $\epsilon_\varphi \equiv m_\varphi / m_S$ and the Källén function $\lambda(x, y, z) \equiv [x - (y - z)^2] [x - (y + z)^2]$. Corrections of $O(\epsilon_\varphi^2)$ extend the result provided in \cite{DEramo:2010keq} where the mediator was taken to be always much lighter than the DM. For all the cases studied in this paper, ALP decays have two monochromatic SM particles in the final state, $\varphi \rightarrow X X$. Examples for the particle $X$ include photons, gluons, electrons and bottom quarks as discussed in \Sec{sec:ALP}. Since the ALP is a (pseudo-)scalar particle, the angular distribution of its decay products is isotropic in its rest frame. The corresponding differential distribution of $X$ particles reads
\be
\frac{d N_X}{d E_X^\prime \, d\cos\theta^\prime} = \delta\left(E_X^\prime - \frac{m_\varphi}{2}\right) \ .
\label{eq:dNXrest}
\ee
The angle $\theta^\prime$ is defined as the direction of the emitted $X$ particles in the ALP rest frame with respect to the line of flight of the $\varphi$ produced via semi-annihilations. In this section, quantities referred to the ALP rest frame are always denoted with a prime. The total number of emitted $X$ particles is a Lorentz invariant quantity so the following normalization condition must be satisfied in any reference frame
\be
\int d E_X \, d\cos\theta \frac{dN_X}{d E_X \, d\cos\theta} = 2 \ .
\label{eq:normID}
\ee
In particular, it is straightforward to check the validity of this condition in the ALP rest frame with the differential distribution given in \Eq{eq:dNXrest}.

We derive now the differential energy distribution of the $X$ particles in the GF. This is connected with the ALP rest frame, where the distribution is given in \Eq{eq:dNXrest}, by a Lorentz boost parametrised by the factors $\beta_L \equiv p_\varphi / E_\varphi$ and $\gamma_L \equiv (1 - \beta_L^2)^{-1/2} = E_\varphi / m_\varphi$. The $X$ kinematical variables in the two frames are connected by the relation
\be\label{eq:boostedE}
E^\textup{(GF)}_X=\gamma_L(E'_X+\beta_L\,p_X'\cos\theta')\,\,.
\ee
Relying on such equality, the injection spectrum in the GF is thus computed to be
\be\label{eq:ID1}
\frac{dN_X}{dE_X} =  \int d\cos\theta^\prime dE'_X \frac{dN_X}{dE'_X \, d\cos\theta^\prime} \delta\left(E_X-E^\textup{(GF)}_X \right)\,\,.
\ee
The Dirac delta function enforces the energy value given in \Eq{eq:boostedE}.

The final state energy is proportional to the DM mass which can vary over several orders of magnitude. Furthermore, the spectrum will also depend on the ALP mass. It is convenient to introduce additional dimensionless variables
\begin{align}
\epsilon_X \equiv \frac{2m_X}{m_\varphi}\,,&&x_\varphi \equiv \frac{2E'_X}{m_\varphi}\,,&&x_S\equiv\frac{E_X}{m_S}\,\,.
\end{align}
We can recast \Eq{eq:ID1} for the injection spectra of $X$ particles as follows
\be\label{eq:ID2}
\frac{dN_X}{dx_S} = 2 \int d\cos\theta^\prime dx_\varphi \frac{dN_X}{dx_\varphi \, d\cos\theta'} \, \delta\left(2x_S-\frac34\left(x_\varphi\left(1+\frac{\epsilon_\varphi^2}{3}\right)+\zeta \sqrt{x_\varphi^2-\epsilon_X^2}  \cos\theta'\right)\right)\,\,,
\ee
where we define the quantity $\zeta \equiv\frac43 \frac{p^\ttiny{(GF)}_\varphi}{m_S}$. As aforementioned, ALP primary decay products are distributed isotropically in the ALP rest frame
\be
\frac{dN_X}{dx_\varphi \, d\cos\theta'} = \frac{1}{2} \frac{dN_X}{dx_\varphi} \ .
\ee
The differential energy spectrum is monochromatic as prescribed by \Eq{eq:dNXrest}. In the last part of this derivation, we leave the dependence $\frac{dN_X}{dx_\varphi}$ implicit because, as we will explain below, our derivation is actually more general and not only applicable to find the distribution of the ALP decay primary products. Performing the angular integral via the Dirac delta function, we obtain the final version of the spectrum
\begin{subequations}
\begin{align}
\label{eq:ID3a}\frac{dN_X}{dx_S}&=\frac{4}{3 \zeta}\int_{x_\varphi^\textup{min}}^{x_\varphi^\textup{max}} \frac{dx_\varphi}{\sqrt{x_\varphi^2-\epsilon_X^2}} \frac{dN_X}{dx_\varphi}\,\,,\\
x_\varphi^\textup{min}&=\frac{8 x_S \left(3 + \epsilon _{\varphi }^2 \right) - 3 \zeta  \sqrt{64 x_S^2 - \epsilon _X^2 \left[ \left(3 + \epsilon _{\varphi }^2 \right) - 9 \zeta^2 \right]}}{\left(3 + \epsilon _{\varphi }^2 \right) - 9 \zeta^2}\,\,,\\
x_\varphi^\textup{max}&=\text{Min}\left\{1,\frac{8 x_S \left(3 + \epsilon _{\varphi }^2 \right) + 3 \zeta  \sqrt{64 x_S^2 - \epsilon _X^2 \left[ \left(3 + \epsilon _{\varphi }^2 \right) - 9 \zeta^2 \right]}}{\left(3 + \epsilon _{\varphi }^2 \right) - 9 \zeta^2} \right\}\,\,.
\end{align}
\end{subequations}
The expression in \Eq{eq:ID3a} is the injection spectra for any possible final state $X$ produced via two-body decays $\varphi \rightarrow X X$. As anticipated, this result actually has a much broader validity. Our derivation was based on \Eq{eq:ID1} that connects particle number distributions in the two reference frames without the assumption that $X$ comes from ALP two-body decays with the monochromatic distribution given in \Eq{eq:dNXrest}. However, we need the assumption that $X$ particles are distributed isotropically in the ALP rest frame. This is true for all the cases considered in this work. Therefore, we can use \Eq{eq:ID3a} as our general master formula for the computation of injection spectra of both ALP direct decay products and stable SM states produced via their subsequent evolution.

At this point, what we need to do next depends on whether $X$ is stable and on what specific SM state we aim to detect. If $X$ is unstable, we have to follow the subsequent evolution (e.g., final state radiation and/or hadronization) and evaluate the spectrum of the searched for stable SM state. For stable $X$ particles, we could either search to detect them or follow their subsequent evolution (e.g., final state photons emitted from electrons/positrons). 
 
\subsection{Gamma ray injection spectra}
 
Among possible detectable SM stable states, we choose to focus on gamma ray photons. The only missing ingredients to compute the injection spectra are the differential energy distributions of photons in the ALP rest frame arising from the two-body decays. We consider three exemplary scenarios with $X = (\gamma,\,g,\,e)$. They correspond to the scenarios already discuss in \Sec{sec:ALP} for the ALP cosmology with the only exception of the bottom quark. The reason why we omit this case from the present discussion is understood from the ALP mass window under investigation. Only at the edge of the considered mass window, $2 m_b < m_\varphi \leq 10 \, {\rm GeV}$, ALPs are kinematically allowed to decay into bottom/antibottom pairs. In this situation, where $\epsilon_b \simeq 1$, we saturate the residual amount of available energy to be released in radiation with a resulting negligible $\gamma$ ray spectra. For lighter ALPs, fermion/antifermion decays are forbidden and the ALP decays to gluons via radiatively induced couplings as discussed in \Sec{sec:ALP}. Even if they are loop suppressed, they can be considered prompt for the purposes of indirect detection since the overall rate is set only by DM-ALP interactions. As a consequence, this case is equivalent to $X = g$. The three remaining cases that we discuss exhibit the key features described below.

\begin{itemize}
\item $\varphi\to\gamma\gamma$. The process can take place for any value of $m_\varphi$ since final states are massless. This scenario is worth an accurate discussion for two main reasons. First, it allows for a straightforward derivation of the injection spectrum: the differential photon spectrum in the ALP rest frame is a Dirac delta function, corresponding to monochromatic lines at $m_\varphi/2$. Given the Lorentz boost \Eq{eq:boostedE} relating the two reference frames, we can obtain the explicit relationship for the differential radiation energy distribution per single semi-annihilation event in the GF

\be\label{eq:IDbox}
\frac{dN_\gamma}{dE_\gamma}=\frac{N_\gamma}{\bfp_\varphi}\Theta(E_\gamma-E_\gamma^-)\Theta(E_\gamma^+-E_\gamma)\,\,,
\ee

where the two Heaviside step functions limit the support of the spectrum to be within $[E_\gamma^-,E_\gamma^+]$, with $E_\gamma^\pm=(1\pm\beta_L)E_\varphi/2$.  For large mass hierarchies $m_S\gg m_\varphi$, this translates into $[0,3m_S/4]$. The spectrum is consistently normalized to the number of emitted photons $N_\gamma = 2$. We understand the second reason why this case deserves a thorough study on its own: the gamma rays spectrum per single reaction assumes the quite peculiar shape of a box function centered in $E_\varphi/2$, with the strength of the signal dictated by $m_S$ while the width depends on $m_\varphi$. The features of such scenarios have been intensively studied in \cite{Ibarra:2012aa}. Furthermore, if we smear \Eq{eq:IDbox} with a Gaussian function resembling the \textit{Fermi} Large Area Telescope (LAT) energy resolution \cite{Rando:2009yq} with $\sigma(E)/E\simeq0.1$ we obtain the curves reported in \Fig{fig:boxspectrum}. For nearly degenerate masses (solid green curve), $m_S\simeq m_\varphi$, the one-step semi-annihilation signal approaches the shape of a Dirac delta function since the ALP is produced almost at rest. The resulting phenomenology is that of a direct decay of a particle at rest, leading to $\gamma$ lines at $E_\gamma=m_\varphi/2\simeq m_S/2$. We compare this result with the line signal produced by direct annihilation of $10\,\GeV$ DM particles into photons (dot-dashed black spike). On the other hand, as the DM-ALP mass splitting increases, it is immediately evident that semi-annihilation effects result in a soft photon spectrum with a quite strong broadening. The signal suppression is exaggerated for $m_S\geq100\,\GeV$ -- where the correction from $m_\varphi$ becomes negligible -- and the shape becomes that of a unique plateau extended over more orders of magnitude.

 \begin{figure}[t]
\centering
\includegraphics[width=0.5\textwidth]{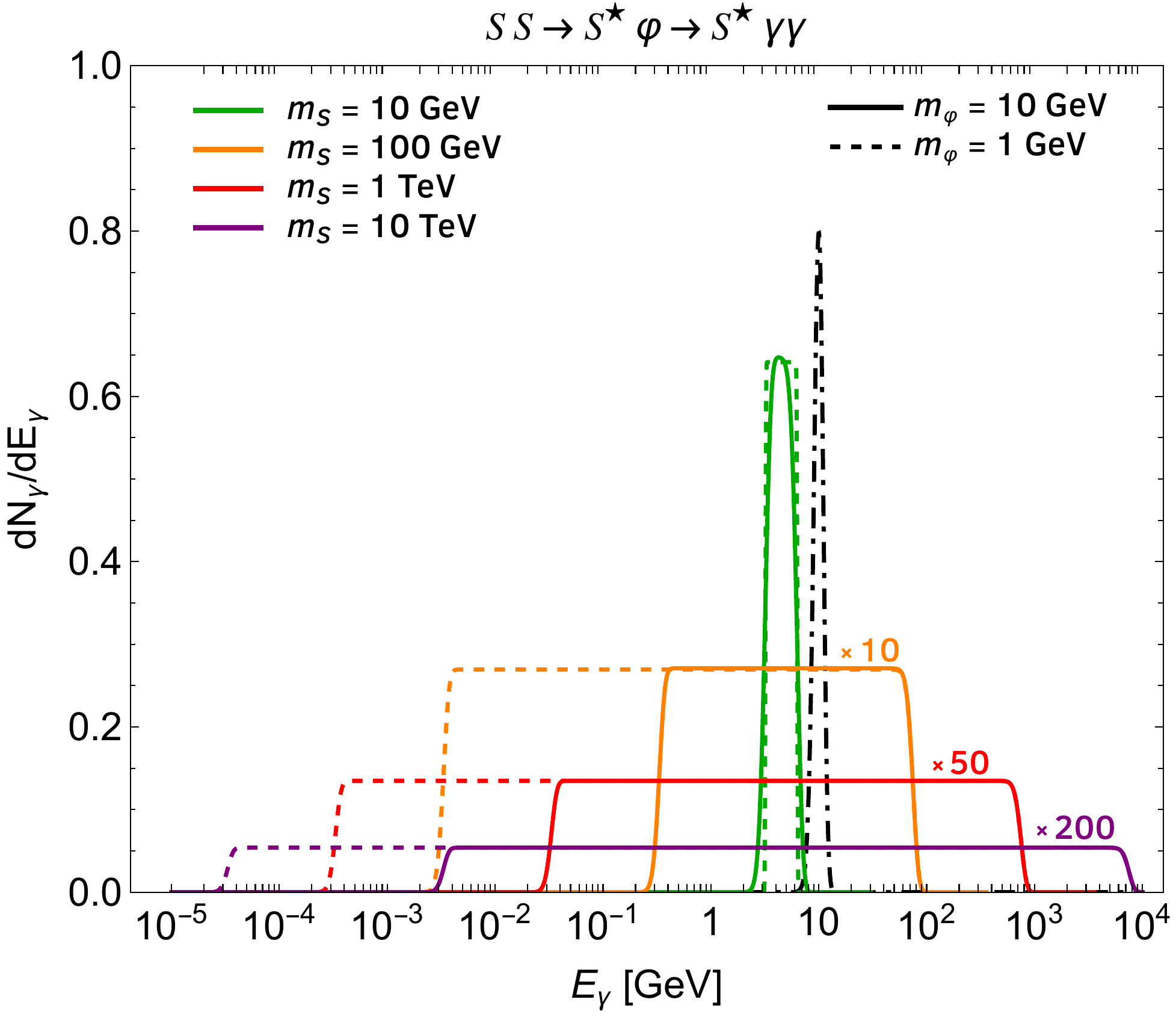}
\caption{Box-shaped $\gamma$-ray injection spectra from one-step cascade semi-annihilation ending directly into photons. The displayed curves are the convolution of \Eq{eq:IDbox} with a gaussian smearing with sensitivity $\sigma(E)/E=0.1$. The dot-dashed black line corresponds to a photon line for DM annihilation with $m_{DM}=10\,\GeV$ with the same smearing.}
\label{fig:boxspectrum}
\end{figure}

\item $\varphi\to gg$. For the same reason as the $\gamma\gamma$ channel, this decay is also always kinematically available. In this case, however, the process leading to the production of gamma rays is more involved since gluons eventually hadronize, leading to a shower of particles and the photon spectrum at production must be computed numerically.

\item $\varphi\to e^+e^-$. This process is always kinematically available in the ALP mass range of interest. Here, the photon is emitted per effect of Final State Radiation (FSR) and the injection spectrum can be computed analytically (see Appendix A of \cite{Mardon:2009rc}).
\end{itemize}

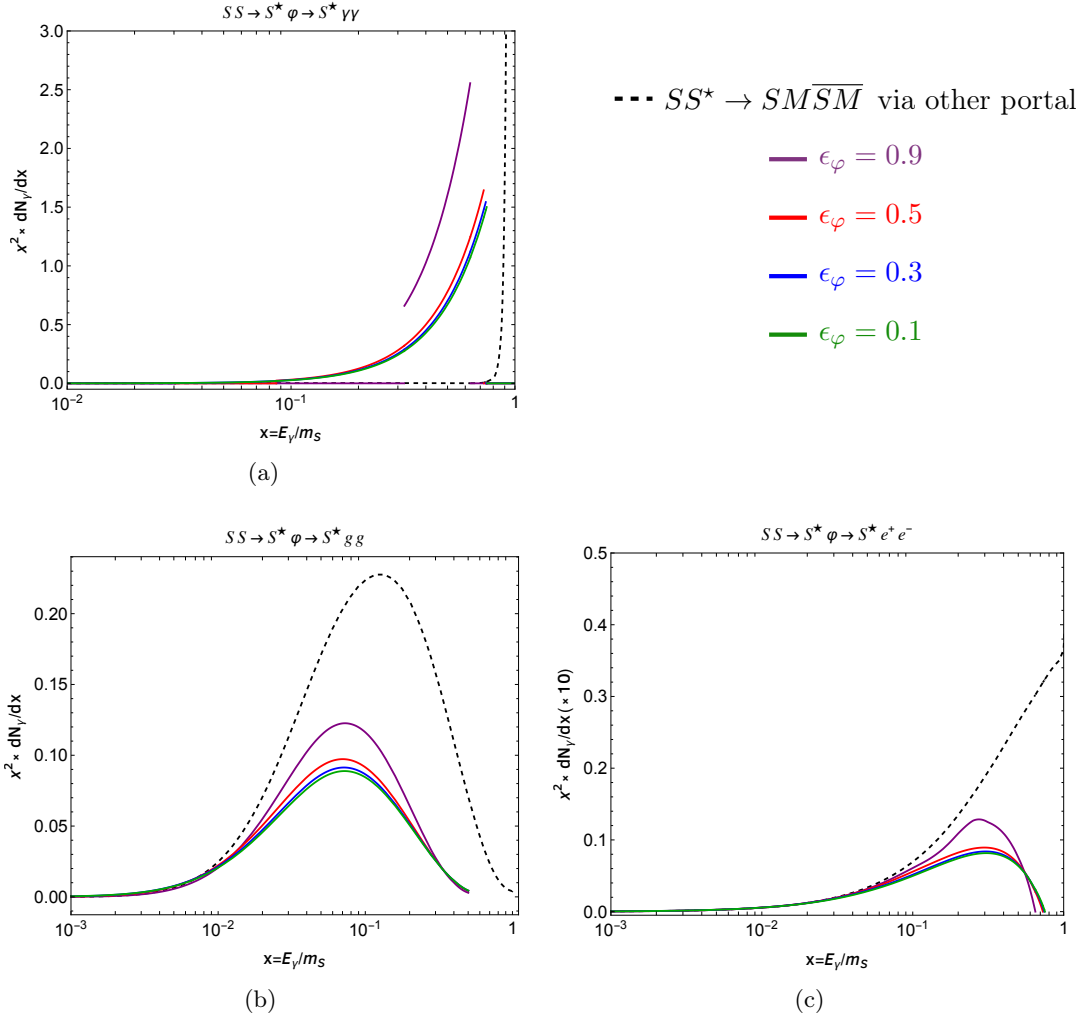
\begin{figure}[t]
    \centering
    \subfloat[\label{fig:IDgamma}]{\includegraphics[width=0.45\textwidth]{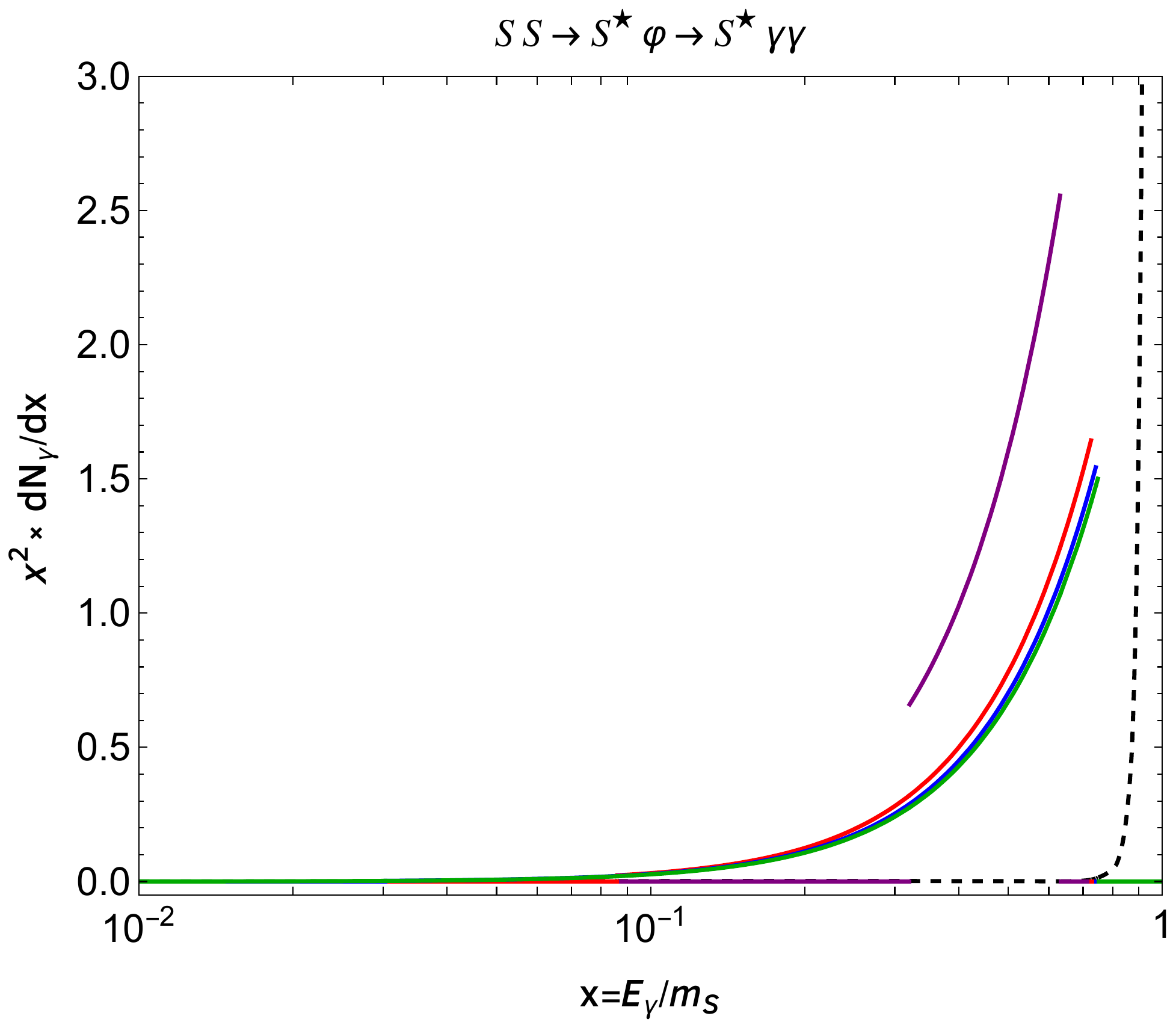}}\hspace{1.95em}
       \subfloat{
        \begin{minipage}{0.4\textwidth}
            \vspace{-20em}
            \centering
            \begin{tabular}{c}
                 \begin{tikzpicture}
                     \draw[black,dashed, ultra thick] (-1.5,0)--(-1,0) node[right]{$SS^\star\to SM\overline{SM}\,\,\,\text{via other portal}$};
                 \end{tikzpicture} \\
                 \begin{tikzpicture}
                     \draw[Purple, ultra thick] (-1.5,-0.5)--(-1,-0.5) node[right]{$\epsilon_\varphi=0.9$};
                 \end{tikzpicture} \\
                 \begin{tikzpicture}
                     \draw[red, ultra thick] (-1.5,-1)--(-1,-1) node[right]{$\epsilon_\varphi=0.5$};
                 \end{tikzpicture} \\
                 \begin{tikzpicture}
                     \draw[blue, ultra thick] (-1.5,-1.5)--(-1,-1.5) node[right]{$\epsilon_\varphi=0.3$};
                 \end{tikzpicture} \\
                 \begin{tikzpicture}
                     \draw[ColorT, ultra thick] (-1.5,-2)--(-1,-2) node[right]{$\epsilon_\varphi=0.1$};
                 \end{tikzpicture}
            \end{tabular}
            \vspace{-5em}
        \end{minipage}
    }\addtocounter{subfigure}{-1}\\
    \subfloat[\label{fig:IDgluon}]{\includegraphics[width=0.455\textwidth]{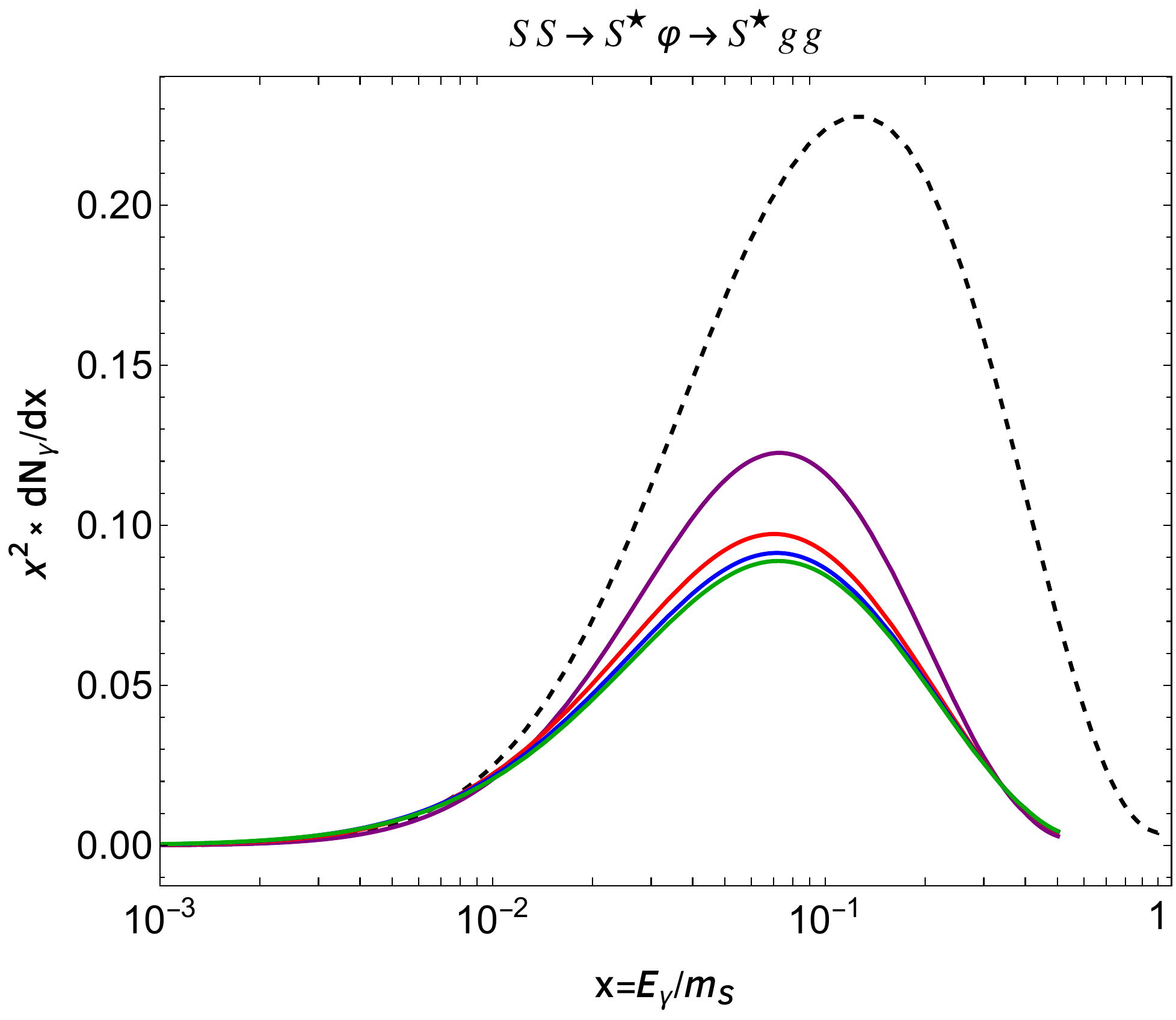}}\quad
    \subfloat[\label{fig:IDel}]{\includegraphics[width=0.455\textwidth]{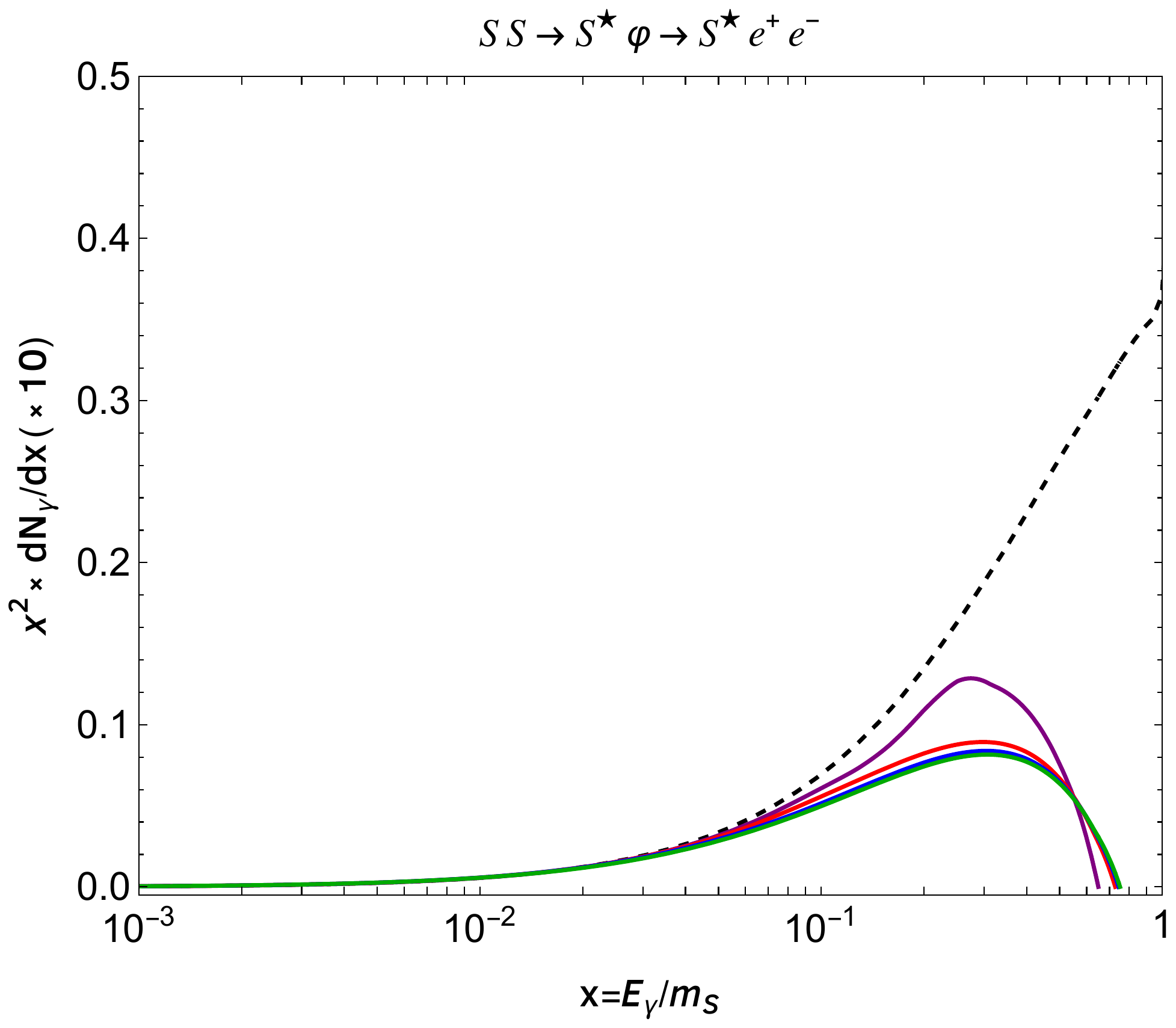}}
    \caption{Gamma ray spectra resulting from one-step semi-annihilation chain processes ending into \emph{photons} (\ref{fig:IDgamma}), \emph{gluons} (\ref{fig:IDgluon}) and \emph{electrons} (\ref{fig:IDel}). Different colored curves correspond to different values of the mass ratio $\epsilon_\varphi \equiv m_\varphi / m_S$. For comparison, we show as dashed black lines also spectra for DM annihilations with $m_{DM}=10\,\GeV$.}
    \label{fig:IDspectra}
\end{figure}

For each of the necessary numerical inputs for the production spectra we rely on the existing code provided by Ref.~\cite{Cirelli:2010xx}. These functions are given for annihilation processes, thusly, in order to properly use them we need to set the mass of the DM annihilating particle to $m_\varphi/2$. An alternative public code has been released by Ref.~\cite{Elor:2015bho}, where they generically treat a chain of \emph{n} intermediate decays sparked by DM annihilations. If using this code, we should account for an extra factor of $1/2$ to correctly lower final states degeneracy.

The injection spectra resulting from one-step cascade semi-annihilations ending in different SM states is computed via \Eq{eq:ID2} by fixing the ALP to DM mass ratio $\epsilon_\varphi$. Another potentially relevant quantity is the final state to ALP mass ratio $\epsilon_X$. However, our final states are either massless (photons, gluons) or much lighter than the ALP (electrons), and we can always neglect $O(\epsilon_X^2)$ in \Eq{eq:ID3a}. The spectral shape is uniquely determined by the value of $\epsilon_\varphi$ which clearly has a degenerate meaning. Varying $\epsilon_\varphi$ can be both interpreted as fixing the ALP mass and changing $m_S$ and viceversa, with masses varying in their ranges. \Fig{fig:IDspectra} shows how photon injection spectra for one-step semi-annihilations turn out to be broadened and softer. For $\epsilon_\varphi\to1$ the spectrum tends to be degenerate with that of a direct decay of a particle of mass $m_\varphi\simeq m_S$. On the other hand, as $\epsilon_\varphi\ll1$, ALP mass effects become negligible and the end point of the spectrum is found at $E_\gamma=3/4 m_S$.

\section{One explicit UV complete theory}
\label{sec:UV}

All the results presented thus far are model-independent and do not rely on any assumptions about the UV complete theory. The relic density calculation in \Sec{sec:relic}, the analysis of ALP cosmology in \Sec{sec:ALP}, and the indirect detection spectra in \Sec{sec:gammaray} all involve energy scales well below the cutoff of the EFT introduced in \Sec{sec:EFT}. Consequently, any UV complete model that matches onto this EFT will yield the same phenomenology, underscoring the versatility and robustness of our framework. To conclude, we present a concrete example of a microscopic theory that remains valid at arbitrarily high energies and, upon integrating out heavy degrees of freedom, reproduces our low-energy EFT. Specifically, we construct a UV-complete theory in which the ALP arises as a Nambu-Goldstone boson of a spontaneously broken Abelian symmetry, featuring low-energy couplings to SM gauge bosons.

The new degrees of freedom consist of two complex scalars, $\Phi$ and $S$, both of which are SM singlets, and a Dirac fermion $\Psi$ with vector-like quantum numbers under \SMG. We introduce a global $U(1)_\varphi$ symmetry under which the new fields carry the following Abelian charges:  
\be
Q_\Phi = -3 \ , \qquad Q_S = +1 \ , \qquad Q_{\Psi_L} = \xi_\Psi  \ , \qquad Q_{\Psi_R} = \xi_\Psi - 3 \ .
\ee
Here, $\Psi_{L,R} = P_{L, R} \Psi$ are obtained using the chiral projectors $P_{L, R} = (1 \mp \gamma^5) / 2$ acting on the Dirac fermion $\Psi$. The parameter $\xi_\Psi$ is an arbitrary real number that does not affect the final results. All SM fields remain neutral under this symmetry. 

The most general Lagrangian compatible with these requirements reads
\be
\mathcal{L}_{\rm UV} = \mathcal{L}_{\rm SM} + |\partial_\mu S|^2+|\partial_\mu \Phi|^2 + \bar\Psi i \slashed{D} \Psi - V^{(\rm UV)}(H, S, \Phi) - \left[ y_\Psi \Phi^\dag \bar\Psi_L \Psi_R + {\rm h.c.} \right] \,.
\label{eq:LUV}
\ee
The first contribution is the usual SM Lagrangian, which also includes scalar potential terms with self-interactions of the Higgs doublet $H$. Besides the canonically normalized kinetic terms for the new fields, the remaining interactions fall into two categories. First, scalar potential interactions involving the three scalar fields in the theory—the SM Higgs doublet and the two new singlets—are allowed whenever they respect gauge invariance and the $U(1)_\varphi$ symmetry. The only renormalizable non-potential interaction that respects these symmetries is the Yukawa operator involving $\Phi$ and $\Psi$.

Our objective is to match this theory onto the low-energy EFT presented in \Sec{sec:EFT}. To achieve this, we must determine the mass spectrum and integrate out the heavy degrees of freedom, which first requires identifying the vacuum state of the theory. For this purpose, we need the explicit expression of the scalar potential and we find it convenient to classify the potential into distinct components:
\be
V^{(\rm UV)}(H, S, \Phi) = V^{(\rm UV)}_S(S) + V^{(\rm UV)}_\Phi(\Phi) + V^{(\rm UV)}_{\rm mix}(H, S, \Phi) \,.
\ee  
The self-interactions of the SM Higgs doublet $H$ are already accounted for in the $\mathcal{L}_{\rm SM}$ term of \Eq{eq:LUV}, while the DM mass and self-interactions match those given in \Eq{eq:potentialZ3} with the exception of the $S^3$ term that is forbidden by the $U(1)_\varphi$ symmetry. The portion of the potential involving only the complex scalar $\Phi$ reads:
\be
V^{(\rm UV)}_\Phi(\Phi) = - \mu_\Phi^2 \Phi^\dagger \Phi + \frac{\lambda_\Phi}{4} (\Phi^\dagger \Phi)^2 \,.
\ee  
We choose $\lambda_\Phi > 0$ to guarantee a potential bounded from below and take $\mu_\Phi^2 > 0$ to induce spontaneous symmetry breaking of $U(1)_\varphi$ via a non-vanishing vev for $\Phi$. Next, we consider the mixing terms involving two or more distinct scalar fields. All quartic terms formed by multiplying a scalar field and its Hermitian conjugate with analogous combinations of other scalars remain invariant. Additionally, the trilinear combination $\Phi S^3$ respects the symmetry structure of our theory. The most general set of mixing terms is
\be
V^{(\rm UV)}_{\rm mix}(H, S, \Phi) = \lambda_{HS} \, H^\dag H \, S^\dag S + \lambda_{\Phi H} \, \Phi^\dag \Phi \, H^\dag H + \lambda_{\Phi S} \, \Phi^\dag \Phi \, S^\dag S + \left[\kappa_\star \Phi S^3 + {\rm h.c.} \right] \,.
\ee  
The Higgs portal coupling $\lambda_{HS}$ is assumed to be small, as discussed in \Sec{sec:EFT}. Similarly, the quartic couplings $\lambda_{\Phi H}$ and $\lambda_{\Phi S}$ are kept sufficiently small to avoid introducing significant corrections to the mass terms of $H$ and $S$ when $\Phi$ acquires a non-zero vev and breaks the $U(1)_\varphi$ symmetry. The trilinear interaction proportional to $\kappa_\star$ plays a pivotal role, as it generates the cubic DM self-interaction at low energies.

The moment has come to take action, so we now expand the field $\Phi$ around its vev
\be
\Phi = \left( \frac{v_\Phi + \rho}{\sqrt{2}} \right) \exp\left[ i \frac{\varphi}{v_\Phi} \right] \ .
\ee
The radial mode $\rho$ takes a mass of order $v_\Phi$ and we can neglect its low-energy dynamics. The phase $\varphi$ is the massless Nambu-Goldstone boson arising from the spontaneous breaking of the $U(1)_\varphi$ symmetry. We focus on the Nambu-Goldstone low-energy interactions
\be
\mathcal{L}_\varphi = \frac{1}{2} (\partial_\mu \varphi)^2 - \frac{v_\Phi}{\sqrt{2}} \left[\kappa_\star  \exp\left[ i \frac{\varphi}{v_\Phi} \right] S^3 + y_\Psi \exp\left[-  i \frac{\varphi}{v_\Phi} \right] \bar\Psi_L \Psi_R + {\rm h.c.} \right] \ .
\ee
The mass of $\Psi$ is also of order $v_\Phi$ and we can integrate it out as well. This procedure is performed conveniently by redefining the fermion field itself via the chiral rotations\footnote{We can write this transformation compactly with the Dirac field as $\Psi \; \rightarrow \exp\left[ i \frac{\varphi}{2 v_\Phi} \gamma^5 \right] \Psi$.}
\be
\Psi_L \; \rightarrow \exp\left[ - i \frac{\varphi}{2 v_\Phi} \right] \Psi_L \ , \qquad \qquad 
\Psi_R \; \rightarrow \exp\left[ i \frac{\varphi}{2 v_\Phi} \right] \Psi_R \ .
\ee
This transformation effectively remove the couplings between the fermion $\Psi$ and the Nambu-Goldstone boson $\varphi$. However, this does not make all the interactions for $\varphi$ vanish because the above field redefinition is anomalous. For concreteness, we assume that the fermion $\Psi$ is charged under the fundamental of the color group $SU(3)_c$ and it carries also non-vanishing hypercharge $Y_\Psi$. After the fermion chiral rotations above we find
\be
\mathcal{L}^{N \partial}_\varphi \supset  \frac{\varphi}{8 \pi v_\Phi} \left[\alpha_G  G_{\mu\nu} \widetilde G^{\mu\nu} +
2 Y_\Psi^2 \alpha_B  B_{\mu\nu} \widetilde B^{\mu\nu} \right] + \frac{v_\Phi}{\sqrt{2}} \left[\kappa_\star  \exp\left[ i \frac{\varphi}{v_\Phi} \right] S^3 + {\rm h.c.} \right] \ .
\label{eq:EFTUV}
\ee
We implicitly assume that a non-vanishing mass term $m_\varphi$ for the ALP is generated by some soft breaking contribution. We also appreciate how the low energy theory displays a remnant \ZT symmetry under which the complex scalar field $S$ transforms according to $S\to e^{2i\pi/3}S$. This is the symmetry that makes the scalar field $S$ stable and a viable DM candidate. The interactions provided in \Eq{eq:EFTUV} are expressed in what we called the non-derivative basis in \Sec{sec:EFTNder}. We are free to set $\wc_G = 1$, and the remaining Wilson coefficients are matched as follows 
\be
f_\varphi = v_\Phi \ , \qquad \qquad \wc_B = 2 Y_\Psi^2 \ , \qquad \qquad A = 3 \sqrt{2} \, \kappa_\star v_\Phi \ .
\label{eq:matchingND}
\ee

We notice how integrating out the heavy degrees of freedom also induces other operators such as $\lambda_5 S^4 S^\dag / (4! v_\Phi)$. The matching procedure yields the result $\lambda_5 = 6 \sqrt{2} k_\star \lambda_{\Phi S} / \lambda_\Phi$. This operator could potentially affect the relic density since it changes the overall number of DM particles by one unit. In particular, the transition amplitude is suppressed by only one power of the ALP decay constant exactly as the one for semi-annihilation. However, this process is expected to be subdominant because of the need of having one extra particle on the external leg. The assumption we made of negligible mixing terms, $\lambda_{\Phi S} \ll 1$, makes this contribution even more negligible. 

One can recover the same theory in the derivative basis by performing the $\varphi$ dependent redefinition $S\to\exp\left[-i  \frac{\varphi}{3 v_\Phi}\right]S$. This operation  cancels out scalar potential interactions between $\varphi$ and $S$ in \Eq{eq:EFTUV} and induces the following ALP derivative couplings 
\be
\mathcal{L}^{\partial}_\varphi \supset \frac{\partial_\mu\varphi}{3 v_\Phi}S^\dagger i\dd^\mu S+\frac{1}{9 v_\Phi^2} S^\dag S (\partial_\mu \varphi)^2 \ .
\label{eq:changebasis}
\ee
We notice how we also generate the dimension 6 ALP derivative interaction  
\be
\delta\mathcal{L}_{\rm INT}^{(6)} = \frac{\wc^{\prime\,2}_S}{4f_\varphi^2}|\partial_\mu\varphi|^2S^\dag S \ .
\label{eq:EFTdim6}
\ee
The matching between the result obtained from the UV complete theory in \Eq{eq:changebasis} and the EFT defined in \Eq{eq:Lag} augmented with the operator in \Eq{eq:EFTdim6} is achieved via the identification $\wc_S = \wc^\prime_S = 2/3$.

We conclude this section with a brief aside. The EFT developed in \Sec{sec:EFT} limits our analysis to operators up to dimension 5. This restriction means that we cannot make reliable predictions for the $2 \to 2$ annihilation processes $SS^\star \leftrightarrow \varphi\varphi$ unless we also specify the size of the dimension 6 operator in \Eq{eq:EFTdim6}. The UV-complete theory enables us to compute the cross section for this process and demonstrates that it vanishes. This is straightforward in the non-derivative basis, where all interactions for the field $\varphi$ are contained in \Eq{eq:EFTUV}, and we can clearly see that this process is forbidden at tree level. Although the result can also be derived in the derivative basis, it is less trivial to observe its non-occurrence. The first operator in \Eq{eq:changebasis} contributes to the $SS^\star \leftrightarrow \varphi\varphi$ annihilation amplitude when inserted twice, and the second term provides an equal and opposite effect to the dimension 5 coupling, as explicitly computed in \Eq{eq:XSALPairs} of \App{app:DMsupp}. This example further strengthens our relic density analysis, where we computed the DM abundance by considering only the semi-annihilation channel $SS \to S^\star \varphi$.

\section{Conclusions and Outlook}
\label{sec:final}

After decades of compelling astrophysical and cosmological evidence, the particle physics identity of DM remains an open question. Numerous candidates have been proposed, many driven by strong theoretical motivations beyond merely explaining the DM abundance. Among this diverse array of possibilities, light and weakly coupled pseudo Nambu-Goldstone bosons stand out due to their robust theoretical foundations. The QCD axion is particularly appealing, given its deep connection to the strong CP problem. Beyond the axion, general ALPs offer cold DM candidates produced through the misalignment mechanism. Notably, the relevance of ALPs in the invisible universe extends beyond their role as DM candidates.

In this work, we explored the role of an ALP as a portal to scalar DM. We adopted an EFT perspective that does not require specifying any UV completion. A unique feature of this scenario is the potential redundancy between the ALP derivative interaction with the spin-one DM current. We examined this issue in \Sec{sec:EFT}, highlighting its connection to the underlying global symmetry that ensures the cosmological stability of the DM field. For a viable framework, we stabilized the DM field via a $\mathbb{Z}_3$ symmetry. The illustration in Tab.~\ref{tab:pheno} concisely summarizes the phenomenology, emphasizing also the key differences compared to the more commonly studied case of fermion DM stabilized by a $\mathbb{Z}_2$ symmetry.

The relic density analysis in \Sec{sec:relic}, combined with the investigation of ALP cosmology in \Sec{sec:ALP}, presents a consistent picture where the DM abundance is generated in the early universe via freeze-out of semi-annihilations. This same process can occur today in astrophysical regions with high DM density, such as the center of the Milky Way, and produce detectable cosmic rays. We explored this in \Sec{sec:gammaray}, with a particular focus on the final state gamma rays. The resulting spectra are enriched compared to the pure annihilating DM case, owing to the intermediate ALP. A key phenomenological feature of this framework is that both the DM relic density and the indirect detection rates are insensitive to the ALP couplings to SM fields. For a consistent relic density analysis, the only requirement is that the ALP be sufficiently coupled to the visible sector to remain in equilibrium during the DM freeze-out epoch. In terms of indirect detection, ALP couplings to SM fields only influence the selection of the ALP branching ratio, thereby shaping the gamma ray spectrum.

Our results motivate further explorations along several directions. A potential avenue for future research could involve relaxing the assumption of flavor-diagonal couplings to SM fermions to investigate the consequences of flavor-violating interactions. It has been demonstrated how such interactions can produce detectable axion dark radiation~\cite{DEramo:2021usm} or DM~\cite{Panci:2022wlc,Aghaie:2024jkj}, and it would be intriguing to study their role within the ALP portal to scalar DM. Additionally, our analysis was based on the specific hierarchy $m_\varphi < m_S$. Exploring larger values for the ALP mass, $m_\varphi > 3 m_S$, the DM production via the three-body ALP decay $\varphi \rightarrow SSS$ becomes kinematically allowed. There are also open avenues to explore for ALP portal to scalar DM scenarios that deviate significantly from the study presented in this work. One such direction involves considering lower ALP masses, where the terrestrial bounds are particularly stringent, making freeze-out production highly implausible. However, when focusing on the freeze-in region, the resulting phenomenology becomes more intriguing. This is easily understood in the non-derivative basis, where all DM interactions are proportional to $S^3$. In particular, freeze-in production must proceed through $2 \rightarrow 3$ scattering processes. Another possible exploration would involve relaxing the assumption of a non-Abelian stabilizing symmetry and instead focusing on the more popular $\mathbb{Z}_2$ symmetry. In this case, one would need to introduce $\mathbb{Z}_2$-invariant scalar potential terms proportional to $S^2$ to ensure that ALP-DM interactions cannot be redefined away. This term induces a mass splitting between the real and imaginary parts of $S$, leading to a DM scenario that warrants further investigation. We briefly addressed a plausible microscopic origin in \Sec{sec:UV}, and it seems worthwhile to explore other options with different patterns of low-energy ALP couplings. We leave the detailed exploration of these avenues for future work.

\acknowledgments
The authors thank Mathias Becker, Federico Cima, Alessandro Lenoci, Silvia Manconi, and Luca Vecchi for useful discussions. This work is supported by the Italian MUR Departments of Excellence grant 2023-2027 ``Quantum Frontiers'', and by Istituto Nazionale di Fisica Nucleare (INFN) through the Theoretical Astroparticle Physics (TAsP) project. F.D. acknowledges support from the European Union’s Horizon 2020 research and innovation programme under the Marie Skłodowska-Curie grant agreement No 860881-HIDDeN. 

\appendix

\section{Collision terms for the Boltzmann equation}
\label{app:coll}

In this Appendix, we present the explicit form of the collision term $\mathcal{C}_\alpha$ that accounts for the process $\alpha$. This contribution appears in the Boltzmann equation for the DM number density given by Eq.~\eqref{eq:genBE}. We maintain a fully general treatment, considering a process involving an arbitrary number of DM particles in both the initial and final states. In addition, the process involves other degrees of freedom $\mathcal{B}$ from the primordial bath that are in thermal equilibrium. We refer to such a generic process as follows
\be
\underbrace{\mathcal{S}(P_1)... \mathcal{S}(P_N)}_{N} \;
\underbrace{\mathcal{B}_1(K_{1}) ... \mathcal{B}_A(K_A)}_A
\, \leftrightarrow \, 
\underbrace{\mathcal{S}(P_{N+1})... \mathcal{S}(P_{N+M})}_M \;  
\underbrace{\mathcal{B}_{A+1}(K_{A+1}) ... \mathcal{B}_{A+B}(K_{A+B})}_B \ .
\label{eq:procgen}
\ee
The field $\mathcal{S}$ can be either a DM particle or antiparticle, $\mathcal{S} = \{S, S^\star\}$. It is convenient to count the number of $S$ and $S^\star$ particles separately in order to properly account for the combinatorial factors. Thus we consider $n$ DM particles $S$ and $N - n$ antiparticles $S^\star$ colliding with $A$ bath particles (not necessarily identical). The final state contains $m$ and $M - m$ DM particles and antiparticles, and also $B$ bath degrees of freedom. Each particle's four-momentum is indicated in parentheses. Four-momenta are represented by uppercase characters, the norm of spatial momenta by the corresponding lowercase characters (e.g., $P$ and $p$, respectively). The collision term for the process in Eq.~\eqref{eq:procgen} is given by\footnote{See, e.g., Refs.~\cite{Bernstein:1988bw,DEramo:2020gpr} for a complete derivation.}
\be
\algn{
\mathcal{C}_\alpha = & \, \frac{(m - n)}{m! n! (M-m)! (N-n)!} \int \prod_{i=1}^{N+M} d\Pi_i \prod_{j=1}^{A+B} d \mathcal{K}_j  \;
(2\pi)^4 \, \delta^{(4)}(P_\text{fin} - P_\text{init}) \, \times \\ &   
\left[ \obar{|\mathcal{M}_{\alpha \rightarrow}|^2} \times \prod_{i=1}^N f_S(p_i) \prod_{j=1}^A f^{\rm eq}_{ \mathcal{B}_j}(k_j) 
\prod_{i=N+1}^{N+M} \left(1 \pm f_S(p_i) \right) \prod_{j=A+1}^{A+B} \left(1 \pm f^{\rm eq}_{ \mathcal{B}_j}(k_j)\right)  + \right. \\ &  \left. 
\quad - \obar{|\mathcal{M}_{\alpha  \leftarrow}|^2} \times \prod_{i=N+1}^{N+M} f_S(p_i) \prod_{j=A+1}^{A+B}  f^{\rm eq}_{ \mathcal{B}_j}(k_j) 
\prod_{i=1}^N \left(1 \pm f_S(p_i) \right) \prod_{j=1}^A \left(1 \pm f^{\rm eq}_{ \mathcal{B}_j}(k_j)\right) \right] .}
\label{eq:Cprocgen}
\ee
All integrations are over the Lorentz invariant phase space measures $d\Pi_i = d^3 p_i / (2 E^i_S (2\pi)^3)$ and $d\mathcal{K}_j = g_j \, d^3 p_j / (2 E_j (2\pi)^3)$, where $g_j$ is the number of internal degrees of freedom for each bath particle.\footnote{There is no analogous factor in the integration measure $d\Pi_i $ since $g_S = 1$. One should not be deceived by the fact that $S$ is a complex scalar, as in this treatment we track particles and antiparticles separately.} We find it convenient to define $P_\text{init} = \sum_{i = 1}^N P_i + \sum_{j=1}^A K_j$ and $P_\text{fin} = \sum_{i = N+1}^{N+M} P_i + \sum_{j=A+1}^{A+B} K_j$ for the ease of notation, and the 4-dimensional Dirac delta function ensures the conservation of energy and momentum. The overall factor $(m-n)$ accounts for the net number variation of $S$ particles, and the factorial are symmetry factors that account for identical DM particles and antiparticles in the initial and final states. If there are identical bath particles in the process, additional symmetry factors must be included for them as well. The second and third lines in Eq.~\eqref{eq:Cprocgen} accounts for the probabilities of the direct and inverse processes, respectively. We assume no matter/antimatter asymmetry in the dark sector and therefore we can take $f_S = f_{S^\star}$. The squared matrix elements are averaged over both initial and final states, and we account for quantum statistical effects for bosons (Bose enhancement, $+$ sign) and fermions (Pauli blocking, $-$ sign). 

The effective interactions of the ALP that we consider in this work preserve CP symmetry, meaning the two squared matrix elements are identical: $\obar{|\mathcal{M}_{\alpha \leftarrow}|^2} = \obar{|\mathcal{M}_{\alpha \rightarrow}|^2} \equiv \obar{|\mathcal{M}_{\alpha}|^2}$. Quantum degeneracy effects are negligible in the early universe and it is valid to describe all degrees of freedom via the Maxwell-Boltzmann statistics with the approximation $(1\pm f_i)\simeq1$. The bath particles $\mathcal{B}_j$ are in both kinetic and chemical equilibrium, and their equilibrium Maxwell-Boltzmann distribution function is given by $f^{\rm eq}_{\mathcal{B}_j} = \exp[-E_{\mathcal{B}_j} / T]$. In contrast, DM particles and antiparticles are only in kinetic equilibrium at the freeze-out epoch. Consequently, the most we can do is to express the distribution function as $f_S = (n_S / n_S^{\rm eq}) f^{\rm eq}_S$. Here, the DM equilibrium number density is evaluated within the Maxwell-Boltzmann statistics, and it reads explicitly
\be
n_S^{\rm eq} = \frac{m_S^2\,T}{2\pi^2} \,K_2[m_S/T] \ .
\label{eq:nSeq}
\ee
Furthermore, conservation of energy allows us to identify 
\be
 \prod_{i=1}^N f^{\rm eq}_S(p_i) \prod_{j=1}^A f^{\rm eq}_{ \mathcal{B}_j}(k_j) = \prod_{i=N+1}^{N+M} f^{\rm eq}_S(p_i) \prod_{j=A+1}^{A+B} f^{\rm eq}_{ \mathcal{B}_j}(k_j) \ .
\ee
Combining all together, the expression in Eq.~\eqref{eq:Cprocgen} becomes
\be
\algn{
\mathcal{C}_\alpha = & \, \frac{(m - n)}{m! n! (M-m)! (N-n)!} \left[ \left( \frac{n_S}{n_S^{\rm eq}} \right)^N  - \left( \frac{n_S}{n_S^{\rm eq}} \right)^M \right] \, \times \\ & 
 \int \prod_{i=1}^{N+M} d\Pi_i \prod_{j=1}^{A+B} d \mathcal{K}_j  \; \prod_{i=1}^N f^{\rm eq}_S(p_i) \prod_{j=1}^A f^{\rm eq}_j(k_j)  \;
(2\pi)^4 \, \delta^{(4)}(P_\text{fin} - P_\text{init})   
\obar{|\mathcal{M}_\alpha|^2} \ .}
\label{eq:Cprocgen2}
\ee

For the purpose of this work, we are only interested in process with two initial state DM (anti-)particles: annihilations ($\{N, n, M, m, A, B\} = \{2, 1, 0, 0, 0, F\}$) and semi-annihilations ($\{N, n, M, m, A, B\} = \{2, 2, 1, 0, 0, F\}$). For both cases we have $\{N, A\} = \{2, 0\}$, and the index $M$ can be values either $0$ or $1$ for annihilations and semi-annihilations, respectively. Once one specializes to these two case, it is useful to introduce the thermal average of the scattering cross section times the Møller velocity
\be
\algn{
\langle \sigma_\alpha  \vmol \rangle = & \, \frac{1}{(n_S^{\rm eq})^2}  
\int \frac{d^3 p_1}{2 E_S(p_1) (2 \pi)^3} \frac{d^3 p_2}{2 E_S(p_2) (2 \pi)^3} f^{\rm eq}_S(p_1) f^{\rm eq}_S(p_2)\;  \, \times \\  & \, \qquad \qquad  
\obar{|\mathcal{M}_\alpha|^2}  (2\pi)^4 \, \delta^{(4)}(P_\text{fin} - P_\text{init})  \, d \Pi \prod_{j=1}^{F} d \mathcal{K}_j  \ ,}
\ee
where the phase space integral $d \Pi$ for the final state DM particle is present only for semi-annihilations and therefore
\be
d \Pi = \left\{  \begin{array}{lcccl} 
1 & \qquad & & & \text{annihilation} \\
\frac{d^3 p_3}{2 E_S(p_3) (2 \pi)^3} & & & \qquad & \text{semi-annihilation} 
\end{array}
\right. \ .
\ee
The integral over the initial state momenta can be performed by standard methods~\cite{Gondolo:1990dk, DEramo:2017ecx}
\be
\langle \sigma_\alpha \vmol  \rangle =  
\frac{ \int_{s_{\rm min}}^\infty ds \, \sqrt{s} \left(s - 4 m_S^2\right) \, \sigma_\alpha(s) \, K_1\left[\sqrt{s} / T\right]}{8 m_S^4 T K_2[m_S/T]^2} \ ,
\label{eq:thav}
\ee
where the integration over the Mandelstam variable $s$ has a lower integration extreme that depends on the particle spectrum $s_{\rm min} = {\rm max}\left\{ 4 m_S^2, ((1 - \eta) m_S + \sum_j m_{\mathcal{B}_j})^2 \right\}$ with $\eta = 1$ for annihilations and $\eta = 0$ for semi-annihilations. Putting everything together, we can finally derive the explicit expressions for the collision operators of these two processes
\begin{subequations}
\begin{align}
\label{eq:CollAnn} \text{Annihilation:} \qquad \qquad & \mathcal{C}_{S S^\star \rightarrow  \mathcal{B}_1 \ldots \mathcal{B}_F} = - \langle \sigma_{S S^\star \rightarrow  \mathcal{B}_1 \ldots \mathcal{B}_F}  \vmol \rangle   \left[ n_S^2  - (n_S^{\rm eq})^2 \right] \ , \\
\label{eq:CollSemi} \text{Semi-annihilation:}   \qquad \qquad & \mathcal{C}_{S S \rightarrow S^\star \mathcal{B}_1 \ldots \mathcal{B}_F} = - \langle \sigma_{S S \rightarrow S^\star \mathcal{B}_1 \ldots \mathcal{B}_F}  \vmol \rangle  \left[ n_S^2  - n_S n_S^{\rm eq} \right]  \ .
\end{align}
\end{subequations}

\section{Cross sections for dark matter (semi-)annihilations}
\label{app:DMsupp}

We present explicit calculations for the cross sections of all potentially relevant DM number-changing processes. For each case, we perform the calculations in both the derivative and non-derivative field bases, ensuring that the results are basis-independent. We start from a generic binary collisions that can lead to arbitrary $N$-body final states
\be
\mathcal{S}(P_1) \, \mathcal{S}(P_2) \, \leftrightarrow \,  \mathcal{F}_1(K_1) \, \ldots \, \mathcal{F}_N(K_N) \ .
\label{eq:genscat}
\ee
The conventions for the four-momenta and spatial momenta are the same as in the previous Appendix. The field $\mathcal{S}$ can be either a DM particle or antiparticle, and final state particles $\mathcal{F}$ include both DM or SM degrees of freedom.  The specific goal here is to derive the expression for the total cross section $\sigma_\alpha(s)$ of a given process $\alpha$ as a function of the (square of the) center-of-mass energy, $s = (P_1 + P_2)^2$. We begin by recalling the general expression for the scattering cross section
\be\label{eq:sigmagen}
\sigma_\alpha = \frac{1}{4\sqrt{(P_1\cdot P_2)^2-m_S^4}}\int \obar{|\mathcal{M}_\alpha|^2} 
\underbrace{(2\pi)^4\delta^4(P_1+P_2-{\textstyle{\sum_i^N}}K_i) \prod_{i=1}^N g_i \frac{\text d^3k_i}{2E_i (2\pi)^3}}_{\text d\Pi^{(N)}} \ .
\ee
To remain consistent with the previous notation we keep the squared scattering matrix element $\avgM$ averaged over both initial and final states; this is the reason for the degeneracy factors of $g_i$ appearing in the Lorentz invariant $N$-body phase space factor $\text d\Pi^{(N)}$. Our main focus in this work is on  DM (semi-)annihilations with two-body final states. The squared matrix elements never carry an explicit dependence on the scattering angle for the interactions considered in this work and therefore the phase-space integration is straightforward. We write the total cross section for DM (semi-)annihilations 
\be
\sigma_{\mathcal{S} \mathcal{S} \rightarrow \mathcal{F}_1 \mathcal{F}_2}(s) = \frac{\obar{|\mathcal{M}_{\mathcal{S} \mathcal{S} \rightarrow \mathcal{F}_1 \mathcal{F}_2}|^2}}{16 \pi \, s \, \sqrt{1-4 m_S^2/s}} \sqrt{1 - 2 \frac{m_{\mathcal{F}_1}^2 + m_{\mathcal{F}_2}^2}{s} + \frac{(m_{\mathcal{F}_1}^2 - m_{\mathcal{F}_2}^2)^2}{s^2} } \ .
\label{eq:XSbinary}
\ee

\begin{itemize}

\item \textbf{Annihilations to standard model fields.} Binary collisions in which DM annihilates with its own antiparticle to produce a final state consisting of two SM particles are forbidden at tree level. This result is manifest in the non-derivative basis, where the only DM interaction with other fields is the quartic term $S^3 \varphi$ (see \Eq{eq:VND}). The same result can be verified in the derivative basis, although the calculation is less trivial due to the quadratic interactions between the ALP spacetime derivative and the DM spin-one current (see \Eqs{eq:Lag}{eq:JmuS}). The corresponding Feynman diagram is shown in Fig.~\ref{fig:SStoSMSM}. Regardless of the SM final state, once we impose conservation of four-momentum at the DM interaction vertex this amplitude is proportional to $(P_1 + P_2) \cdot (P_1 - P_2)$. Upon putting the external states on-shell, we obtain
\be
(P_1 + P_2) \cdot (P_1 - P_2)\xrightarrow[P_1^2=P_2^2=m_S^2]{\textup{on-shell}}0\,\,.
\ee
We conclude that these processes never contribute to DM freeze-out.

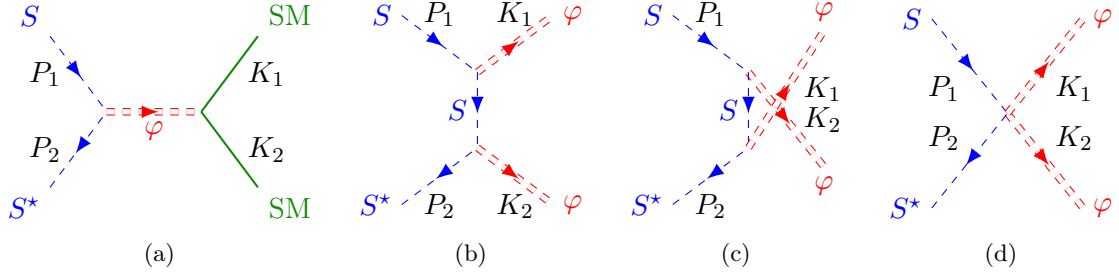
\begin{figure}[t]
\centering
         \subfloat[\label{fig:SStoSMSM}]{\begin{tikzpicture}
         \draw[s](-0.75,1) node[above left]{$\color{blue}{S}$}--node[midway,left=1mm]{$P_1$}(0,0);
         \draw[sb] (-0.75,-1) node[below left]{$\color{blue}{S^\star}$}--node[midway, left=1mm]{$P_2$} (0,0);
         \draw[a] (0,0) -- (1.25,0)node[midway,below]{$\color{red}{\varphi}$};
         \draw[gen] (1.25,0) -- node[midway,right=1mm]{$K_2$} (2,-1.)node[below right]{\color{ColorT}{SM}};
         \draw[gen] (1.25,0) --node[midway,right=1mm]{$K_1$} (2.,1)node[above right]{\color{ColorT}{SM}};
          \end{tikzpicture}} \quad
         \subfloat[\label{fig:annt}]{\begin{tikzpicture}
         \draw[s](-1.,0.75) node[ left]{$\color{blue}{S}$}--node[midway,above=1mm]{$P_1$}(0,0)[baseline];
         \draw[a] (0,0) -- node[midway,above=1mm]{$K_1$}(1,0.75)node[ right]{$\color{red}{\varphi}$};
         \draw[s] (0,0) -- (0,-1)node[midway,left]{$\color{blue}{S}$};
         \draw[a] (0,-1) -- node[midway,below=1mm]{$K_2$}(1,-1.75)node[ right]{$\color{red}{\varphi}$};
         \draw[sb] (-1,-1.75)node[ left]{$\color{blue}{S^\star}$} -- node[midway,below=1mm]{$P_2$}(0,-1);
          \end{tikzpicture}} \quad 
         \subfloat[\label{fig:annu}]{\begin{tikzpicture}
         \draw[s](-1.,0.75) node[left]{$\color{blue}{S}$}--node[midway,above=1mm]{$P_1$}(0,0)[baseline];
         \draw[a] (0,0) --node[midway,right=1mm]{$K_2$} (1,-1.25)node[below]{$\color{red}{\varphi}$};
         \draw[s] (0,0) -- (0,-1)node[midway,left]{$\color{blue}{S}$};
         \draw[a] (0,-1) --node[midway,right=1mm]{$K_1$} (1,0.55)node[above]{$\color{red}{\varphi}$};
         \draw[sb] (-1,-1.75)node[left]{$\color{blue}{S^\star}$} --node[midway,below=1mm]{$P_2$} (0,-1);
          \end{tikzpicture}} \quad 
          \subfloat[\label{fig:annvert}]{\begin{tikzpicture}
         \draw[s](-1.,1.25) node[left]{$\color{blue}{S}$}--node[midway,below left]{$P_1$}(0,0);
         \draw[a] (0,0) --node[midway,above right]{$K_2$} (1,-1.25)node[right]{$\color{red}{\varphi}$};
         \draw[a] (0,0) -- node[midway,below right]{$K_1$}(1,1.25)node[right]{$\color{red}{\varphi}$};
         \draw[sb] (-1,-1.25)node[left]{$\color{blue}{S^\star}$} -- node[midway,above left]{$P_2$}(0,0);
          \end{tikzpicture}}
          \caption{Feynman diagrams for DM annihilations to different final states: SM fermions or gauge bosons in the derivative basis (\ref{fig:SStoSMSM}); ALPs in the derivative basis from dimension 5 interactions (\ref{fig:annt} and \ref{fig:annu}); ALPs in the derivative basis from dimension 6 interactions (\ref{fig:annvert}).}
          \label{fig:SSannihilations}
\end{figure}
 
\item \textbf{Annihilations to ALP pairs.} Another potential DM annihilation process that could contribute to freeze-out involves a final state with two ALPs. However, one has to be careful when including this process, as the cross section calculation requires the inclusion of two ALP fields. If we consider only the contributions from the EFT under investigation defined in \Eq{eq:Lag}, the corresponding Feynman diagrams are those shown in Figs.~\ref{fig:annt} and \ref{fig:annu}. The resulting transition amplitude proportional to $f_\varphi^{-2}$ mirrors that of an additional contribution arising from a dimension 6 contact interaction with two ALPs, which we have not included in the EFT. The symmetries of the low-energy EFT also allow for the presence of the dimension 6 operator defined in \Eq{eq:EFTdim6} that leads to the Feynman diagram depicted in \Fig{fig:annvert}. We start the calculation from the contributions due to the double insertion of the dimension 5 contact interaction and we account for both the $t$ and $u$ channels
\begin{subequations}
\begin{align}
\label{eq:MALPs5t} \M^{(5 t)}_{SS^\star\to\varphi\varphi} & = \bigg(\frac{\wc_S}{2f_\varphi}\bigg)^2 (m_S^2 - t) \ , \\
\label{eq:MALPs6u} \M^{(5 u)}_{SS^\star\to\varphi\varphi} & = \bigg(\frac{\wc_S}{2f_\varphi}\bigg)^2 (m_S^2 - u) \ , \\
\label{eq:MALPs5} \M^{(5)}_{SS^\star\to\varphi\varphi} & =  \M^{(5 t)}_{SS^\star\to\varphi\varphi} +  \M^{(5 u)}_{SS^\star\to\varphi\varphi} = \bigg(\frac{\wc_S}{2f_\varphi}\bigg)^2 (s - 2 m_\varphi^2)\ .
\end{align}
\end{subequations}
In addition to $s$, we introduce the two other Mandelstam variables $t = (P_1 - K_1)^2$ and $u = (P_1 - K_2)^2$. The on-shell conditions impose $s + t + u = 2 m_S^2 + 2 m_\varphi^2$. Finally, the amplitude from the dimension 6 operator results in
\be
 \M^{(6)}_{SS^\star\to\varphi\varphi} = - \frac{\wc^{\prime\,2}_S}{4f_\varphi^2}(s - 2 m^2_\varphi) \ .
\ee
We can write the total amplitude as follows
\be
\M_{SS^\star\to\varphi\varphi} = \M^{(5)}_{SS^\star\to\varphi\varphi} + \M^{(6)}_{SS^\star\to\varphi\varphi} =  \frac{\wc_S^2 - \wc_S^{\prime\,2}}{4 f_\varphi^2} (s - 2 m_\varphi^2) \ .
\ee
This result justifies our choice of normalization in Eq.~\eqref{eq:EFTdim6}, as the amplitude for this process vanishes identically when the Wilson coefficients satisfy $\wc_S = \wc^\p_S$. The reasonableness of this choice is addressed in \Sec{sec:UV}, where we present an explicit UV completion in which this equality is naturally satisfied. The current discussion is framed within the derivative basis for the ALP couplings. If one instead attempts to compute the amplitude in the non-derivative basis, it is straightforward to conclude that the process cannot occur at tree level because the ALP interacts with three DM fields simultaneously (see \Eq{eq:VND}). This might appear to contradict the non-vanishing amplitude we found in the derivative basis. We can resolve this apparent discrepancy by noting that when we connected the two field bases in \Sec{sec:EFT}, we only retained contributions up to dimension 5 and ensured that the field redefinition eliminated the ALP derivative coupling. At the same time, the same field redefinition introduces the dimension 6 operator in Eq.~\eqref{eq:EFTdim6} with coefficient $\Delta \wc^{\prime\,2}_S = - \wc_S^2$ and the resulting matrix element is consistent with the result found in the derivative basis. This demonstrates that the scattering amplitude is indeed basis-independent. The resulting cross section reads
\be
\sigma_{SS^\star\to\varphi\varphi}(s) = \frac{\left| \wc_S^2 - \wc_S^{\prime\,2} \right|^2}{512 \pi} \frac{s}{f_\varphi^4} 
\frac{(1 - 2 m_\varphi^2/s)^2 \sqrt{1 - 4 m_{\varphi}^2 / s} }{\sqrt{1-4 m_S^2/s}} \ .
\label{eq:XSALPairs}
\ee
where we applied \Eq{eq:XSbinary} and included an additional factor of $1/2$ to account for two identical particles in the final state. We do not include this contribution in our freeze-out analysis since the associated cross section is suppressed compared to semi-annihilations by two additional powers of $f_\varphi$.

\item \textbf{Semi-annihilations to ALP.} We evaluate the cross section for the semi-annihilation $S S \rightarrow S^\star \varphi$ in both field bases defined in Sec.~\ref{sec:EFT}. Starting from the derivative basis, the scattering amplitude is triggered by the $\varphi$ derivative coupling to the scalar spin-1 current (where the internal propagator is crucially off-shell) and the cubic DM self-interaction. This process receives contributions from $s,\,t,\,u$ channels. The corresponding Feynnamn diagrams are shown in \Fig{fig:semis}, \ref{fig:semit}, \ref{fig:semiu}, and the amplitudes for each channel explicitly result in
\begin{subequations}
\begin{align}
\M_{SS\to S^\star\varphi}^{(\partial s)} = & \, i \, \frac{\wc_S A}{f_\varphi}\frac{K_2 \cdot(-K_1 - (P_1 + P_2))}{s - m_S^2} = - \frac{i}{2} \frac{\wc_S A}{f_\varphi}  \ , \\
\M_{SS\to S^\star\varphi}^{(\partial t)} = & \, i \,\frac{\wc_S A}{f_\varphi}\frac{K_2 \cdot(P_2 - (P_1 - K_1))}{t - M_S^2} = - \frac{i}{2} \frac{\wc_S A}{f_\varphi}  \ , \\
\M_{SS\to S^\star\varphi}^{(\partial u)} = & \, i \,\frac{\wc_S A}{f_\varphi}\frac{K_2 \cdot (P_1 -(P_2 - K_1)) }{u - M_S^2} = - \frac{i}{2} \frac{\wc_S A}{f_\varphi}  \ .
\end{align}
\end{subequations}

The last equalities hold when we put on-shell the external particles. We notice how the three channels give identical contributions and therefore the total  amplitude reads
\be
\M_{SS\to S^\star\varphi}^{(\partial)} = \M_{SS\to S^\star\varphi}^{(\partial s)} + \M_{SS\to S^\star\varphi}^{(\partial t)}  + \M_{SS\to S^\star\varphi}^{(\partial u)} =  - i \frac{3}{2} \frac{\wc_S A}{f_\varphi} \ .
\ee
The calculation of the semi-annihilation scattering amplitude is even easier in the non-derivative basis where the relevant interaction is that of \Eq{eq:VND}. There is only one Feynman diagram for the semi-annihilation process in this basis, which corresponds to the quartic vertex in \Fig{fig:semivertex}, and the resulting scattering amplitude reads
\be
\M_{SS\to S^\star\varphi}^{(N \partial)} = - \lambda_{S\varphi}\,\,.
\ee
The matching condition given in \Eq{eq:VND} ensures that the scattering amplitude does not depend on the basis. The resulting cross section is
\be
\sigma_{SS\to S^\star\varphi}(s)=\frac{\lambda _{S\varphi}^2}{16 \pi \, s \, \sqrt{1-4 m_S^2/s}} \,  
\sqrt{1 - 2 \frac{m_S^2 + m_\varphi^2}{s} + \frac{(m_S^2 - m_\varphi^2)^2}{s^2}}  \ .
\label{eq:Xsecsemi}
\ee

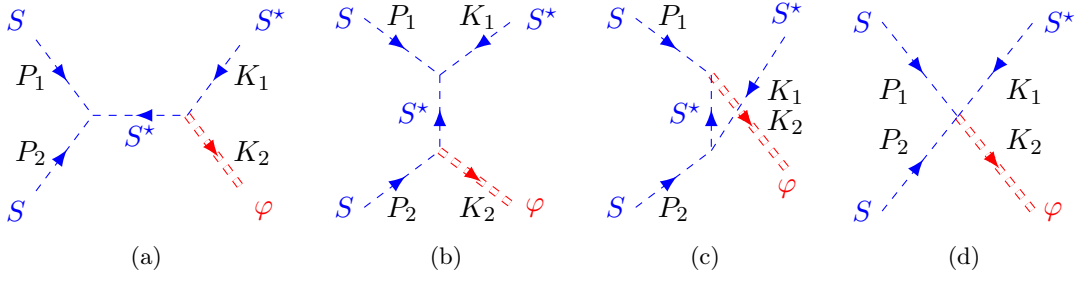
\begin{figure}[t]
\centering
        \subfloat[\label{fig:semis}]{\begin{tikzpicture}
         \draw[s](-0.75,1) node[above left]{$\color{blue}{S}$}--node[midway,left=1mm]{$P_1$}(0,0);
         \draw[s] (-0.75,-1) node[below left]{$\color{blue}{S}$}--node[midway, left=1mm]{$P_2$} (0,0);
         \draw[sb] (0,0) -- (1.25,0)node[midway,below]{$\color{blue}{S^\star}$};
         \draw[a] (1.25,0) -- node[midway,right=1mm]{$K_2$} (2,-1.)node[below right]{$\color{red}{\varphi}$};
         \draw[sb] (1.25,0) --node[midway,right=1mm]{$K_1$} (2.,1)node[above right]{$\color{blue}{S^\star}$};
          \end{tikzpicture}}\quad
\subfloat[\label{fig:semit}]{\begin{tikzpicture}
         \draw[s](-1.,0.75) node[ left]{$\color{blue}{S}$}--node[midway,above=1mm]{$P_1$}(0,0)[baseline];
         \draw[sb] (0,0) -- node[midway,above=1mm]{$K_1$}(1,0.75)node[ right]{$\color{blue}{S^\star}$};
         \draw[sb] (0,0) -- (0,-1)node[midway,left]{$\color{blue}{S^\star}$};
         \draw[a] (0,-1) -- node[midway,below=1mm]{$K_2$}(1,-1.75)node[ right]{$\color{red}{\varphi}$};
         \draw[s] (-1,-1.75)node[ left]{$\color{blue}{S}$} -- node[midway,below=1mm]{$P_2$}(0,-1);
          \end{tikzpicture}}\quad
\subfloat[\label{fig:semiu}]{\begin{tikzpicture}
        \draw[s](-1.,0.75) node[left]{$\color{blue}{S}$}--node[midway,above=1mm]{$P_1$}(0,0)[baseline];
         \draw[a] (0,0) --node[midway,right=1mm]{$K_2$} (1,-1.25)node[below]{$\color{red}{\varphi}$};
         \draw[sb] (0,0) -- (0,-1)node[midway,left]{$\color{blue}{S^\star}$};
         \draw[sb] (0,-1) --node[midway,right=1mm]{$K_1$} (1,0.55)node[above]{$\color{blue}{S^\star}$};
         \draw[s] (-1,-1.75)node[left]{$\color{blue}{S}$} --node[midway,below=1mm]{$P_2$} (0,-1);          \end{tikzpicture}}\quad
          \subfloat[\label{fig:semivertex}]{\begin{tikzpicture}
         \draw[s](-1.,1.25) node[left]{$\color{blue}{S}$}--node[midway,below left]{$P_1$}(0,0);
         \draw[a] (0,0) --node[midway,above right]{$K_2$} (1,-1.25)node[right]{$\color{red}{\varphi}$};
         \draw[sb] (0,0) -- node[midway,below right]{$K_1$}(1,1.25)node[right]{$\color{blue}{S^\star}$};
         \draw[s] (-1,-1.25)node[left]{$\color{blue}{S}$} -- node[midway,above left]{$P_2$}(0,0);
          \end{tikzpicture}}
          \caption{DM semi-annihilation $S S \to S^\star \varphi$: Feynman diagrams in the derivative basis (\ref{fig:semis}, \ref{fig:semit}, and \ref{fig:semiu}) and non-derivative basis (\ref{fig:semivertex}).}
        \label{fig:semi}
\end{figure}

\end{itemize}

\section{ALP decays and scatterings}
\label{app:ALP}

We provide computational details for the analysis presented in \Sec{sec:ALP}. Specifically, we evaluate the rates for processes relevant to ALP thermalization in the early universe: decays to gauge bosons and fermions, and scatterings with bath particles. Decays to DM pairs are not considered here since we always work in the parameter space region $m_S > m_\varphi$. Before discussing specific thermalization channels, we review the formalism that we employ to perform the calculations in this Appendix.

We provided already the expression in \Eq{eq:sigmagen} for the cross section of any binary collision. Concerning decays, the ALP partial width for an arbitrary final state results in 
\be
\Gamma_{\varphi \rightarrow \mathcal{F}_1 \ldots \mathcal{F}_N} = \frac{1}{2 m_\varphi} \int 
\obar{|\mathcal{M}_{\varphi \rightarrow \mathcal{F}_1 \ldots \mathcal{F}_N}|^2} 
\underbrace{(2\pi)^4\delta^4(P-{\textstyle{\sum_i^N}} K_i) \prod_{i=1}^N g_i \frac{\text d^3k_i}{2E_i (2\pi)^3}}_{\text d\Pi^{(N)}} \ . 
\label{eq:gendecay}
\ee
Here, we denote the four-momentum of the ALP with the capital letter $P$. We keep our conventions consistent throughout the paper: the four-momenta of the final state particles are the same as in \Eq{eq:genscat}, and we denote $\avgM$ the squared matrix element averaged over both initial and final states. When ALP decays to SM fields are kinematically allowed, the final state is always a pair of gauge bosons or fermions. We can restrict our focus on two-body decays that are monochromatic and write the partial decay with as follows
\be
\Gamma_{\varphi \rightarrow \mathcal{F}_1 \mathcal{F}_2} = \frac{\sum_{\mathcal{F}} \obar{|\mathcal{M}_{\varphi \rightarrow \mathcal{F}_1 \mathcal{F}_2}|^2} }{16 \pi m_\varphi} \sqrt{1 - 2 \frac{m_{\mathcal{F}_1}^2 + m_{\mathcal{F}_2}^2}{m_\varphi^2} + \frac{(m_{\mathcal{F}_1}^2 - m_{\mathcal{F}_2}^2)^2}{m_\varphi^4}}  \ .
\label{eq:twobodydecay}
\ee
We notice how the sum over the final states is necessary since according to our conventions the squared matrix element $\avgM$ is averaged over the final states.

The only missing ingredients to compute decay and scattering rates are the explicit expressions for the matrix elements. We provide the calculations for all the decay channels discussed in \Sec{sec:ALP} and compute the cross sections only for ALP production processes which are mediated by interactions with SM fermions. For couplings to SM gauge bosons the situation is rather intricate due to IR divergences generated by the exchange of massless gauge fields. We do not present new calculations on proper treatments of thermal effects but resort to the current literature for gluon~\cite{DEramo:2021psx,DEramo:2021lgb,Salvio:2013iaa} and photons~\cite{Braaten:1991dd,Bolz:2000fu,Cadamuro:2011fd} couplings.  As already stated in the main text, the choice for the field basis is important due to threshold corrections. We choose to work in the derivative basis where these corrections are absent.

\begin{itemize}

\item \textbf{ALP decays to gauge bosons.} We consider gluons and photons in the final state, and the Feynman diagrams for these two processes are shown in \Figs{fig:phitogluons}{fig:phitophotons}. The calculation is analogous for the two cases with the only differences being group theory numerical factors. The transition amplitude for the decay results
\be
\mathcal{M}_{\mathcal{\varphi \rightarrow VV}} = i \, \frac{\wc_V \alpha_V}{2 \pi f_\varphi} \epsilon_{\mu\nu\rho\sigma} \, K_1^\mu K_2^\rho \, \epsilon^{\nu\star}(K_1) \epsilon^{\sigma\star}(K_2) \ , \qquad \qquad \qquad V = G, \gamma \ . 
\ee
The Wilson coefficient for the coupling to gluons is defined in \Eq{eq:Lag}, the analogous interaction with photon results from the matching and it gives $\wc_\gamma = \wc_W + \wc_B$. The polarization tensor for gluons in the final state carry color indices and we leave them implicit. We square and sum over the final spin and color (if any) states
\be
\sum_{\mathcal{F}} \obar{|\mathcal{M}_{\varphi \rightarrow V V}|^2} = \mathcal{D}_V \, \frac{\wc_V^2 \alpha_V^2}{2 \pi^2 f_\varphi^2} \left[ (K_1 \cdot K_2)^2 - K_1^2 K_2^2 \right] \ .
\ee
The multiplicative factor $\mathcal{D}_V$ is the dimension of the adjoint representation of the corresponding gauge group and accounts for all possible color degrees of freedom in the final state, $\mathcal{D}_G = 8$ and $\mathcal{D}_\gamma = 1$. We put the massless gauge boson on-shell, $K_1 \cdot K_2 = m_\varphi^2 / 2$, and we include in the partial decay width an additional factor of $1/2$ that accounts for identical final state particles
\be
\Gamma_{\mathcal{\varphi \rightarrow VV}} = \mathcal{D}_V \, \frac{\wc_V^2 \alpha_V^2 m_\varphi^3}{256 \pi^3 f_\varphi^2} \ , \qquad \qquad \qquad V = G, \gamma \ . 
\label{eq:gammaALPgauge}
\ee

\begin{figure}[t]
\centering
           \subfloat[\label{fig:phitogluons}]{\begin{tikzpicture}
         \draw[a](-1.25,0) node[left]{\color{red}{$\varphi$}}--(0,0);
         \draw[g] (0,0) -- (1.25,1.25)node[right]{\color{ColorT}{$g$}};
         \draw[g] (0,0) -- (1.25,-1.25)node[right]{\color{ColorT}{$g$}};
          \end{tikzpicture}} \qquad\qquad
          \subfloat[\label{fig:phitophotons}]{\begin{tikzpicture}
         \draw[a](-1.25,0) node[left]{\color{red}{$\varphi$}}--(0,0);
         \draw[v] (0,0) -- (1.25,1.25)node[right]{\color{ColorT}{$\gamma$}};
         \draw[v] (0,0) -- (1.25,-1.25)node[right]{\color{ColorT}{$\gamma$}};
          \end{tikzpicture}} \qquad\qquad
         \subfloat[\label{fig:phitofermions}]{\begin{tikzpicture}
         \draw[a](-1.25,0) node[left]{\color{red}{$\varphi$}}--(-0,0);
         \draw[f] (0,0) -- (1.25,1.25)node[right]{\color{ColorT}{$\psi$}};
         \draw[fb] (0,0) -- (1.25,-1.25)node[right]{\color{ColorT}{$\bar\psi$}};
          \end{tikzpicture}}
\caption{Feynman diagrams for ALP decays to gluons (\ref{fig:phitogluons}), photons (\ref{fig:phitophotons}), fermions (\ref{fig:phitofermions}).}
\label{fig:ALPdecay}
\end{figure}
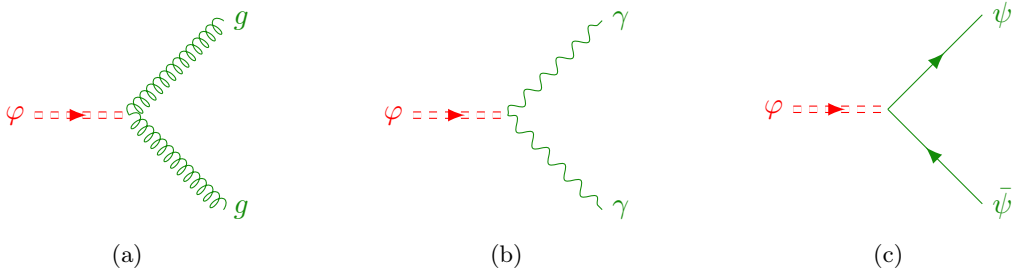

\item \textbf{ALP decays to fermions.} We perform the calculation of the ALP partial decay width into pairs of massive SM fermions in the derivative field basis. The result for this specific partial decay width would be of course the same if we rotate SM fermions to move to the non-derivative basis. It is straightforward to check this equivalence by using the Dirac equation of motion in the derivative basis amplitude. Given the ALP mass range considered in this work, we compute the transition amplitudes with the interactions provided in \Eq{eq:JmuDirac} for Dirac fields. The matrix element for this class of decay reads
\be
\mathcal{M}_{\varphi \rightarrow \psi \bar{\psi}} = \frac{\wc_\psi}{2 f_\varphi} (- i P_\mu) \, \bar{u}(K_1) \gamma^\mu \gamma^5 v(K_2) \ , \qquad \qquad \qquad \psi = q, \ell \ . 
\ee
We squared the amplitude and sum over all possible final states
\be
\frac{\sum_{\mathcal{F}} \obar{|\mathcal{M}_{\varphi \rightarrow \psi \bar{\psi}}|^2}}{N_c^\psi} = \wc_\psi^2 \frac{2 (P \cdot K_1) (P \cdot K_2) + (m_\psi^2 - K_1 \cdot K_2) P^2}{f_\varphi^2}  = 2 \, \wc_\psi^2 \frac{ m_\varphi^2 m_\psi^2}{f_\varphi^2}   \ .
\ee
Here, the number of fermion colors $N_c^\psi$ is equal to $3$ ($\psi = q$) or $1$ ($\psi = \ell$), and in the last equality we put the external particles on-shell.The resulting decay width reads
\be
\Gamma_{\varphi\to \psi\bar\psi} = N_c^\psi \frac{\wc_\psi^2 m_\varphi m_\psi^2}{8 \pi f_\varphi^2} \sqrt{1 - \frac{4 m_\psi^2}{m_\varphi^2}} \ . 
\label{eq:ALPdecaypsi}
\ee
This channel is kinematically allowed as long as the $\varphi$ mass is at least twice the mass of the daughter fermion, $m_\varphi>2m_\psi$. Furthermore, the direct proportionality to the fermion mass squared determines a hierarchy between partial widths into different two different fermions $\psi$ and $\psi^\prime$ scaling as $\left(m_\psi / m_{\psi'}\right)^2$.

\item \textbf{ALP scattering with fermions I: Particle/antiparticle annihilation.} ALP production via scatterings mediated by dimension 5 interactions with SM fermions has two main contributions. Here, we analyze the one that we dub particle/antiparticle annihilations. The two Feynman diagrams in Figs.~\ref{fig:annihilationt} and \ref{fig:annihilationu} describe this process when the other interaction playing an active role is electromagnetism. Therefore, this process is present for all SM fermions but neutrinos. For the specific case when the fermion $\psi$ is a quark, there are analogous diagrams with the final state photon replaced by a gluon. The squared matrix elements and the resulting cross sections have been computed in the literature relevant to axion dark radiation~\cite{DEramo:2018vss,Arias-Aragon:2020qtn,Green:2021hjh,DEramo:2024jhn}. If the temperature of interest is above the weak scale, the other electroweak gauge bosons play also an active role~\cite{Arias-Aragon:2020shv}.

We compute the squared matrix element for particle/antiparticle annihilations with a photon in the final state. The equivalent expression for the diagrams with a gluon is straightforwardly connected only via color algebra factors. The novelty with respect to existing calculation in the axion dark radiation literature is that we keep the ALP mass $m_\varphi$ finite. Indeed, it is important to take into account both $m_\varphi$ and $m_\psi$ effects in the calculation since the hierarchy between the two scales is variable. For some cases, such as for electron/positron annihilations, the ALP mass is the largest scale in the process. The amplitudes for the $t$ and $u$ channels read
\begin{subequations}
\begin{align}
\M_{\psi \bar\psi \rightarrow \gamma \varphi} = &\, \M^{(t)}_{\psi \bar\psi \rightarrow \gamma \varphi} + \M^{(u)}_{\psi \bar\psi \rightarrow \gamma \varphi} \ , \\
\M^{(t)}_{\psi \bar\psi \rightarrow \gamma \varphi} = &\, i \left( \frac{\wc_\psi}{2 f_\varphi} \, Q_\psi e \right) \epsilon_\mu^\star(K_3, \lambda) K_{4 \nu} \;
\bar{v}(K_2)  \gamma^\nu \gamma^5 \frac{\slashed{K_1} - \slashed{K_3} + m_\psi}{t - m_\psi^2} \gamma^\mu u(K_1) \ ,\\
\M^{(u)}_{\psi \bar\psi \rightarrow \gamma \varphi} = & \, i \left( \frac{\wc_\psi}{2 f_\varphi} \, Q_\psi e \right) \epsilon_\mu^\star(K_3, \lambda) K_{4 \nu} \;
\bar{v}(K_2) \gamma^\mu \frac{\slashed{K_1} - \slashed{K_4} + m_\psi}{u - m_\psi^2}  \gamma^\nu \gamma^5 u(K_1)  \ .
\end{align}
\end{subequations}
We sum the two contributions, take the squared absolute value, and put the external legs on-shell. After averaging over both initial and final states, we find
\be
\begin{split}
\overline{ \left| \mathcal{M}_{\psi \bar\psi \rightarrow \gamma \varphi} \right|^2} = & \, \frac{\wc_\psi^2 Q_\psi^2 e^2}{2 f_\varphi^2} \frac{m_\psi^2}{(m_\psi^2-t)^2 (m_\varphi^2+m_\psi^2-s-t)^2} \times \\ &
\left[ 
m_{\psi}^2 \left( -3 s^2 m_{\varphi}^2 + m_{\varphi}^4 (5 s + 2 t) - 3 m_{\varphi}^6 + s^2 (s + 2 t) \right) + \right. \\ & \left.  
- t \left( m_{\varphi}^4 + s^2 \right) \left( -m_{\varphi}^2 + s + t \right) - m_{\psi}^4 \left( m_{\varphi}^4 + s^2 \right) \right] \ .
\end{split}
\label{eq:MsquaredPSIPAIR}
\ee
We keep our conventions consistent throughout the paper and we always present squared matrix elements averaged over both initial and final states. The expression above is completely general and it does not rely upon any assumption about the mass spectrum. It is convenient to obtain the approximate results which are valid when there is a mass hierarchy between the masses of $\psi$ and $\varphi$. We consider both hierarchies and the results of these expansions are
\begin{subequations}
\begin{align}
\overline{ \left| \M_{\psi \bar\psi \rightarrow \gamma \varphi}\right|^2} \underset{m_\varphi \ll m_\psi}{\approx} & \, \frac{\wc_\psi^2 Q_\psi^2 e^2}{2 f_\varphi^2} 
\frac{m_\psi^2 s^2}{( m_{\psi}^2 - t) (s + t - m_{\psi}^2)}
\ , \\
\overline{ \left| \M_{\psi \bar\psi \rightarrow \gamma \varphi}\right|^2}  \underset{m_\psi \ll m_\varphi}{\approx} & \, \frac{\wc_\psi^2 Q_\psi^2 e^2}{2 f_\varphi^2} 
\frac{m_\psi^2 (s^2 + m_{\varphi}^4)}{t \left(m_{\varphi}^2 - s - t \right)}
 \ .
\end{align}
\end{subequations}
The first expression correctly reproduces known results in the literature~\cite{DEramo:2018vss,DEramo:2024jhn}. The second expression, valid in the limit of a negligible fermion mass, vanishes for a massless fermion because the transition amplitude is proportional to \( m_\psi \). This behavior can be understood by tracing the fermion flow in the Feynman diagrams shown in Figs.~\ref{fig:annihilationt} and \ref{fig:annihilationu}, where it becomes evident that a factor of \( m_\psi \) must be picked up from the internal fermion propagators. Alternatively, one can switch to the non-derivative basis and observe that the operator mediating ALP-fermion interactions is itself proportional to the fermion mass.
  
With the squared matrix element in hand, it is straightforward to evaluate the total cross section. We exploit Lorentz invariant to perform the phase space integral in the center of mass frame and write the final result in a manifestly Lorentz invariant form as a function of the Mandelstam variable $s$. The integral over the azimuthal angle is trivial and it gives an overall factor of $2 \pi$. The scattering angle $\theta$ is related to the Mandelstam variable $t$ via the relation
\be\label{eq:tannihilation}
t = m_\psi^2 - \frac{1}{2} (s-m_\varphi^2)\left(1 - \sqrt{1 - \frac{4 m_\psi^2}{s}} \cos\theta \right) \ .
\ee
We make use of \Eq{eq:tannihilation} to make manifest the dependence of the squared matrix element on the center of mass scattering angle, and this allows us to perform the phase space integral in that specific frame. The final result is a function of $s$ only and therefore Lorentz invariant
\be
\begin{split}
& \, \sigma_{\psi \bar\psi \rightarrow \gamma \varphi}(s) =  \frac{\wc_\psi^2 Q_\psi^2 e^2}{4 \pi f_\varphi^2} \frac{m_\psi^2}{s (s - 4 m_\psi^2) (s - m_\varphi^2)} \times \\ &
\quad \quad \times \left[ (s^2 - 4 m_\varphi^2 m_\psi^2 + m_\varphi^4 ) \tanh ^{-1} \left(\sqrt{1-\frac{4 m_\psi^2}{s}}\right) - s\,m_\varphi^2 \sqrt{1-\frac{4 m_\psi^2}{s}} \, \right] \ .
\end{split}
\ee
The annihilation of a fermion-antifermion pair into a photon is the dominant process when $\psi$ is a charged lepton. This process also occurs when $\psi$ is a quark, with an identical cross section since the initial state is a color singlet and averaging over the number of colors does not introduce additional numerical factors. However, it is largely superseded by an analogous process in which the photon is replaced by a gluon, due to the hierarchy between the electromagnetic and strong coupling constants. The corresponding Feynman diagrams have the same structural form, differing only in their color algebra factors. To obtain the cross section for the gluon-mediated process, two modifications are required. First, the squared matrix element in \Eq{eq:MsquaredPSIPAIR} must be rescaled accordingly: we sum over all possible color degrees of freedom and we get an overall factor of $d_{\mathcal{A}} \, T_{\mathcal{R}} = 8 \times 1/2 = 4, $\footnote{Here, we use $d_{\mathcal{A}}$ and $T_{\mathcal{R}}$ to denote the dimension of the adjoint representation of the gauge group and the Dynkin index of the representation $\mathcal{R}$ under which the fermion field transforms, respectively. For quarks charged under the SM strong interactions, these take the values $d_{\mathcal{A}} = 8$ and $T_{\mathcal{R}} = 1/2$.} and the average over all possible colors give another factor of $3 \times 3 \times 8 = 72$ in the denominator. For these reasons, the squared matrix element summed and averaged over both initial and final states results in
\be
\begin{split}
\overline{ \left| \mathcal{M}_{q \bar q \rightarrow g \varphi} \right|^2} = & \, \frac{\wc_\psi^2 g^2_s}{36 f_\varphi^2} \frac{m_\psi^2}{(m_\psi^2-t)^2 (m_\varphi^2+m_\psi^2-s-t)^2} \times \\ &
\left[ 
m_{\psi}^2 \left( -3 s^2 m_{\varphi}^2 + m_{\varphi}^4 (5 s + 2 t) - 3 m_{\varphi}^6 + s^2 (s + 2 t) \right) + \right. \\ & \left.  
- t \left( m_{\varphi}^4 + s^2 \right) \left( -m_{\varphi}^2 + s + t \right) - m_{\psi}^4 \left( m_{\varphi}^4 + s^2 \right) \right] \ .
\end{split}
\label{eq:MsquaredPSIPAIRgluon}
\ee
We also replaced the fermion electric charge with the gauge coupling of strong interactions $g_s$. The calculation of the total cross section is straightforward. We integrate the above expression over the final state phase exactly as we just did for the process with the photon, and we include an additional factor of $8$ to account for the gluon colors in the final state. The resulting cross section reads
 \be
\begin{split}
& \, \sigma_{q \bar q \rightarrow g \varphi}(s) =  \frac{\wc_\psi^2 g_s^2}{9 \pi f_\varphi^2} \frac{m_\psi^2}{s (s - 4 m_\psi^2) (s - m_\varphi^2)} \times \\ &
\quad \quad \times \left[ (s^2 - 4 m_\varphi^2 m_\psi^2 + m_\varphi^4 ) \tanh ^{-1} \left(\sqrt{1-\frac{4 m_\psi^2}{s}}\right) - s\,m_\varphi^2 \sqrt{1-\frac{4 m_\psi^2}{s}} \, \right] \ .
\end{split}
\ee

\begin{figure}[t]
\centering
          \subfloat[\label{fig:annihilationt}]{\begin{tikzpicture}
          \draw[f](-1.,0.75) node[ left]{$\color{ColorT}{\psi}$}--node[midway,above=2mm]{$K_1$}(0,0)[baseline];
          \draw[fb] (-1,-1.75)node[ left]{$\color{ColorT}{\bar\psi}$} -- node[midway,below=1mm]{$K_2$}(0,-1);
          \draw[f] (0,0) -- (0,-1)node[midway,left]{$\color{ColorT}{\psi}$};
          \draw[v] (0,0) -- node[midway,above=2.mm]{$K_3$}(1,0.75)node[ right]{$\color{ColorT}{\gamma}$};
          \draw[a] (0,-1) -- node[midway,below=1mm]{$K_4$}(1,-1.75)node[ right]{$\color{red}{\varphi}$};
          \end{tikzpicture}}\quad
          \subfloat[\label{fig:annihilationu}]{\begin{tikzpicture}
          \draw[f](-1.,0.75) node[left]{$\color{ColorT}{\psi}$}--node[midway,above=1mm]{$K_1$}(0,0)[baseline];
          \draw[f] (0,0) -- (0,-1)node[midway,left]{$\color{ColorT}{\psi}$};
          \draw[fb] (-1,-1.75)node[left]{$\color{ColorT}{\bar\psi}$} --node[midway,below=1mm]{$K_2$} (0,-1); 
          \draw[a] (0,0) --node[midway,right=1.5mm]{$K_4$} (1,-1.25)node[below]{$\color{red}{\varphi}$};
          \draw[v] (0,-1) --node[midway,right=1.5mm]{$K_3$} (1,0.55)node[above]{$\color{ColorT}{\gamma}$};
          \end{tikzpicture}} \quad
        \subfloat[\label{fig:comptons}]{\begin{tikzpicture}
         \draw[f] (-0.75,1) node[left]{$\color{ColorT}{\psi}$}--node[midway, left=1mm]{$K_1$} (0,0);
         \draw[v](-0.75,-1) node[left]{$\color{ColorT}{\gamma}$}--node[midway,left=1mm]{$K_2$}(0,0);
         \draw[f] (0,0) -- (1.25,0)node[midway,above]{$\color{ColorT}{\psi}$};
         \draw[f] (1.25,0) -- node[midway, right=1mm]{$K_3$} (2,1.)node[right]{$\color{ColorT}{\psi}$};
         \draw[a] (1.25,0) --node[midway, right=1mm]{$K_4$} (2.,-1)node[right]{$\color{red}{\varphi}$};
         \path (0,-1.25) node[below,opacity=0] {$$}; 
          \end{tikzpicture}}\quad
          \subfloat[\label{fig:comptonu}]{\begin{tikzpicture}
       \draw[f] (-0.75,1) node[left]{$\color{ColorT}{\psi}$}--node[midway, left=1mm]{$K_1$} (0,0);
         \draw[v](-0.5,-1) node[left]{$\color{ColorT}{\gamma}$}--node[midway,left=8.5mm]{$K_2$}(1.25,0);
         \draw[f] (0,0) -- (1.25,0)node[midway,above]{$\color{ColorT}{\psi}$};
         \draw[a] (0,0) -- node[midway, right=8.5mm]{$K_4$}(1.75,-1)node[right]{$\color{red}{\varphi}$};
         \draw[f] (1.25,0) -- node[midway, right=1mm]{$K_3$} (2,1.)node[right]{$\color{ColorT}{\psi}$};
         \path (0,-1.25) node[below,opacity=0] {$$}; 
          \end{tikzpicture}}   
          \caption{Feynman diagrams illustrating ALP thermalization via fermion scatterings. \Figs{fig:annihilationt}{fig:annihilationu} depict fermion-antifermion pair annihilations, while \Figs{fig:comptons}{fig:comptonu} correspond to Compton-like scatterings. There are also processes with all quarks replaced by antiquarks. If the SM fermion $\psi$ is a quark, analogous processes occur with photons replaced by gluons.}
\label{fig:DFSZthermal}
\end{figure}
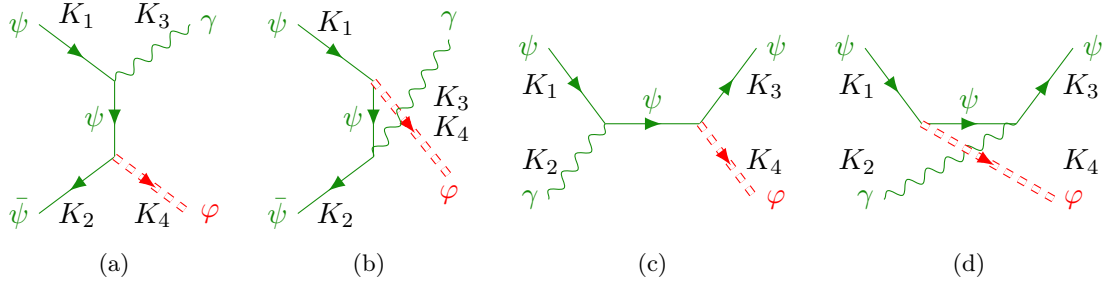

\item \textbf{ALP scattering with fermions II: Compton-like scattering.} Another process contributing to ALP thermalization, which we refer to as Compton-like scattering, is shown in Figs.~\ref{fig:comptons} and \ref{fig:comptonu}. The Feynman diagrams depict the contribution for a fermion $\psi$, but an analogous process occurs when $\psi$ is replaced by its antiparticle $\bar{\psi}$. If $\psi$ is a quark, there are diagrams with an external gluon as well. As before, we first perform the calculation for a final-state photon and then obtain the corresponding result for a gluon by rescaling with color algebra factors.  

The amplitudes for the $s$- and $u$-channel processes are given by  
\begin{subequations}
\begin{align}
\M_{\psi \gamma \rightarrow \psi \varphi} = &\, \M^{(s)}_{\psi \gamma \rightarrow \psi \varphi} + \M^{(u)}_{\psi \gamma \rightarrow \psi \varphi} \ , \\
\M^{(s)}_{\psi \gamma \rightarrow \psi \varphi} = & \, i  \left( \frac{\wc_\psi}{2 f_\varphi} \, Q_\psi e \right) \epsilon_\mu(K_2, \lambda) K_{4 \nu} \;
\bar{u}(K_3)  \gamma^\nu \gamma^5 \frac{\slashed{K_1} + \slashed{K_2} + m_\psi}{s - m_\psi^2} \gamma^\mu u(K_1) 
\ ,\\
\M^{(u)}_{\psi \gamma \rightarrow \psi \varphi} = & \, i \left( \frac{\wc_\psi}{2 f_\varphi} \, Q_\psi e \right) \epsilon_\mu(K_2, \lambda) K_{4 \nu} \;
\bar{u}(K_2) \gamma^\mu \frac{\slashed{K_1} - \slashed{K_4} + m_\psi}{u - m_\psi^2}  \gamma^\nu \gamma^5 u(K_1) \ .
\end{align}
\end{subequations}
We sum the total amplitude and average over all possible initial and final states
\be
\begin{split}
\overline{ \left| \mathcal{M}_{\psi \gamma \rightarrow \psi \varphi} \right|^2} = & \, \frac{\wc_\psi^2 Q_\psi^2 e^2}{2 f_\varphi^2} \frac{m_\psi^2}{(m_\psi^2-s)^2 (m_\varphi^2+m_\psi^2-s-t)^2} \times \\ &
\left[
-m_{\psi }^2 \left(- 3t^2 m_{\varphi }^2 + m_{\varphi }^4 (2s+5t) - 3m_{\varphi }^6 + t^2 (2s+t)\right)  + \right. \\ & \left.
+ s \left(m_{\varphi }^4 + t^2\right) \left(-m_{\varphi }^2 + s + t\right)  
+ m_{\psi }^4 \left(m_{\varphi }^4 + t^2\right)
\right] \ .
\end{split}
\label{eq:MsquaredPSICOMPTON}
\ee
As a check of our calculations, we notice how the squared matrix element in \Eq{eq:MsquaredPSICOMPTON} is related to the one in \Eq{eq:MsquaredPSIPAIR} by crossing symmetry. In other words, the squared matrix element for Compton-like scattering can be obtained from the one of pair annihilation by performing the replacement $s \leftrightarrow t$ with a further overall minus sign. The approximate expressions valid when one particle has a mass significantly larger than the other read
\begin{subequations}
\begin{align}
\overline{ \left| \M_{\psi \gamma \rightarrow \psi \varphi}\right|^2} \underset{m_\varphi \ll m_\psi}{\approx} & \, 
\frac{\wc_\psi^2 Q_\psi^2 e^2}{2 f_\varphi^2}  \frac{m_\psi^2 t^2}{( s - m_{\psi}^2) (s + t - m_{\psi}^2)} \ , \\
\overline{ \left| \M_{\psi \gamma \rightarrow \psi \varphi}\right|^2}  \underset{m_\psi \ll m_\varphi}{\approx} & \, 
\frac{\wc_\psi^2 Q_\psi^2 e^2}{2 f_\varphi^2}  \frac{m_\psi^2 (t^2 + m_{\varphi}^4)}{s \left(s + t - m_{\varphi}^2 \right)}  \ .
\end{align}
\end{subequations}
The first one agrees with results in the literature~\cite{DEramo:2018vss,DEramo:2024jhn}. The phase space integration is performed as in the previous case and the total cross section results in
\be
\begin{split}
\sigma_{\psi \gamma \rightarrow \psi \varphi}(s) = &\,  \frac{\wc_{\psi }^2 Q_{\psi }^2 e^2 }{32 \pi f_{\varphi }^2}
\frac{\sqrt{\lambda(s, m_\psi, m_\varphi)} \, m_\psi^2 }{s^2 (s - m_\psi^2)^3} \times \\ &
\left[ 4 s^2 \frac{\lambda(s, m_\psi, m_\varphi) + m_\varphi^4}{\sqrt{\lambda(s, m_\psi, m_\varphi)}}  \coth^{-1}\left( \frac{s + m_\psi^2 - m_\varphi^2}{\sqrt{\lambda(s, m_\psi, m_\varphi)}} \right) +    \right. \\ & \left. 
+ m_{\varphi}^2 \left(7 s^2 + 2 \, s\, m_{\psi}^2 - m_{\psi}^4 \right) - \left(s - m_{\psi}^2\right)^2 \left(3 s - m_{\psi}^2\right) \right] \ ,
\end{split}
\ee
where we remind the Källén function $\lambda(x, y, z) \equiv [x - (y - z)^2] [x - (y + z)^2]$.

We conclude with the generalization of the previous result when the gauge boson in the initial state is a gluon. As discussed previously, there is no need to perform new calculations but we just need to rescale the squared matrix element in \Eq{eq:MsquaredPSICOMPTON} by accounting for the color degrees of freedom. For this case, the scaling works exactly as in the previous case: summing over all color degrees of freedom brings a multiplicative factor of $d_{\mathcal{A}} \, T_{\mathcal{R}} = 8 \times 1/2 = 4$, and to average over both initial and final states we divide this number by $3 \times 3 \times 8 = 72$. The squared matrix element reads
\be
\begin{split}
\overline{ \left| \mathcal{M}_{q g \rightarrow q \varphi} \right|^2} = & \, \frac{\wc_\psi^2 g_s^2}{36 f_\varphi^2} \frac{m_\psi^2}{(m_\psi^2-s)^2 (m_\varphi^2+m_\psi^2-s-t)^2} \times \\ &
\left[
-m_{\psi }^2 \left(- 3t^2 m_{\varphi }^2 + m_{\varphi }^4 (2s+5t) - 3m_{\varphi }^6 + t^2 (2s+t)\right)  + \right. \\ & \left.
+ s \left(m_{\varphi }^4 + t^2\right) \left(-m_{\varphi }^2 + s + t\right)  
+ m_{\psi }^4 \left(m_{\varphi }^4 + t^2\right)
\right] \ .
\end{split}
\label{eq:MsquaredPSICOMPTONgluon}
\ee
The phase space integration is identical to the previous case, and it is important to remember to sum also over the final color state. This operation brings and additional factor of $3$, and the total cross section reads 
\be
\begin{split}
\sigma_{q g \rightarrow q \varphi}(s) = &\,  \frac{\wc_{\psi }^2 g_s^2}{192 \pi f_{\varphi }^2}
\frac{\sqrt{\lambda(s, m_\psi, m_\varphi)} \, m_\psi^2 }{s^2 (s - m_\psi^2)^3} \times \\ &
\left[ 4 s^2 \frac{\lambda(s, m_\psi, m_\varphi) + m_\varphi^4}{\sqrt{\lambda(s, m_\psi, m_\varphi)}} \coth^{-1}\left( \frac{s + m_\psi^2 - m_\varphi^2}{\sqrt{\lambda(s, m_\psi, m_\varphi)}} \right) +    \right. \\ & \left. 
+ m_{\varphi}^2 \left(7 s^2 + 2 \, s\, m_{\psi}^2 - m_{\psi}^4 \right) - \left(s - m_{\psi}^2\right)^2 \left(3 s - m_{\psi}^2\right) \right] \ .
\end{split}
\ee

\end{itemize}

\bibliographystyle{JHEP}
\bibliography{ALPortalS3}

\end{document}